\documentclass{aa}
\pdfoutput=1
\usepackage{graphicx, multirow, tablefootnote, longtable}
\usepackage{txfonts, listings}
\usepackage{natbib}
\usepackage{xcolor, tikz}
\usepackage[percent]{overpic}
\usepackage[colorlinks=true, citecolor=blue]{hyperref}
\usepackage{natbib}
\bibpunct{(}{)}{;}{a}{}{,}
\defcitealias{Queiroz2018}{Q18}
\defcitealias{Queiroz2020}{Q20}
\defcitealias{Santiago2016}{S16}
\usepackage{tablefootnote}
\definecolor{lime}{HTML}{A6CE39}

\DeclareRobustCommand{\orcidicon}{%
    \begin{tikzpicture}
    \draw[lime, fill=lime] (0,0) 
    circle [radius=0.16] 
    node[white] {{\fontfamily{qag}\selectfont \tiny ID}};
    \draw[white, fill=white] (-0.0625,0.095) 
    circle [radius=0.007];
    \end{tikzpicture}
    \hspace{-2mm}
}

\begin{document}

%  \title{{\tt StarHorse} results for spectroscopic surveys with {\it Gaia} EDR3: Scanning local samples with t-SNE and ages}

 \title{{\tt StarHorse} results for spectroscopic surveys + {\it Gaia} DR3: \\
 Chrono-chemical populations in the solar vicinity, the genuine thick disk, and young-alpha rich stars}

   \author{A. B. A. Queiroz\href{https://orcid.org/0000-0001-9209-7599}{\orcidicon}\inst{1,2,3},
   F. Anders\href{https://orcid.org/0000-0003-4524-9363}{\orcidicon}\inst{4,5,6},
   C. Chiappini\href{https://orcid.org/0000-0003-1269-7282}{\orcidicon}\inst{1,3},
   A. Khalatyan\inst{1},
   B. X. Santiago\inst{7,3},\\
   S. Nepal\href{https://orcid.org/0000-0002-8557-5684}{\orcidicon}\inst{1,2},
   M. Steinmetz\href{https://orcid.org/0000-0001-6516-7459}{\orcidicon}\inst{1},
   C. Gallart\href{https://orcid.org/0000-0001-6728-806X}{\orcidicon}\inst{8,9},
   M. Valentini\inst{1},
   M. Dal Ponte\inst{7,3},
   B. Barbuy\href{https://orcid.org/0000-0001-9264-4417}{\orcidicon}\inst{10},\\
   A. P\'erez-Villegas\href{https://orcid.org/0000-0002-5974-3998}{\orcidicon}\inst{11},
   T. Masseron\href{https://orcid.org/0000-0002-6939-0831}{\orcidicon}\inst{8,9},
   Jos\'e G. Fern\'andez-Trincado\href{https://orcid.org/0000-0002-0900-9760}{\orcidicon}\inst{12},
   S. Khoperskov\inst{1},\\
   I. Minchev\href{https://orcid.org/0000-0002-5627-0355}{\orcidicon}\inst{1},
   E. Fern\'andez-Alvar\href{https://orcid.org/0000-0001-8994-0936}{\orcidicon}\inst{8,9},
   Richard R. Lane\href{https://orcid.org/0000-0003-1805-0316}{\orcidicon}\inst{13},
   C. Nitschelm\href{https://orcid.org/0000-0003-4752-4365}{\orcidicon}\inst{14}
}

   \authorrunning{A. Queiroz et al.}
   \titlerunning{{\tt StarHorse} results for spectroscopic surveys with {\it Gaia} DR3}
    \institute{Leibniz-Institut f\"ur Astrophysik Potsdam (AIP), An der Sternwarte 16, 14482 Potsdam, Germany\\
              \email{aqueiroz@aip.de}
    \and{Institut f\"{u}r Physik und Astronomie, Universit\"{a}t Potsdam, Haus 28 Karl-Liebknecht-Str. 24/25, D-14476 Golm, Germany}
    \and{Laborat\'orio Interinstitucional de e-Astronomia - LIneA, Rua Gal. Jos\'e Cristino 77, Rio de Janeiro, RJ - 20921-400, Brazil}
    \and{Dept. de Física Quàntica i Astrofísica (FQA), Universitat de Barcelona (UB),  C Martí i Franqués, 1, 08028 Barcelona, Spain}
    \and{Institut de Ciències del Cosmos (ICCUB), Universitat de Barcelona (UB), C Martí i Franqués, 1, 08028 Barcelona, Spain}
    \and{Institut d'Estudis Espacials de Catalunya (IEEC), C Gran Capità, 2-4, 08034 Barcelona, Spain}
        \and{Instituto de F\'\i sica, Universidade Federal do Rio Grande do Sul, Caixa Postal 15051, Porto Alegre, RS - 91501-970, Brazil}
        \and{Instituto de Astrof\'isica de Canarias, E-38205 La Laguna, Tenerife, Spain}
        \and{departamento de Astrof\'isica, Universidad de La Laguna, E-38200 La Laguna, Tenerife, Spain}
        \and{Department of Astronomy, Universidade de S\~ao Paulo, S\~ao Paulo 05508-090, Brazil}
        \and{Instituto de Astronom\'ia, Universidad Nacional Aut\'onoma de M\'exico, A. P. 106, C.P. 22800, Ensenada, B. C., M\'exico}
        \and{Instituto de Astronom\'ia, Universidad Cat\'olica del Norte, Av. Angamos 0610, Antofagasta, Chile}
        \and{Centro de Investigaci\'on en Astronom\'ia, Universidad Bernardo O'Higgins, Avenida Viel 1497, Santiago, Chile}
        \and{Centro de Astronom\'ia, Universidad de Antofagasta, Avenida Angamos 601, Antofagasta 1270300, Chile}
 }

   \date{Received \today; accepted xx.yy.20zz}
\abstract
{The {\it Gaia} mission has provided an invaluable wealth of astrometric data for more than a billion stars in our Galaxy. The synergy between {\it Gaia} astrometry, photometry, and spectroscopic surveys give us comprehensive information about the Milky Way. Using the Bayesian isochrone-fitting code {\tt StarHorse}, we derive distances and extinctions for more than 10 million unique stars observed by both {\it Gaia} Data Release 3 as well as public spectroscopic surveys: 557\,559 in GALAH+ DR3, 4\,531\,028 in LAMOST DR7 LRS, 347\,535 in LAMOST DR7 MRS, 562\,424 in APOGEE DR17, 471\,490 in RAVE DR6, 249\,991 in SDSS DR12 (optical spectra from BOSS and SEGUE), 67\,562 in the {\it Gaia}-ESO DR5 survey, and 4\,211\,087 in the {\it Gaia} RVS part of {\it Gaia} DR3 release. {\tt StarHorse} can extend the precision of distances and extinctions measurements where {\it Gaia} parallaxes alone would be uncertain. 
We use \texttt{StarHorse} for the first time to derive stellar age for main-sequence turnoff and subgiant branch stars (MSTO-SGB), around 2.5 million stars with age uncertainties typically around 30\%, 15\% for only SGB stars, depending on the resolution of the survey. With the derived ages in hand, we investigate the chemical-age relations. In particular, the $\alpha$ and neutron-capture element ratios versus age in the solar neighbourhood show trends similar to previous works, validating our ages. We use the chemical abundances from local subgiant samples of GALAH DR3, APOGEE DR17 and LAMOST MRS DR7 to map groups with similar chemical compositions and {\tt StarHorse} ages with the dimensionality reduction technique t-SNE and the clustering algorithm HDBSCAN. We identify three distinct groups in all three samples. Their kinematic properties confirm them to be the genuine chemical thick disk, the thin disk and a considerable number of young alpha-rich stars (427), which are also a part of the delivered catalogues. We confirm that the genuine thick disk's kinematics and age properties are radically different from those of the thin disk and compatible with high-redshift (z$\approx$2) star-forming disks with high dispersion velocities. We also find a few extra chemical populations in the GALAH DR3, thanks to the availability of neutron-capture elements.%, which can be associated with the passage of Sagittarius 6.5 Gyr ago.
}
\maketitle

\section{Introduction}\label{intro}

The European Space Agency satellite {\it Gaia} mission \citep{GaiaCollaboration2016a} is continuing to revolutionize and transform Galactic astrophysics in many areas \citep{Brown2021}. 
The latest release from the {\it Gaia}-mission, ({\it Gaia} DR3; \citealt{GaiaCollaboration2022}) is built upon the Early Data Release 3 \citep[EDR3][]{GaiaCollaboration2021} which includes 36 months of observations, with positions and photometry for $1.7\cdot 10^9$ sources, and full astrometric solutions, \citep{Lindegren2021a}, for $1.3\cdot 10^9$ objects. {\it Gaia} DR3 extends EDR3 by delivering multiple data products, for example, low-resolution BP/RP spectra and astrophysical parameters for about 400 million sources \citep{Andrae2022} and about 5 million sources with medium resolution spectra observed with the Radial Velocity Spectrometer (RVS) instrument \citep{Recio-Blanco2022}.
Combining astrometric solutions from {\it Gaia} with large-scale spectroscopic surveys is fundamental for Galactic archaeology because it enables us to access the full phase space and the chemical composition of millions of stars. Such rich information gives us essential clues to the formation and evolution history of the Milky Way \citep{Freeman2002, Matteucci2001, Matteucci2021, Pagel2009}, disentangling the multiple overlapping processes that once took place in our Galaxy, such as mergers, secular evolution, and gas accretion flows.

The synergy between astrometry and spectroscopy resulted in many important discoveries in the different components of our Galaxy. As proof of that, we have the characterization of the halo and the discovery of several accreted dwarf galaxies (e.g. \citealt{Koppelman2018, Mackereth2019, Myeong2019, Limberg2021, Trincado2020b, Trincado2020d, Trincado2022, Horta2020, Ruiz-Lara2022}) and the massive {\it Gaia-}Enceladus merger event \citep{Haywood2018b, Helmi2018, Belokurov2018}. These structures substantially influence the formation of the thick disk and halo (for a review, see \citealt{Belokurov2018, DiMatteo2019, Helmi2020}). 
The chemical duality of the Galactic disk, which was primarily evident in [$\alpha$/Fe] vs. [Fe/H] in the solar neighbourhood, was shown by several authors to designate the chemical thin and thick disks \citep{Adibekyan2011, Bensby2014, Anders2014, Hayden2015}. Further, \citet{Rojas-Arriagada2019, Queiroz2020} show that the same chemical bimodality extends to the inner Galaxy, indicating populations with different formation paths.
Finally, the characterization of the Galactic bulge and bar \citep{Bovy2019, Lian2020, Rojas-Arriagada2020, Queiroz2021} into its chemo-orbital space reveals a diversity of populations coexisting in the inner Galaxy. Recently works that studied the inner Galaxy's metal-poor counterpart show evidence of a pressure-supported component which follows a more spherical distribution than the disk and little to no rotation \citep{Kunder2020, Arentsen2020, Lucey2021, Rix2022}.

To achieve all the aforementioned scientific results is essential to calculate precise distances from the astrometric solutions provided by {\it Gaia}. As shown by \citet{Bailer-Jones2015} it is a limited and dangerous approach to determine distances by inverting the parallax, especially for high astrometric uncertainties and large volumes of the Galaxy.
In \citet[][hereafter Q18]{Queiroz2018}, we first presented the {\tt StarHorse} code: a Bayesian isochrone fitting tool that makes versatile use of spectroscopic, photometric, and astrometric data to determine distances, extinctions, and stellar parameters of field stars. The method was then extensively validated using simulations and external catalogues of asteroseismology, open clusters and binaries. Therefore, in \citet[][hereafter Q20]{Queiroz2020} many efforts were put together to provide catalogues generated from {\tt StarHorse} using {\it Gaia} DR2 data with APOGEE DR16 and other spectroscopic surveys, resulting in an important leap in stellar parameter precision.

In this paper, we provide updated {\tt StarHorse} stellar parameters, distances, and extinctions for major spectroscopic surveys (see Table \ref{summarytable}) combined with the {\it Gaia} DR3 data. The {\tt StarHorse} results for APOGEE DR17 \citep{sdss2021} are already published in the form of a value-added catalogue jointly with SDSS DR17, except for the ages, which are published here for the first time.
The paper focuses on science enabled by sub-samples for which {\tt StarHorse} delivers reasonable age estimates thanks to the exquisite quality of the {\it Gaia} parallaxes. However, the results are limited to a local volume bubble of d$<2$kpc since ages derived by isochronal matching can only be reliable for the main sequence turn-off (MSTO), and subgiant branch (SGB) regimes, the degeneracies between neighbouring isochrones in the Hertzsprung-Rusell diagram are much smaller for these cases.

%Most analyses of chemical substructures have so far focused on kinematic selections and the very metal-poor regime, where the accreted debris stand out from the disc components \citep[e.g.][]{Limberg2021, Limberg2021a, Simpson2021, Ruiz-Lara2022}.
%Most analyses of disk and halo substructures have so far focused on kinematic or chemical selections, where the accreted debris stand out from the disc components \citep[e.g.][]{Limberg2021, Limberg2021a, Simpson2021, Ruiz-Lara2022}.
%Since, in this work, the age information is only available for a volume dominated by the metal-rich thin disk, we must resort to a different approach.
In this work, we take advantage of the rich chemical information delivered by spectroscopic surveys combined with {\tt StarHorse} ages to explore the detection of (known and new) chrono-chemical subgroups more in the line of classical "chemical tagging"; see, e.g. \citealt{Freeman2002, Anders2018, Buder2022}. To this aim, we use three different survey samples to map groups with similar chemical compositions.% using a combination of the dimensionality reduction technique t-SNE and HDBSCAN clustering.

This paper is outlined as follows. In section \ref{method}, we summarize the Bayesian technique {\tt StarHorse}, the references for its newest implementations and main configuration. In section \ref{in2sh}, we describe all datasets we used as input to the {\tt StarHorse} code, astrometric, photometric and spectroscopic data. In section \ref{out2sh}, we discuss the main parameters derived with our method, the newly released {\tt StarHorse} catalogues, which contain more than 10 million stars and 2.5 million nearby stars with ages, and a few validations of the parameters. In section \ref{agerel}, we show relations between the derived ages and some chemical relations. In section \ref{tsne}, we show our results using the chemo-age multi-dimension in the t-sne and HDBSCAN technique. Finally, in section \ref{discuss}, we present our new conclusions and summarize our main results. All the catalogues used in this work are made public in the Leibniz Instititute f\"ur Astrophysik (AIP) database\footnote{\url{data.aip.de}}.

\section{Method}\label{method}

\begin{figure}
\includegraphics[width=0.5\textwidth]{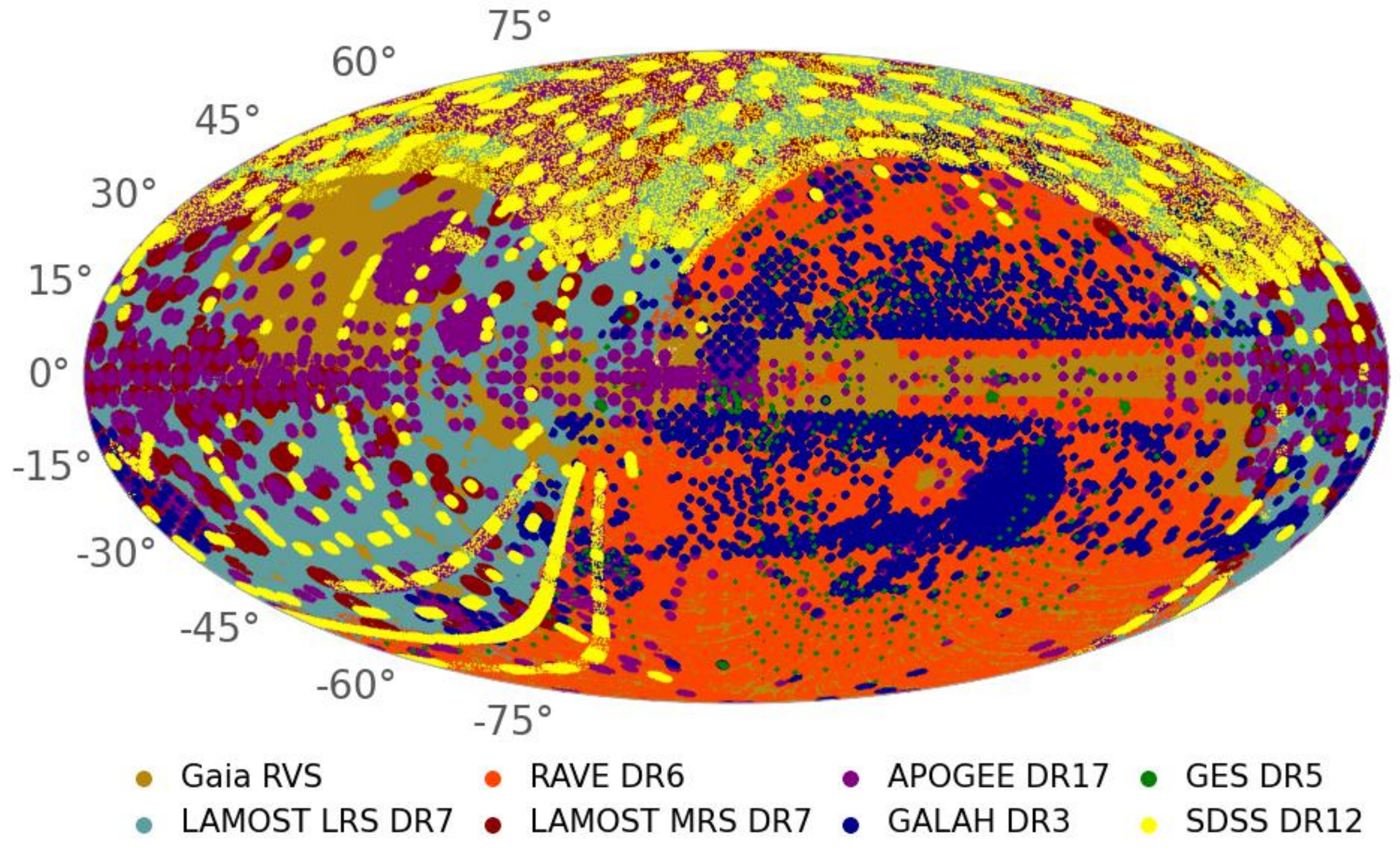}
\caption{Sky distribution of all public spectroscopic surveys for which we derive {\tt StarHorse} parameters.}
\label{fig:footprint}
\end{figure}

\begin{table*}
\centering
\caption{Summary of the datasets for which we deliver {\tt StarHorse} parameters in this work. Upper rows show the main numbers of input catalogue sources, those that survived the quality cuts, those for which the code converged to a solution, and the number of MSTO and subgiants selected on the output. The lower rows show the model configuration and parallax coverage for the final input catalogues.}
\begin{tabular}{lrcrcr}
Survey & $N_{\rm objects}^{\rm cat}$ &  $N_{\rm objects}^{\rm Quality\ cuts}$  &  $N_{\rm stars}^{\rm Converged} $ & $N_{\rm stars}^{MSTO}$ & $N_{\rm stars}^{\rm subgiants}$\\
\hline
LAMOST DR7 LRS  & 6\,179\,327 & 4\,803\,496 & 4\,531\,028 & 1\,206\,381 & 131\,845 \\
LAMOST DR7 MRS & 738\,025 & 457\,359 & 425\,281 & 106\,125  & 15\,547 \\
SDSS DR12 (optical) & 503\,967 & 258\,194 & 249\,991 & 68\,391 & 13\,584 \\
GALAH DR3 & 588\,571 &  581\,149 & 557\,559 & 127\,016 & 26\,318 \\
RAVE DR6   & 517\,095 &  515\,800 & 471\,490 & 94\,765 & 23\,809\\
APOGEE DR17 & 733\,901 &  720\,970 & 562\,424 & 61\,331 & 15\,060 \\
GES DR5 & 114\,324 & 75\,008 & 67\,562 & 11\,542 & 3\,320 \\
{\it Gaia} DR3 RVS & 5\,594\,205 & 4\,833\,548  & 4\,211\,087 & 799\,149  & 178\,719 \\
\hline
\hline
Survey & \multicolumn{2}{c}{Model Resolution} & bestfilter & parallaxes \\
LAMOST DR7 LRS  & \multicolumn{2}{c}{age$_{step}$ = 0.1 ; met$_{step}$ = 0.05} & $Ks_{2MASS}$ & 99\%\\
LAMOST DR7 MRS & \multicolumn{2}{c}{age$_{step}$ = 0.05 ; met$_{step}$ = 0.02} & $Ks_{2MASS}$ & 77\%\\
SDSS DR12 (optical) & \multicolumn{2}{c}{age$_{step}$ = 0.1 ; met$_{step}$ = 0.05} & parallax & 100\%\\
GALAH+DR3 & \multicolumn{2}{c}{ age$_{step} = 0.05$ ; met$_{step} = 0.02$ } & $Ks_{2MASS}$ & 98\%\\
RAVE DR6   &  \multicolumn{2}{c}{age$_{step}$ = 0.1 ; met$_{step}$ = 0.05} & $Ks_{2MASS}$ & 86\%\\
APOGEE DR17 & \multicolumn{2}{c}{age$_{step}$ = 0.05 ; met$_{step}$ = 0.02} & $H_{2MASS}$ &75\%\\
GES DR5 & \multicolumn{2}{c}{age$_{step}$ = 0.05 ; met$_{step}$ = 0.02} & $H_{2MASS}$ &97\%\\
{\it Gaia} DR3 RVS & \multicolumn{2}{c}{age$_{step}$ = 0.1 ; met$_{step}$ = 0.05} & $G_{Gaia}$ &100\%\\
\hline
\end{tabular}
\label{summarytable}
\end{table*}

Isochrone fitting has been extensively used in astronomy to indirectly derive unknown stellar parameters by using known measured stellar properties \citep[e.g.][]{Pont2004, Jorgensen2005, daSilva2006, Naylor2006}. A diversity of methods can be applied to the fitting procedure, \citep[e.g.][]{Burnett2010, Rodrigues2014, Santiago2016, Mints2018, Das2019, Lebreton2020, Souza2020}. Here we use {\tt StarHorse} (\citetalias{Santiago2016,Queiroz2018, Queiroz2020}, \citealt{Anders2019, Anders2022}), a Bayesian isochrone-fitting code that has been optimised for heterogeneous input data (including spectroscopy, photometry, and astrometry). Its results are limited only by observational errors and the accuracy of the adopted stellar evolution models.

{\tt StarHorse} is able to derive distances $d$, extinctions $A_V$ (at $\lambda$ = 542 nm), ages $\tau$, masses $m_\ast$, effective temperatures $T_{\rm eff}$, metallicities [M/H], and surface gravities $\log g$. 
The resulting parameter's uncertainties are directly linked with the set of observables used as input. A complete set of observables comprises multi-band photometry (from blue to mid-infrared wavelengths), parallax, $\log g$, T$_{\rm eff}$, $[M/H]$, and an extinction prior A$_{v}$.
In this work, we use all this information by combining data from public spectroscopic surveys with photometric surveys and {\it Gaia} parallaxes. We then execute the Bayesian technique to quantitatively match the observable set with stellar evolutionary models from the PAdova and TRiestre Stellar Evolution Code \citep[PARSEC][]{Bressan2012}, ranging from 0.025 to 13.73 Gyr in age and $-2.2$ to +0.6 in metallicity.

Since \citetalias{Queiroz2020}, {\tt StarHorse} has seen several upgrades that are explained in Section 3 of \citet{Anders2022}. These upgrades include the implementation of extragalactic and globular cluster priors, a change in the bar-angle prior (to the canonical value of 27 degrees; e.g. \citealt{Bland-Hawthorn2016}), a new 3D extinction prior, and updated evolutionary models that include diffusion (especially important during the evolutionary phases close to the MSTO). Finally, the new catalogues presented here also take advantage of the more precise and additional data products of {\it Gaia} DR3.

\section{Input data}\label{in2sh}

The large set of available spectroscopic surveys gives us detailed information about individual stars, such as chemical abundances, atmospheric parameters and radial velocities. By combining this information with photometry and astrometry, we can constrain models by a small range of limits and effectively derive the best fitting {\tt StarHorse} parameters with low uncertainties. 

We follow a very similar approach to previous {\tt StarHorse} papers (\citetalias{Queiroz2018, Queiroz2020}). In Table \ref{summarytable}, we summarize the input numbers of stars for each spectroscopic survey, the stars remaining after applying a few quality cuts, the resulting number of converged stars, and the following amount of MSTO and SBG stars with available {\tt StarHorse} ages. The quality cuts applied before executing {\tt StarHorse} vary from survey to survey, and a more detailed explanation is given in the following sub-sections. As regard to model grid resolution and the photometric passband that we used as the "bestfilter" are also described in the lower rows of Table \ref{summarytable}. The age$_{step}$ and met$_{step}$ represent the spacing between age and metallicity in the models we use in the \texttt{StarHore} method, for all cases the age$_{step}$ is linear; for higher resolution surveys, we use a thinner model grid. The "bestfilter" is the primary choice from which we draw the possible distance values. For more information, see section 3.2.1 of \citetalias{Queiroz2018}.

In Figure \ref{fig:footprint} we show the sky distribution for all public spectroscopic surveys for which derive {\tt StarHorse} parameters, the area coverage of the surveys is very complementary and focus on the different components of our Galaxy.
Below, we summarize the configurations and calibrations done to all input data, the spectroscopic surveys and the photometric and astrometric catalogues used in this work.

\subsection{Astrometric and photometric input}\label{edr3}

{\tt Gaia} is an astrometric and photometric space mission from ESA launched in 2013 and which since then has delivered parallaxes and proper motions for more than 1 billion sources. Its early third release EDR3  \citep{GaiaCollaboration2021} has astrometric solutions with uncertainties twice as better than its previous DR2 release.
All resulting catalogues given in this paper were produced by combining the spectroscopic surveys with parallaxes from {\it Gaia} EDR3, which is an important new ingredient for the resulting {\tt StarHorse} distances. We use the parallax corrections advertised by \citet{Lindegren2021a}, and the most conservative parallax uncertainty inflation factor derived in the analysis of \citet{Fabricius2021}, see their Fig. 19. Besides these corrections, we crossmatched our catalogues with the {\tt fidelity\_v1} column from \citet{Rybizki2021}, which provides a scalar indicator for astrometric quality. For fidelities $<0.5$, we do not use any parallax information. In the last column of the lower rows of Table \ref{summarytable} we show the coverage percentage of available parallaxes for the input catalogues that pass this condition.

As photometric input, we use infra-red photometry from 2MASS $JHKs$ \citep{Cutri2003} and unWISE $W1W2$ \citep{Schlafly2019}, optical data from PanSTARRS-1 $grizy$ \citep{Scolnic2015}, and SkyMapper DR2 $griz$ \citep{Onken2019}, adopting generous minimum photometric uncertainties (between 0.03 and 0.08 mag). Magnitude shifts were applied to PanSTARRS-1 as in \citetalias{Queiroz2020} using the values from \citep{Scolnic2015} and shift corrections were also applied to SkyMapper passbands according to \citet{Yang2021}.

\subsection{Spectroscopic catalogues}
\label{cats}

We compute posterior ages, masses, temperatures, surface gravities, distances, and extinctions for eight spectroscopic stellar surveys. We have crossmatched all spectroscopic surveys with Gaia EDR3 using the stilts CDSskymatch tool\footnote{http://www.star.bris.ac.uk/ mbt/stilts/sun256/cdsskymatch.html}. We proceeded with a 1.5 arcsec search radius for this, and with the setup as "find=each" this configuration set the best match (best distance) for each row or blank when there was no match. We have not used any previous {\it Gaia} crossmatches done by the spectroscopic surveys with their sources since we can be consistent amongst our results by doing our own crossmatch. The photometric surveys such as PanSTARRS and Skymapper, which are already crossmatched with {\it Gaia} at the {\it Gaia} archive, are downloaded by their source id. For unWise, we have crossmatched all surveys with the same configuration as {\it Gaia} in the stilts CDS-skymatch tool. 
For all catalogues, we use the \citet{Salaris1993} transformation between [Fe/H] and [M/H] for stars with valid [$\alpha$/Fe] values. For those without a reported [$\alpha$/Fe] ratio, we assumed [M/H]$\simeq$[Fe/H]. The data curation applied for each survey is explained in the following subsections and the resulting numbers of stars are given in Table \ref{summarytable}. We want to clarify that from each survey's uncertainty distribution, we usually remove a small fraction with substantial input observable uncertainties compared to the full distribution. We do so because it is computationally very costly to calculate the likelihood for many models inside an extensive uncertainty range. The threshold of acceptable uncertainties to \texttt{StarHorse} changes with the choice of the model grid - high-resolution surveys typically have minor uncertainties, requiring a denser model grid.

\subsubsection{APOGEE DR17}

DR17 \citet{sdss2021} is the final data release of the fourth phase of the Sloan Digital Sky Survey \citep[SDSS IV][]{Blanton2017}. It contains the complete catalogue of the Apache Point Observatory Galactic Evolution Experiment \citep[APOGEE;][]{Majewski2017} survey, which in December 2021 publicly released near infra-red spectra of over 650,000 stars. The APOGEE survey has been collecting data in the northern hemisphere since 2011 and south hemisphere since 2015. Both hemispheres observations use the twin NIR spectrographs with high resolution (R$\approx$22\,500) \citep{Wilson2019} on the SDSS 2.5-m telescope at Apache Point Observatory \citep{Gunn2006} and the 2.5-m du Pont telescope at Las Campanas Observatory \citep[LCO][]{Bowen1973}. The data reduction pipeline is described in \citep{Nidever2015}.
The processed products of APOGEE DR17 are similar to the previous releases \citep{Abolfathi2018,Holtzman2018,Jonsson2020}. We use the temperature, surface gravity and metallicity results from the ASPCAP pipeline \citep{Garcia-Perez2016,Jonsson2020} to produce a new {\tt StarHorse} catalogue as in \citetalias{Queiroz2020}. We use primarily the calibrated parameters indicated in the pipeline, when those are not available we use spectroscopic parameters. For DR17 new synthetic spectral grids were added in ASPCAP, which also account for non-local thermodynamic equilibrium in some elements. This led to the adoption of a different spectral synthesis code Synspec \citep{Hubeny2017}. Parameters reduction is also available with the previous spectral systhesis code TurboSpectrum \citep{Alvarez1998} but in {\tt StarHorse} we only used the given parameters from Synspec. In the appendix \ref{apoabu} we show some differences between the derived abundances in Synspec and TurboSpectrum, which are discussed later in the analysis.

As an input to {\tt StarHorse}, we selected only stars with available H$_{2MASS}$ passband and spectral parameters (FPARAM[0,1,3])\footnote{Output parameter array from ASPCAP stellar parameters fit, where 0, 1, 3 correspond to T$_{\rm eff}$, logg and [M/H]}, which reduces the total number of objects in the initial catalogue from 733\,901 to 720\,970. We then run {\tt StarHorse} with a spacely fine model grid (see lower rows of Table \ref{summarytable}). 22\% of the input did not converge to a solution, meaning that these stars were incompatible with any stellar evolutionary model in our grid. The results for {\tt StarHorse} using the data from APOGEE DR17 are also published in form of a value-added catalogue (VAC) in \citet{sdss2021}.

\subsubsection{GALAH DR3}
The Galactic Archaeology with HERMES survey \citep[GALAH,][]{deSilva2015, Martell2017} is a high-resolution spectroscopic survey that covers mostly a local volume, $d< \approx 2kpc$. Their latest data release, GALAH DR3, was published in November 2020. GALAH data are acquired with the High Efficiency and Resolution Multi-Element Spectrograph (HERMES), where the light is dispersed at $R \approx$ 28\,000, coupled to the 3.9-metre Anglo-Australian Telescope (AAT). HERMES observes in four different wavelengths simultaneously: Blue: 471.5 - 490.0 nm; Green: 564.9 - 587.3 nm; Red: 647.8 - 673.7 nm; IR: 758.5 - 788.7 nm.
We use the recommended catalogue, which contains radial velocities, atmospheric parameters and abundances for a total of 588,571 stars \citep{Buder2021}. The stellar parameters are derived using the spectrum synthesis code Spectroscopy Made Easy (SME) and 1D marcs model atmospheres \citep{Piskunov2017}. GALAH makes available abundances for around 30 different elements, which cover five different nucleosynthetic pathways ($\alpha$-process elements mostly formed by core-collapse supernovae, iron-peak elements formed mainly in type-Ia supernovae, s-process elements formed in the late-life stage of low mass stars, r-process elements formed by the merging of neutron stars, as well as lithium (created by the Big Bang and both created and destroyed in stars; \citealt{Kobayashi2020}). To derive {\tt StarHorse} parameters, we selected only stars with mutually available $T_{\rm eff}$, $\log{g}$ and K$_{2mass}$ passband as input. The coverage of high quality {\it Gaia} parallaxes for this sample is very high since most stars are nearby. Therefore, the resulting distances have very low uncertainties, as seen in Figure \ref{fig:allunc}. For GALAH, we run {\tt StarHorse} with a fine model grid, given the high resolution of the survey, and only 5\% of the input catalogue did not converge.
\subsubsection{LAMOST DR7}
The Large Sky Area Multi-Object Fiber Spectroscopic Telescope \citep[LAMOST,][]{Cui2012, Zhao2012} is a spectroscopic survey covering a large area of the northern hemisphere, including stars and galaxies. LAMOST stellar parameter catalogues can be divided into LAMOST low-resolution (LRS) and medium-resolution (MRS). LAMOST, DR7, has been publicly available since March 2020 and includes the spectra obtained from the pilot survey through the seventh-year regular survey. We downloaded the stellar parameter catalogues both for LRS (6\,179\,327) and MRS (738\,025)\footnote{\url{http://dr7.lamost.org/catalogue}}. A new LAMOST data release, DR8, is available since September 2022 and contains circa 500 thousand new observations in LRS and more 500 thousand in MRS. We will also make \texttt{StarHorse} parameters publicly available for this data release in the near future, but LAMOST DR8 is not part of the analysis in this paper.

Both MRS and LRS stellar parameter catalogues provide atmospheric parameters, metallicity, and projected rotation velocity estimated by the LAMOST Stellar Parameter pipeline \citep[LASP][]{Wu2014}, as well as an estimate of alpha abundances by the method of template matching based on the MARCS synthetic spectra \citep{Decin2004}. For LAMOST MRS, coadd and single exposure spectra have a resolution of $R \approx$ 7\,500. The label-transfer method gives twelve individual element abundances based on a convolutional neural network (CNN). For MRS stellar parameter catalogue we selected all-stars with mutually available $\log{g}$, $T_{\rm eff}$, and 2MASS $K_s$ photometry, and made the following cuts in uncertainty: $\sigma_{T_{\rm eff}}<300 K$; $\sigma_{\log g}<0.5 K$; $\sigma_{[Fe/H]}<0.3 K$. This leaves us with 457\,359 stars as {\tt StarHorse} input. As we did for the high-resolution surveys, we ran LAMOST MRS with the fine model grid, which leads to a convergence rate of 93\%.
The LAMOST LRS parameter catalogue, the largest dataset in this work, consists of A, F, G and K type stars. We selected stars with available $\log{g}$, $T_eff$, 2MASS Ks passband and made the following cuts in uncertainty: $\sigma_{T_{\rm eff}}<500 K$; $\sigma_{\log g}<0.8 K$; $\sigma_{[Fe/H]}<0.5 K$, resulting in 4\,803\,496 stars as input. Using a coarsely spaced grid of models for LAMOST LRS,  {\tt StarHorse} was able to deliver results for 80\% of this input catalogue. 

\subsubsection{SDSS DR12/SEGUE}
The Sloan Extension for Galactic Understanding and Exploration \citep[SEGUE][]{Yanny2009} is a spectroscopic survey that was conducted with the Sloan Foundation 2.5m Telescope \citep{Gunn2006} using the two original low-resolution SDSS fibre spectrographs \citep[$R\approx 2\,000$,][]{Smee2013}. The surveys targeted mostly metal-poor halo and disk stars. The stellar parameters from optical stellar spectra collected with SDSS/SEGUE were processed through the SEGUE Stellar Parameter Pipeline (SSPP), which reports three primary stellar parameters, $T_{\rm eff}$, $\log{g}$, and metallicity. Most stars have $T_{\rm eff}$ in the range between 4\,000 and 10\,000 K and spectral signal-to-noise ratios greater than 10 \citep{Lee2008, Allende-Pietro2008}. In the final data release (SDSS DR12; \citealt{Alam2015}), the pipeline also provided [$\alpha$/Fe] abundance ratios \citep{Lee2011}. From this catalogue, we use the recommended adopted values for $T_{\rm eff}$, $\log{g}$ and [Fe/H], selecting only stars with signal-to-noise ratios greater than 20 and that have all of these parameters available.

\begin{figure*}
\centering
\includegraphics[width=0.95\textwidth]{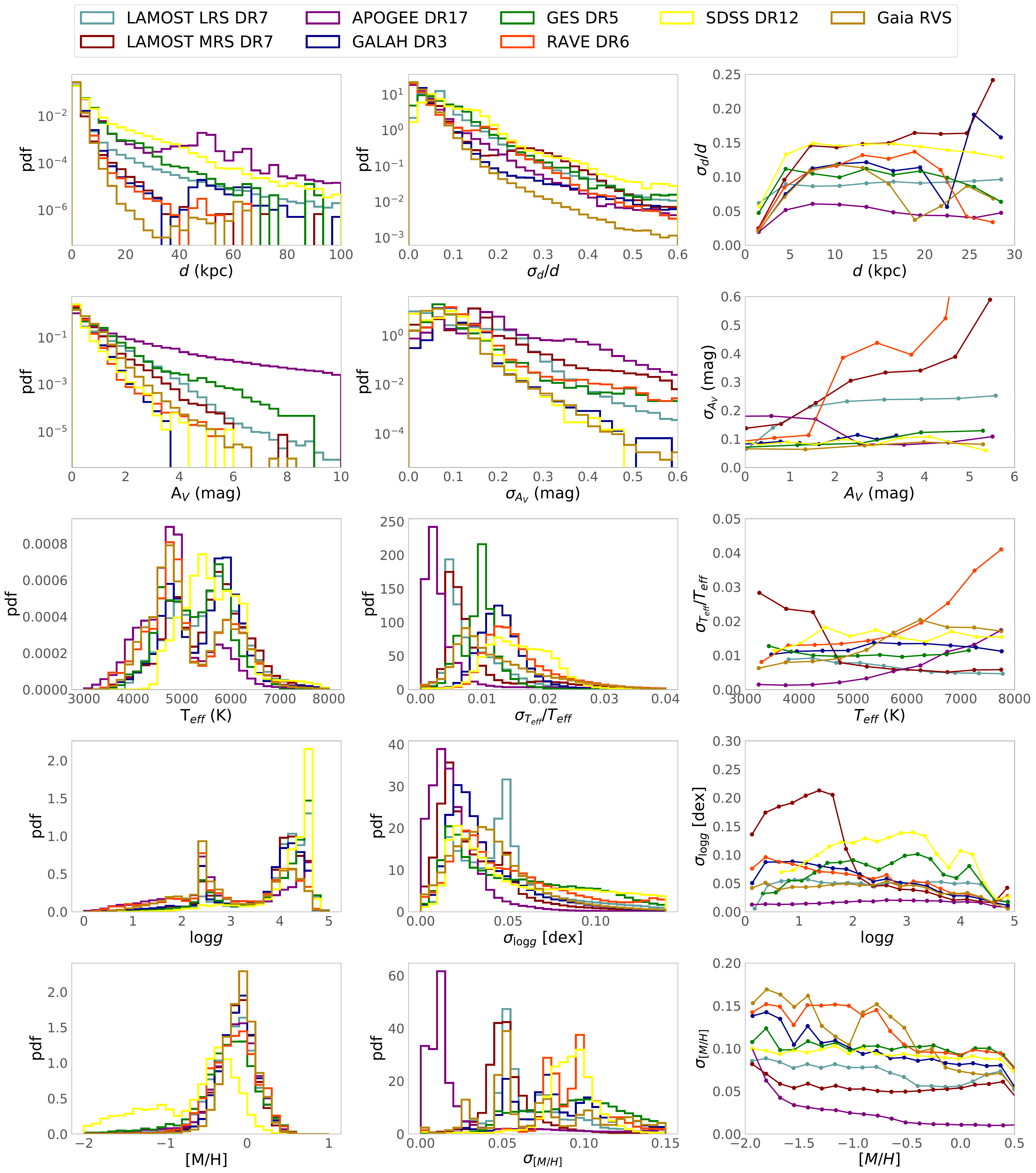}
\caption{Uncertainties distributions for {\tt StarHorse} results. Left and middle panels: probability density functions of {\tt StarHorse} output parameters and their respective uncertainties. The distributions are shown for each spectroscopic survey separately, as indicated in the legend. The upper panels for distance and extinction have their y-axis in logarithm scale to show the extent to larger values. The right panels show the median trend of the dependence of each parameter with its associated uncertainty.}
\label{fig:allunc}
\end{figure*}

\subsubsection{GES DR5}
The {\it Gaia}-ESO survey \citep{Gilmore2012} targets $>$ 10$^5$  stars in all major components of the Milky Way and open clusters of all ages and masses. The survey conducted its observations with the Fibre Large Array Multi Element Spectrograph \citep[FLAMES;][]{Pasquini2002}, which feeds two different instruments covering the whole visual spectral range. The fifth and final data release of GES was made public in May 2022 \citep{Randich2022}. It has significantly increased the number of observed stars, 114\,324, about four times the size of the previous public release, and it also increased the number of derived abundances and cluster parameters. Several working groups focussing on different types of stars and evolutionary stages analysed the GES spectra \citep{Heiter2021}. We downloaded the full catalogue\footnote{https://www.gaia-eso.eu/data-products/public-data-releases/gaia-eso-data-release-dr50}, and used the recommended homogenised atmospheric parameters as  {\tt StarHorse} input. We only use entries with errors smaller than 300 K in temperature,  0.5 dex in surface gravity, and 0.6 dex in iron abundance. To correct the metallicities for the solar scale using the \citet{Salaris1993} formula, we have calculated a global [$\alpha$/Fe] estimate based on the abundances of Si, Ca, and Mg, available for about 58\% of the stars in the catalogue. Compared to the previous {\tt StarHorse} run on GES data, this is an important update because there are many more stars, and we do not exclude the open clusters from our analysis any longer. 

\subsubsection{RAVE DR6}
The final data release of the RAdial Velocity Experiment \citep[RAVE;][]{Steinmetz2006} survey, DR6 \citep{Steinmetz2020} became public in 2020. The spectra from RAVE is acquired with the multi-object spectrograph deployed on 1.2-m UK Schmidt Telescope of the Australian Astronomical Observatory (AAO). The spectra have a medium resolution of R $\approx$ \,7500 and cover the CaII-triplet region (8410-8795$\AA$). We use the final RAVE data release and in particular, the purely spectroscopically derived stellar atmospheric parameters subscripted cal\_madera. In \citetalias{Queiroz2020} we explain the processing of this final RAVE data release in detail and we follow the same procedure for pre-processing this catalogue. The only difference is that this catalogue is now cross-matched with {\it Gaia} EDR3 instead of DR2.

\subsubsection{{\it Gaia} DR3 RVS}

Besides its photometric and astrometric instruments, {\it Gaia} also features a spectroscopic facility, the Radial Velocity Spectrometer (RVS). The instrument observes in the near-infrared (845-872 nm) and has a resolution of $\lambda/\Delta\lambda \approx $11\,500 \citep{Cropper2018}. The third data release of {\it Gaia} contains data of the first 36 months of RVS observations, obtained with the General Stellar Parametriser from the Spectroscopy (GSP-Spec; \citealt{Recio-Blanco2022}) module of the Astrophysical parameters inference system (Apsis; \citealt{Creevey2022}). There are two analysis workflows to process these data: the MatisseGauguin pipeline and an artificial neural network \citep{Recio-Blanco2016}. We work here only with the data analysed by MatisseGauguin, which provides the stellar atmospheric parameters and individual chemical abundances of N, Mg, Si, S, Ca, Ti, Cr, Fe I, Fe II, Ni, Zr, Ce, and Nd for about 5.6 million stars \citep{Recio-Blanco2022}. We downloaded the data from the {\it Gaia} archive's DR3 table of astrophysical parameters. Following a similar procedure to the other spectroscopic surveys, we combine the data with zero point-corrected parallaxes from {\it Gaia} EDR3, and with broad-band photometric data. We do not apply any quality flag cuts when running {\tt StarHorse}, but we only select stars with acceptably small nominal uncertainties ($\sigma_{T_{\rm eff}}<700$ K, $\sigma_{\log g}<1.0$ dex, $\sigma_{\rm [Fe\_H]}<0.6$ dex). We also removed stars with [Fe\_H]$<$-3, since those fall outside the metallicity range covered by the PARSEC stellar model grid used. Regarding parameter calibrations, we applied the suggestions for the calibration of $\log g$, [M/H] and $[\alpha$/Fe] detailed in \citet{Recio-Blanco2022}.

\begin{figure*}
\centering
\includegraphics[width=1.0\textwidth]{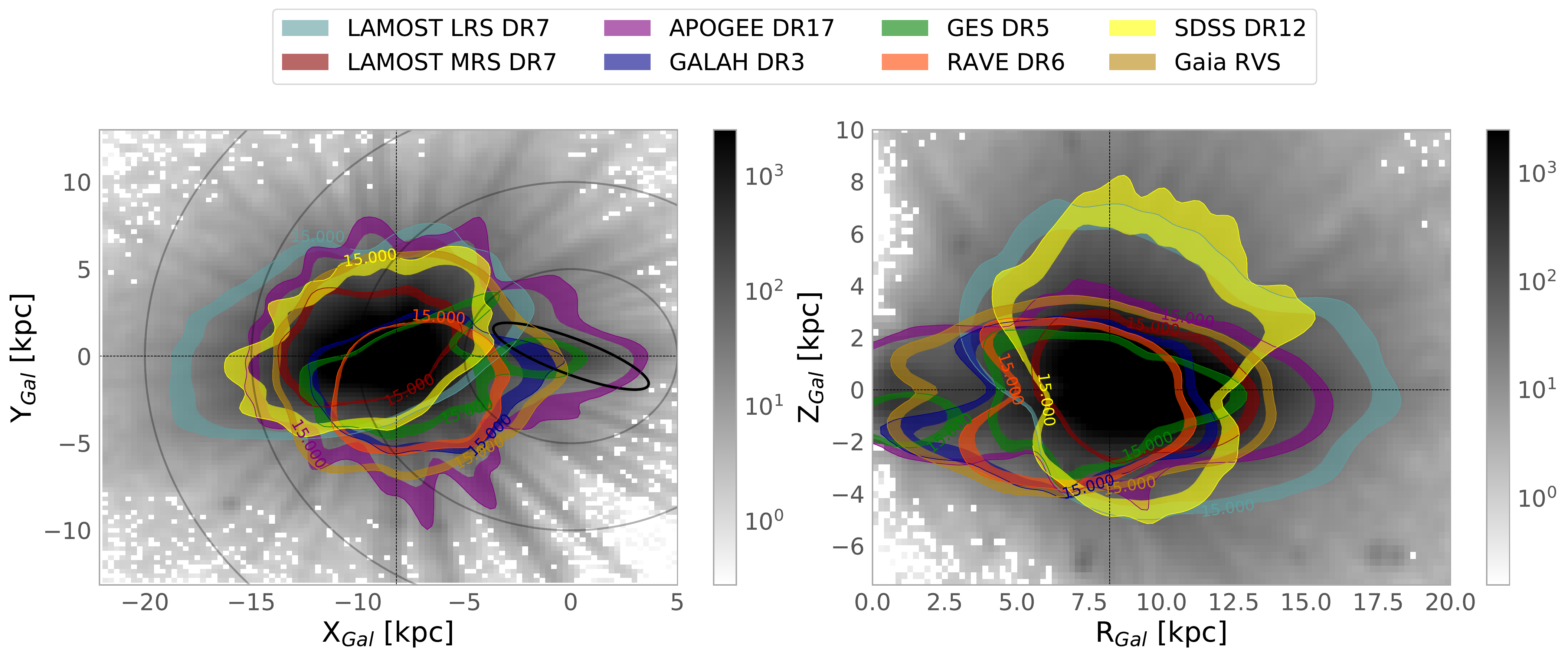}
\caption{Galactic distribution of all spectroscopic surveys for which we present {\tt StarHorse} results in this paper. The grey background density shows the star counts for all surveys combined, while the coloured bands trace the region between iso-contours of $25\,000$ and $15\,000$ stars per pixel for each survey. To guide the eye, grey circles are placed in multiples of 5 kpc around the Galactic centre in the left panel. The approximate location and extent of the Galactic bar is indicated by the black ellipse (minor axis $=$ 2kpc; major axis $=$ 8kpc, inclination $=$ 25$^{\circ}$), and the solar position is marked by the dashed lines.
Left panel: Cartesian $XY$ projection. Right panel: Cylindrical $RZ$ projection.}
\label{fig:summaryXY}
\end{figure*}

\begin{figure*}
\includegraphics[width=\textwidth]{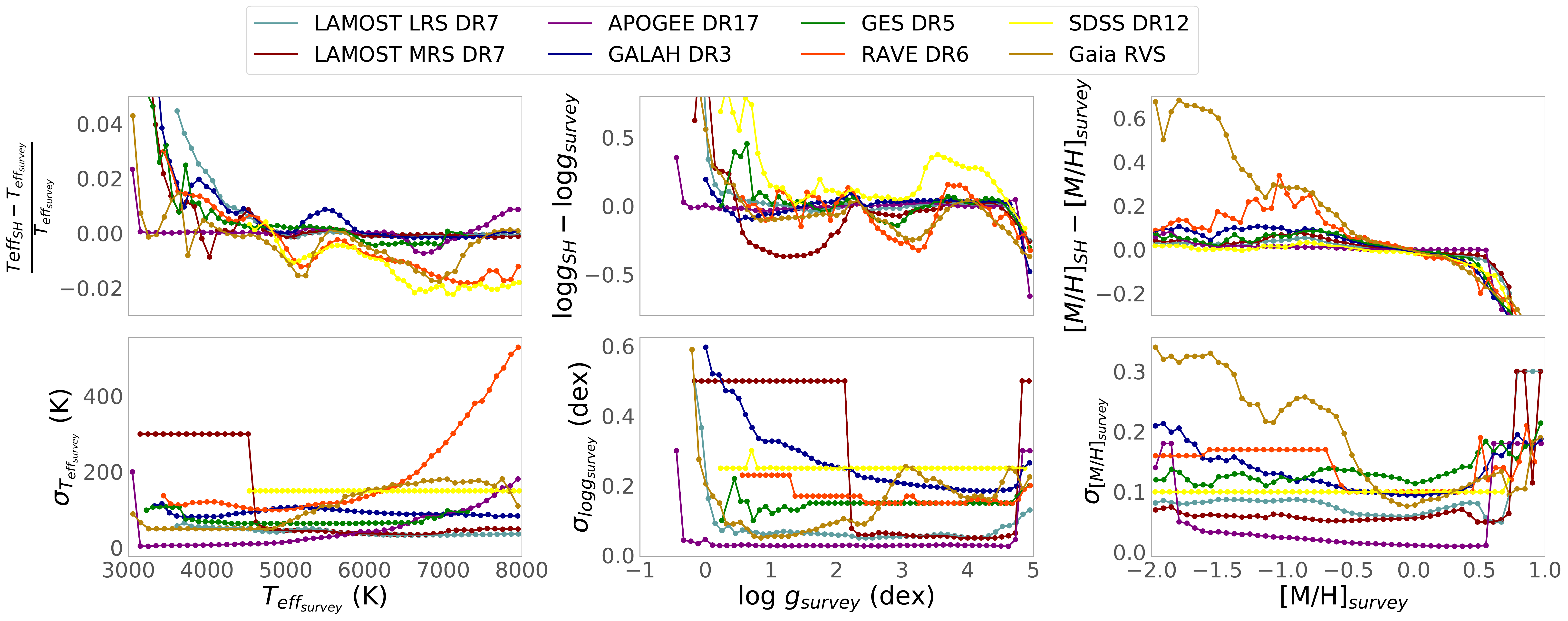}
\caption{Consistency of {\tt StarHorse} input and output parameters. Top panels: Median of the relative discrepancy between input parameters and {\tt StarHorse} output parameters for each survey. Bottom panels: Median of the dependency between input uncertainties versus input parameters.}
\label{fig:inout}
\end{figure*}

\begin{figure*}
\centering
\includegraphics[width=\textwidth]{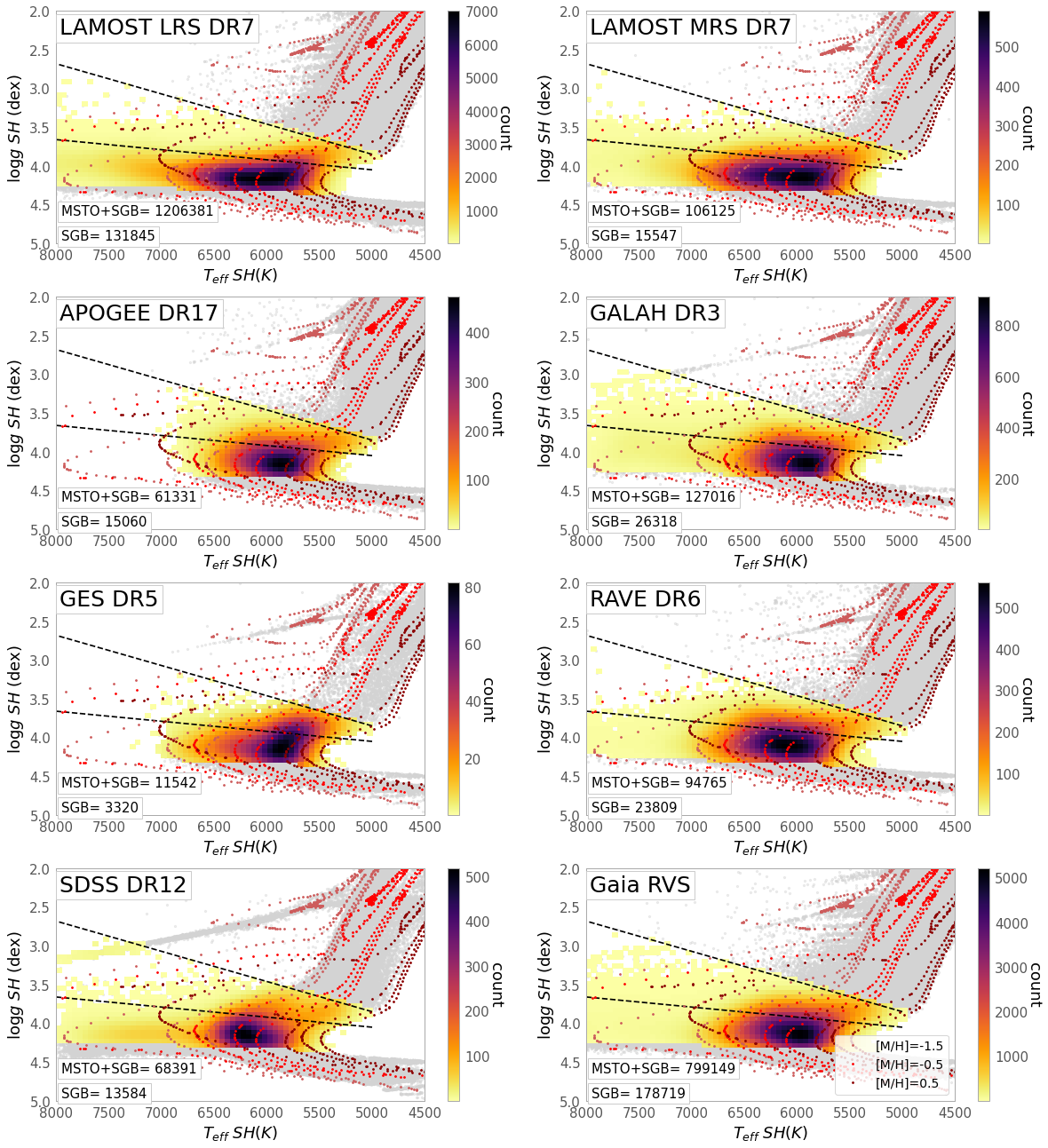}
\caption{{\tt StarHorse} {\it Kiel} diagram for the samples studied in this work. The grey dots in the background of each panel show all the stars in the respective survey, while the colour-coded histograms highlight the MSTO+SGB regime for which we deliver {\tt StarHorse} ages. The dashed lines limit the SGB, for which the computed ages are most precise. PARSEC isochrones are overplotted in red for three different metallicities, as indicated in the lower right corner of the Figure. For each metallicity four different ages are shown for 1, 4, 7 and 10 Gyr.}
\label{fig:CMDsub}
\end{figure*}

\begin{figure*}
\centering
\includegraphics[width=17.0cm]{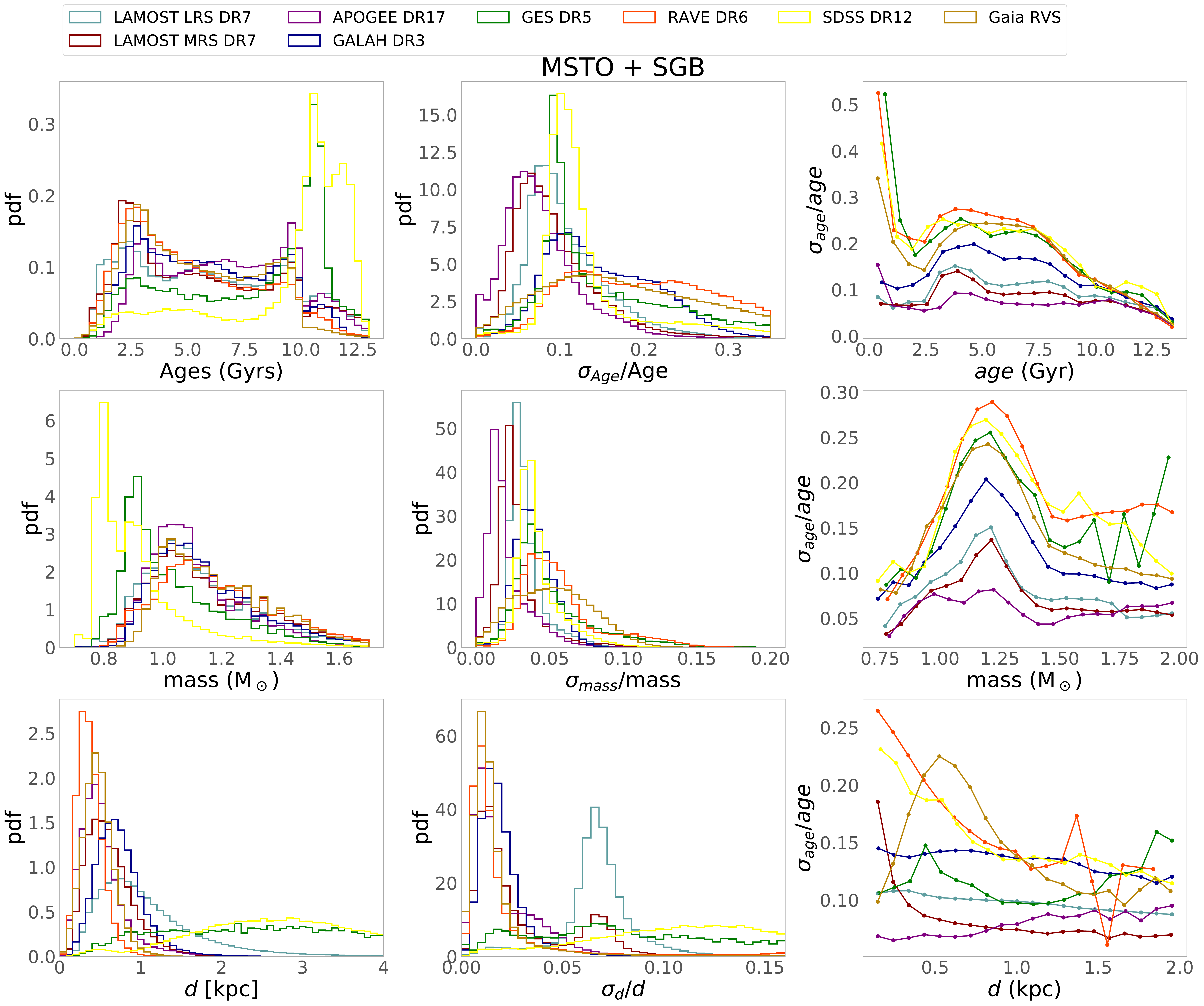}
\caption{Distributions of ages, masses, distances and their uncertainties for the MSTO+SGB samples. The y-axis shows the probability density. All histograms are normalised so that the area under the histogram integrates to 1. The right panels show the mean age uncertainty per bin  of age, mass and distance for each survey.}
\label{fig:ageunce}
\end{figure*}

\section{New {\tt StarHorse} catalogues}
\label{out2sh}
We present a new catalogue set derived from the stellar spectroscopic surveys described in Section \ref{cats} combined with photometry and {\it Gaia} EDR3 parallaxes (Section \ref{edr3}). We provide
percentiles of the posterior distribution functions of masses, effective temperatures, surface gravities, metallicities, distances, and extinctions for each successful converged source according to Table \ref{summarytable}. We deliver the final data in the same format as in \citetalias{Queiroz2020} Table A.1 for each spectroscopic survey used as input. The median value, 50th percentile, should be taken as the best estimate for that given quantity, and the uncertainty can be determined using the 84th and 16th percentiles. In this release, we also make for the first time age determinations for a selection of main-sequence turn-off and sub-giant branch (MSTO+SGB) stars. The given ages follow the same format as the other {\tt StarHorse} parameters, but we flag everything outside our MSTO+SGB selection as -999. All the newly produced {\tt StarHorse} catalogues are available for download from \url{data.aip.de} and through VizieR. Some of the results of \texttt{StarHorse} for APOGEE, GALAH and SDSS12 have already been analysed by recent publications on the study of halo debris \citep{Limberg2021, Perottoni2022, Limberg2022}.

\begin{figure*}
\centering
\includegraphics[width=17.0cm]{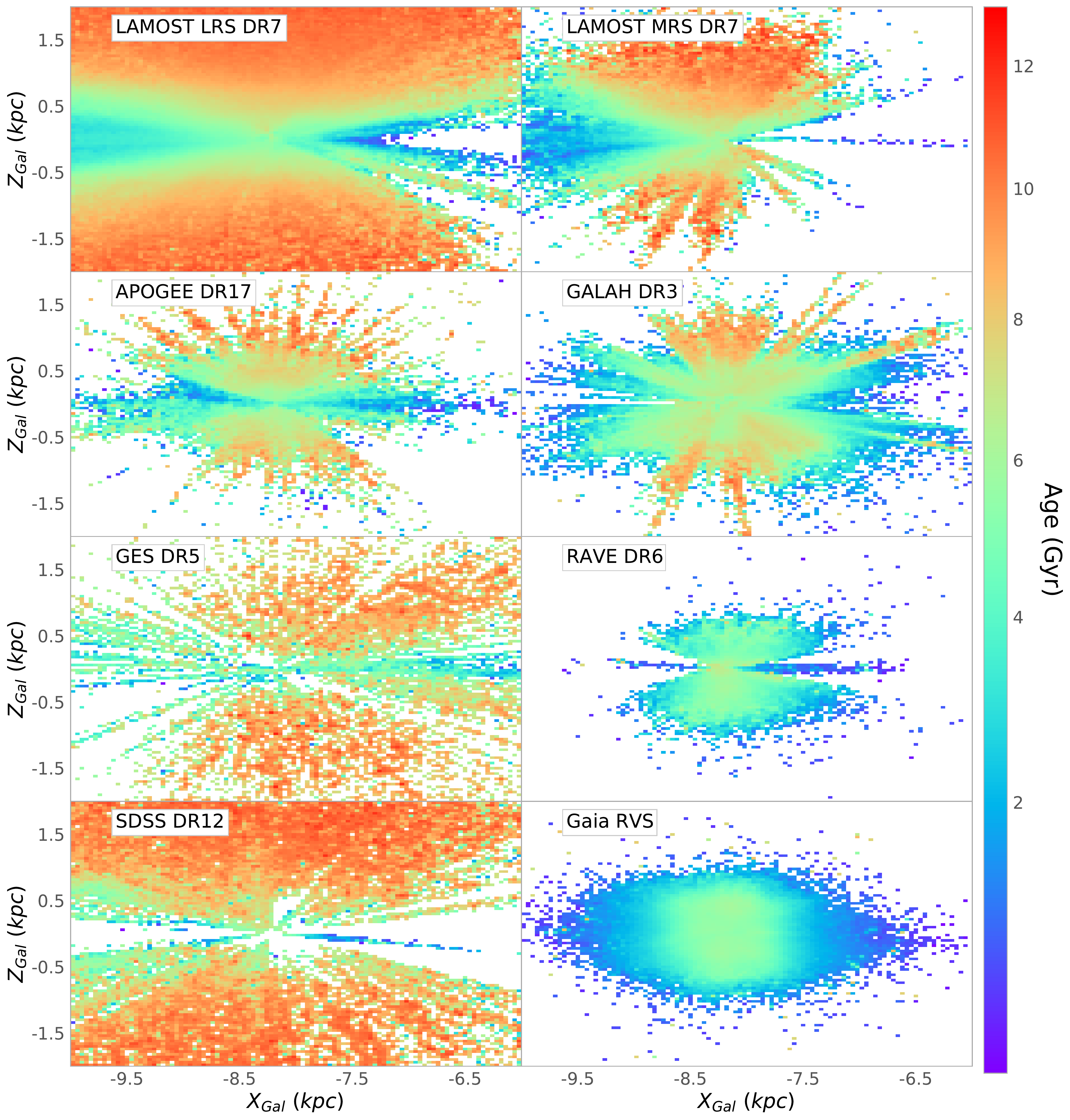}
\caption{Galactocentric X and Z projection of MSTO+SGB samples colour-coded by {\tt StarHorse} ages, the colourbar is in power law scale with $\gamma$=0.7.}
\label{fig:ageXY}
\end{figure*}

\begin{table*}
\centering
\caption{Mean relative error or uncertainty per {\tt StarHorse} output parameter per spectroscopic survey}
\begin{tabular}{lrcrcrcrc}
Survey & $\sigma_{d}/d$ &  $\sigma_{A_V} $  &  $\sigma_{T_{\rm eff}}/T_{\rm eff}$ & $\sigma_{\rm [M/H]}$ & $\sigma_{\log g}$ & $\sigma_{m_\ast}/m_\ast$ & $\sigma_{age}/age_{MSTO+SGB}$ & $\sigma_{age}/age_{SGB}$ \\
     & (\%) & (mag) &  (\%) & (dex) & (dex) & (\%) & (\%) & (\%) \\
\hline
LAMOST DR7 LRS      & 7.5 & 0.082 & 0.8 & 0.067 & 0.053 & 9.6  & 11.0 & 15.6  \\
LAMOST DR7 MRS      & 4.9 & 0.129 & 0.9 & 0.072 & 0.042 & 11.2 & 10.3 &  9.15\\
SDSS DR12 optical   & 10  & 0.075 & 1.6 & 0.093 & 0.080 & 8.0  & 14.8 & 12.0\\
GALAH+ DR3          & 3.6 & 0.092 & 1.3 & 0.092 & 0.041 & 12.1 & 16.6 & 11.0 \\
RAVE DR6            & 5.1 & 0.099 & 1.6 & 0.099 & 0.058 & 8.0  & 23.1 & 19.0 \\
APOGEE DR17         & 4.3 & 0.178 & 0.4 & 0.029 & 0.021 & 12.6 & 8.0 & 6.5 \\
GES DR5             & 5.8 & 0.099 & 1.2 & 0.076 & 0.053 & 11.2 & 16.6 & 16.4\\
{\it Gaia} DR3 RVS  & 3.1 & 0.069 & 1.3 & 0.172 & 0.044 & 17.3  & 20.7 & 12.0 \\
\hline
\hline
\end{tabular}
\label{uncertaintytable}
\end{table*}

\begin{figure*}
\centering
\includegraphics[width=\textwidth]{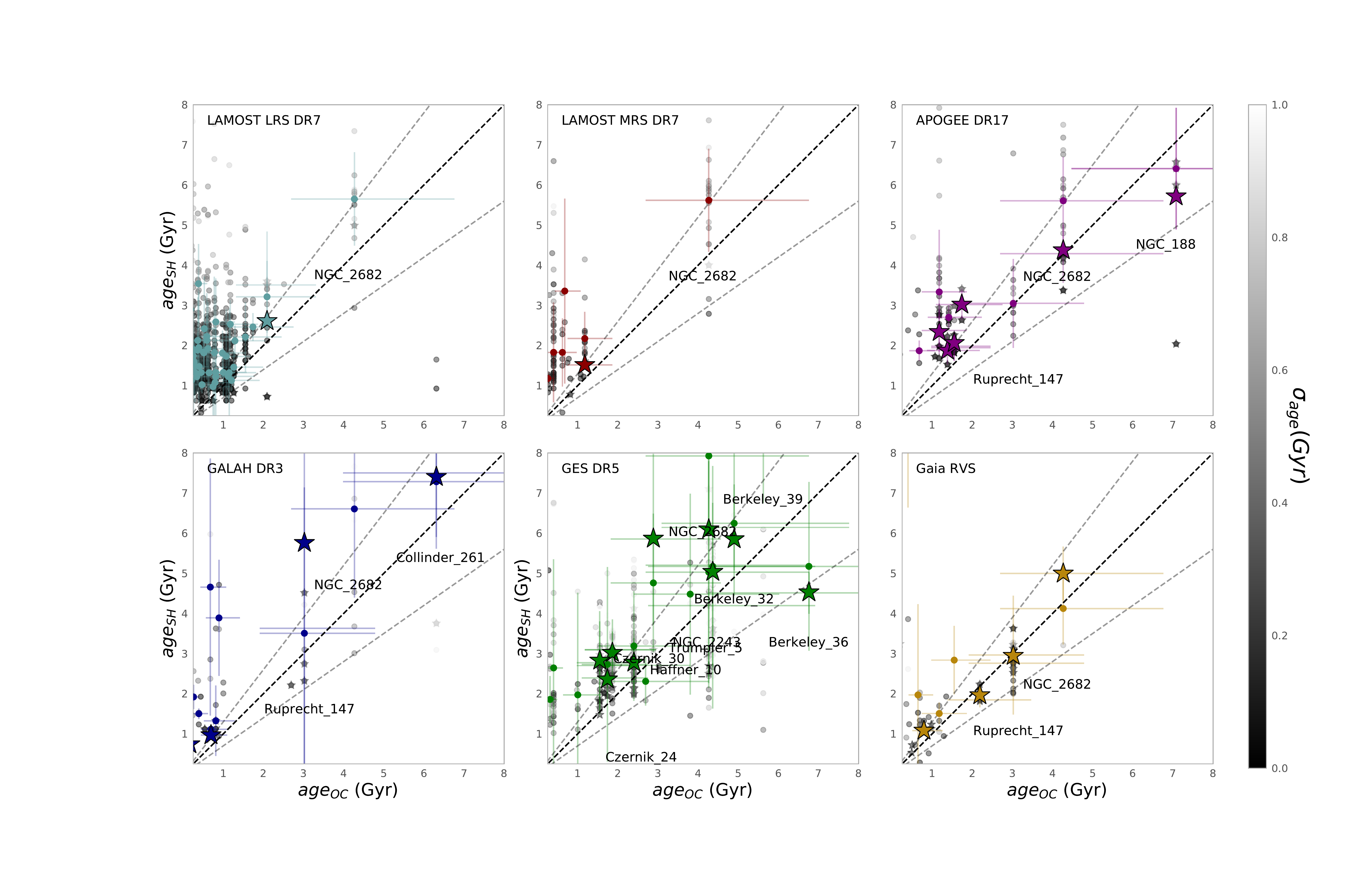}
\caption{Comparison of our MSTO+SGB age estimates with open cluster ages from \citet{Cantat-Gaudin2020}. In each panel, grey dots are individual OC member MSTO stars (membership probability $>95\%$), colour-coded by their posterior age uncertainty. For OCs that contain more than 3 MSTO cluster members, the uniform coloured points and error bars indicate the median age and the 1$\sigma$ quantiles. We indicate as a star symbol only the SGB sample which has smaller {\tt StarHorse} uncertainties. The horizontal error bars reflect the 0.2 dex uncertainties quoted by \citet{Cantat-Gaudin2020}. The only OC that was observed by all surveys is M67 (NGC 2682). We do not show comparisons for RAVE and SDSS since there were less than 5 stars in common with the cluster sample. The dashed black line delineates the identity line, while the grey dashed lines correspond to $\pm$30\% deviation}
\label{fig:ageval1}
\end{figure*}

\begin{figure*}
\centering
\includegraphics[width=18.0cm]{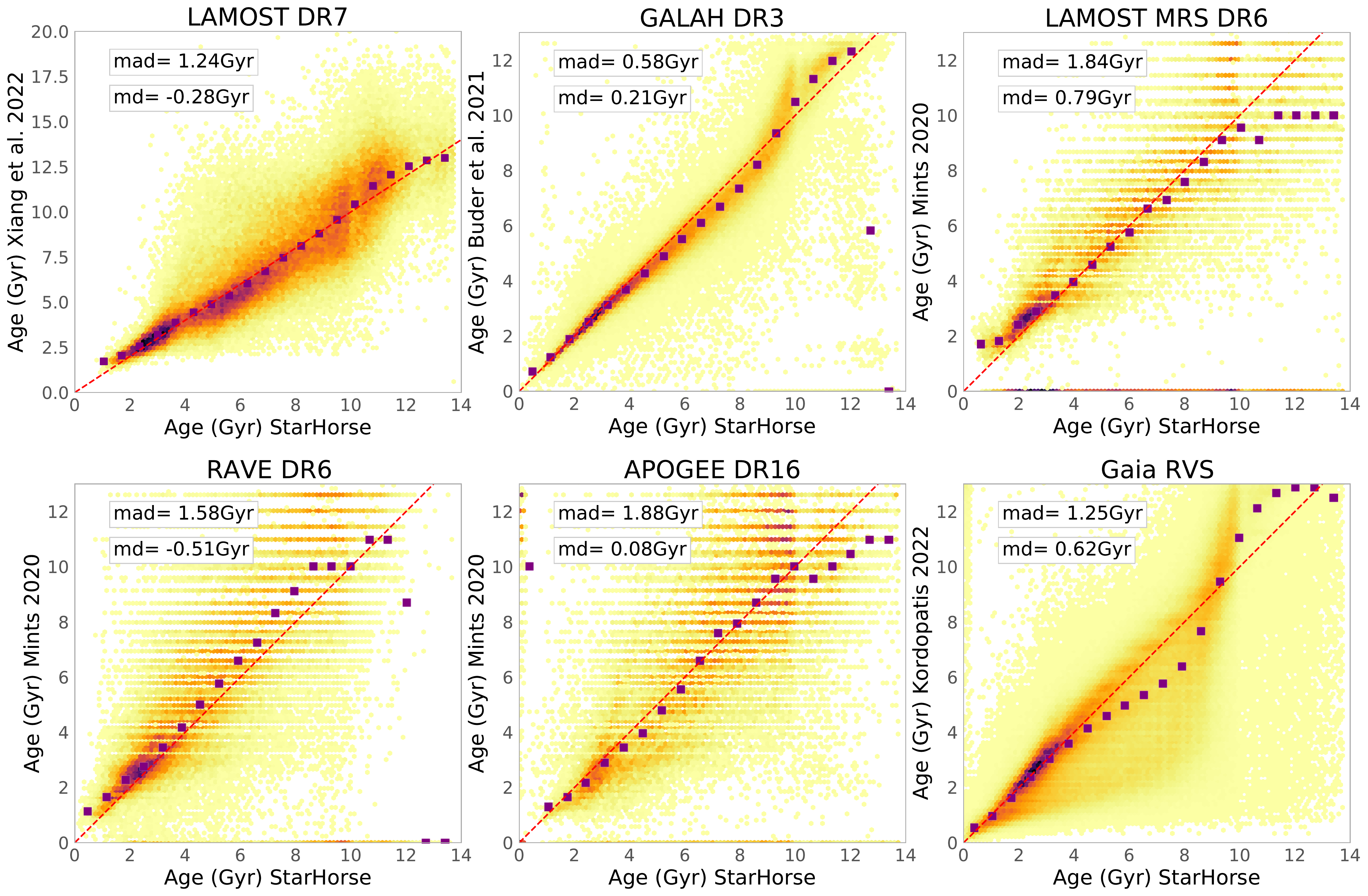}
\caption{Comparison of our MSTO+SGB age estimates with values from the recent literature. From top left to bottom right: comparison to the LAMOST DR7 ages of \citet{Xiang2022}, the GALAH DR3 ages of \citet{Buder2021}, comparison to the LAMOST, RAVE, and APOGEE age estimates of \citet{Mints2020} and  comparison to the ages calculated by \citet{Kordopatis2022} for {\it Gaia} RVS sample. In each panel, we show the number density of stars, and the magenta points indicate the median trends. The global spread for each comparison is shown as the mean absolute deviation, mad, and the global shift as mean deviation, md.}
\label{fig:ageval2}
\end{figure*}

\begin{figure*}
\centering
\includegraphics[width=.49\textwidth]{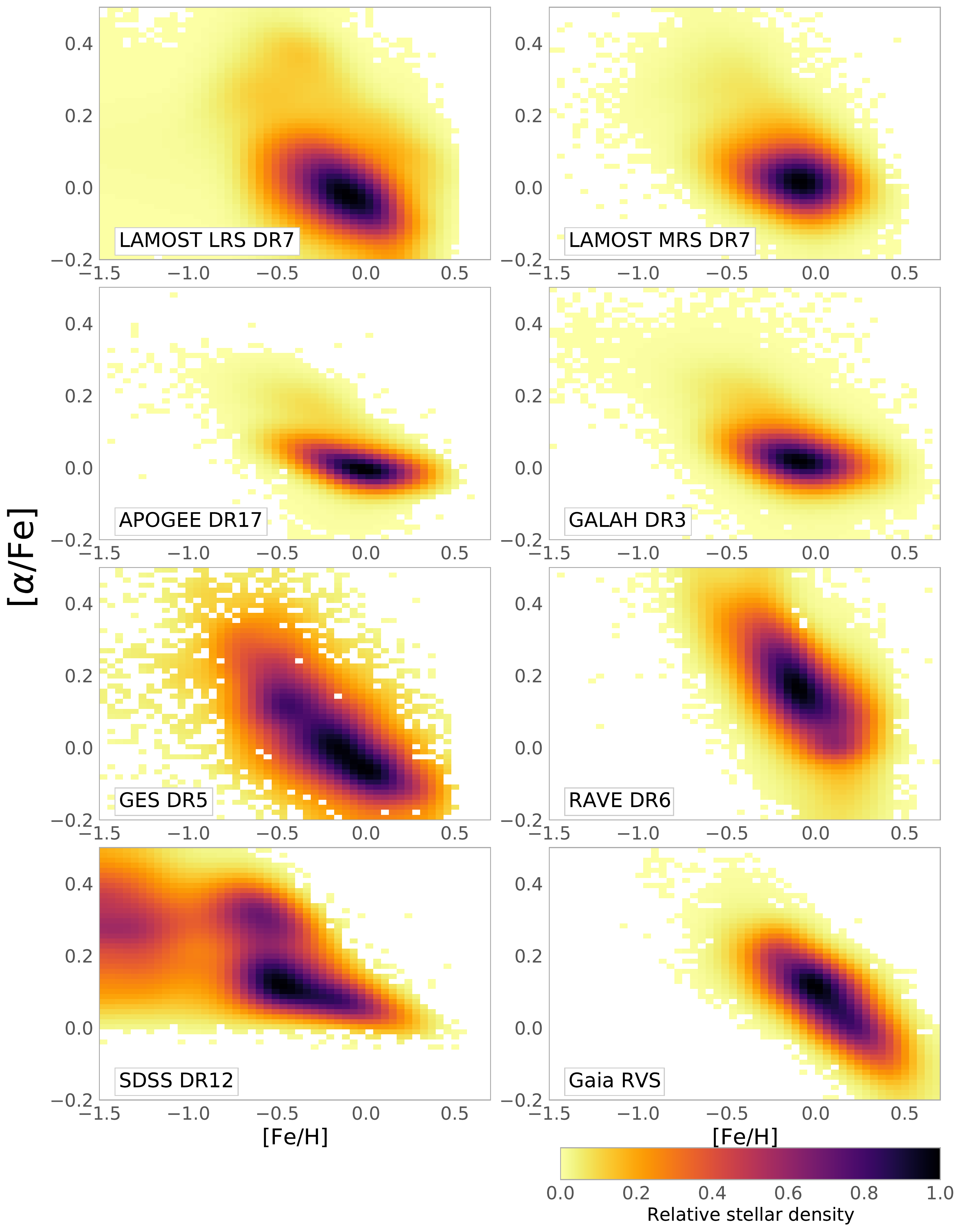}
\includegraphics[width=.49\textwidth]{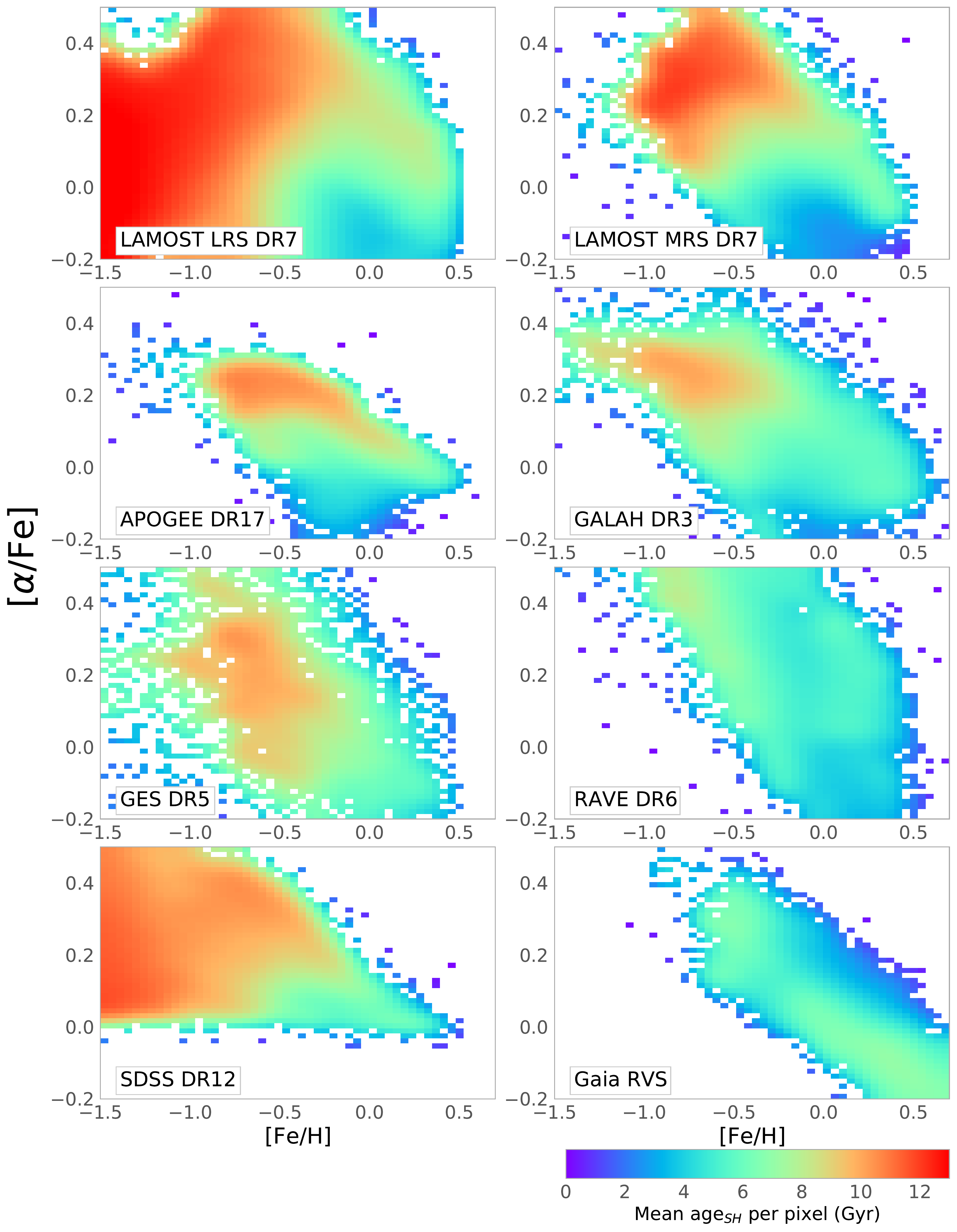}
\caption{[$\alpha$/Fe] vs. [Fe/H] distributions for the MSTO+SGB samples of the analysed surveys. Left: Density distributions (relative to the maximum count in each survey). Right: The same, but coloured by the mean age per pixel.}
\label{fig:alphafe}
\end{figure*}

\begin{figure*}
\centering
\includegraphics[width=14.5cm]{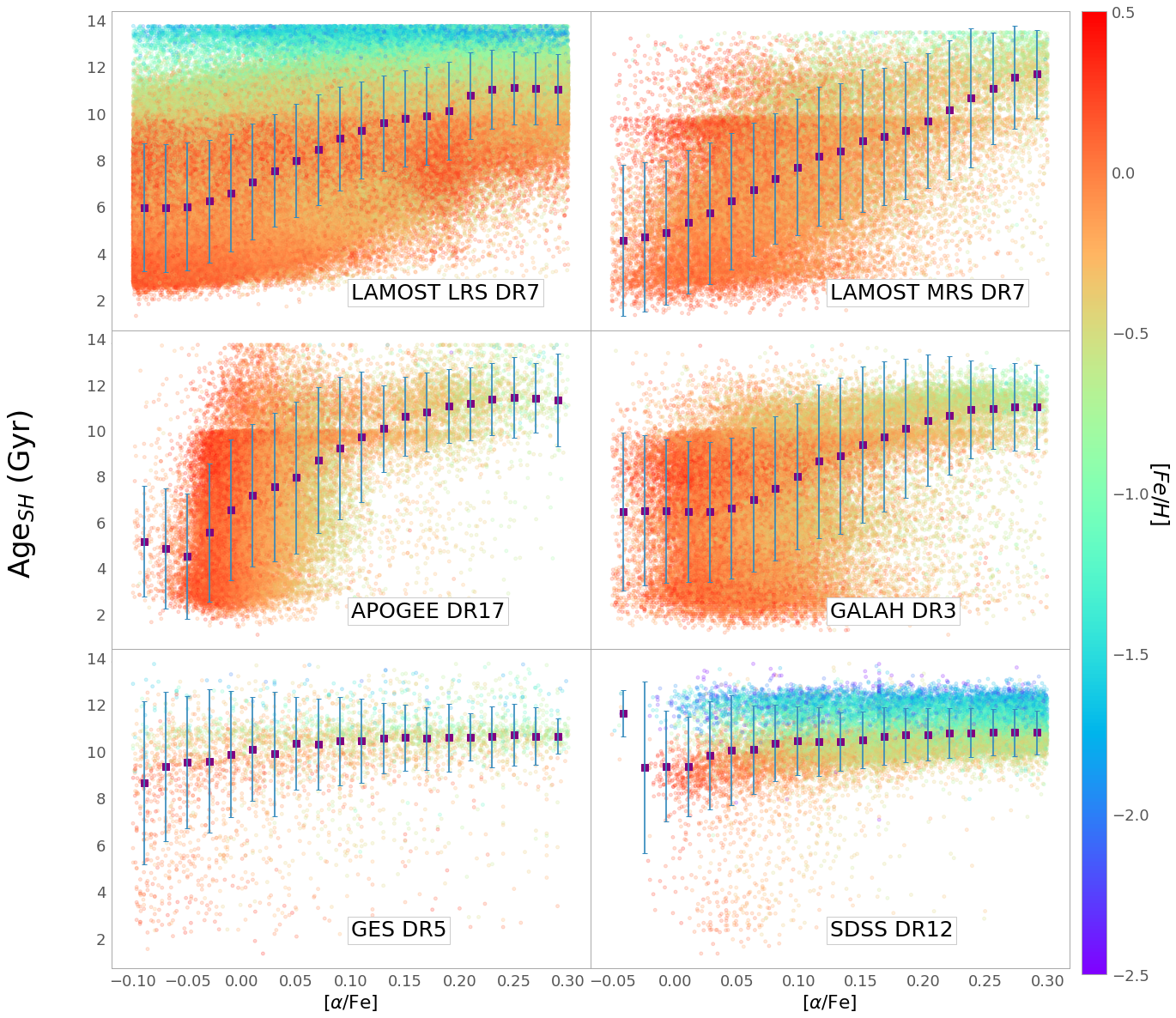}
\caption{[$\alpha$/Fe] vs age distribution for the MSTO+SGB samples of each survey. A cleaning per signal to noise and suggested flags was performed. The purple squares show the median trend per bin in [$\alpha$/Fe] while the error bars show its one $\sigma$ deviation. We only display the surveys that have mean statistical uncertainty in age $<$ 20\% according to Table \ref{uncertaintytable} (no RAVE or \textit{Gaia} RVS), we have also performed a cleaning of flags in the [$\alpha/Fe$] determination from each survey.}
\label{fig:agevsalpha}
\end{figure*}

\subsection{{\tt StarHorse} distances and extinctions}

Precise distances and extinctions are fundamental for Galactic archaeology \citep{Helmi2020}.
By combining spectroscopic and {\it Gaia} data, {\tt StarHorse} achieves precise distances from the inner to the outer Galaxy. As seen in the left panels of Figure \ref{fig:allunc} we get relative errors in the distance of only 15\% for distances as far as 20 kpc, and a mean extinction uncertainty of about 0.2 (mag). Distances and extinctions have also been extensively validated with simulations and external methods in \citetalias{Queiroz2018} and \citetalias{Queiroz2020}, showing internal precision in the distance and extinctions of about 8\% and 0.04 mag, respectively. In Figure \ref{fig:summaryXY} we show the distribution of stars for all surveys for which we compute distances in Galactocentric Cartesian coordinates. This map expresses the extent and capability of the resulting data, which samples very well the solar vicinity,  reaches the inner parts of the Galaxy, covers the outer disk beyond $R_{\rm Gal}=20$ kpc and extends to $|Z_{\rm Gal}|>10$ kpc. We display the distribution of parameters and their uncertainties in Figure \ref{fig:allunc}, and we show the mean uncertainty in each parameter for each survey in Table \ref{uncertaintytable}. The mean relative distance uncertainty for all surveys lies below 10\%, while for GALAH, APOGEE, LAMOST MRS and Gaia DR3 RVS, it is below 5\%. It is noticeable from Figure \ref{fig:allunc} that with the new prior implementation \citep{Anders2022}, some survey distances extend to other galaxies, e.g. APOGEE reaches the Magellanic Clouds, the Sagittarius dwarf galaxy, and some globular clusters. $A_{V}$ varies primarily according to each survey's selection function, and its uncertainty is strongly correlated with photometry, but on average below 0.2 mag. For the most precise determinations of $A_{V}$, one can select the stars with the complete photometry input set (detailed in the {\tt StarHorse} input flags).

\subsection{{\tt StarHorse} $T_{\rm eff}$, $\log{g}$ and metallicity}\label{inout}

Surface temperatures and gravities are also present in the output from the {\tt StarHorse} catalogues. The code uses these parameters as input from the spectroscopic surveys. Therefore these are just slight improvements to the measurements, but this is especially useful for the atmospherical parameters that were not initially calibrated by surveys or have significant uncertainties and caveats. In Figure \ref{fig:inout} we show a comparison between the atmospheric input parameters and the output {\tt StarHorse} parameters for each spectroscopic survey, as well as their input uncertainties. The differences between the high-resolution surveys are minor since their uncertainties are well-constrained. Most surveys show differences in effective temperature between cold stars and hot stars, respectively, with $T_{\rm eff}<4000$ K and $T_{\rm eff}>7000$ K. SDSS DR12 and RAVE show the most significant deviation in input temperature for hot stars, which are usually overestimated with respect to the models by 5\%. The surface gravity is the most deviating parameter. There are considerable differences between input and output for the whole $\log{g}$ range; LAMOST MRS has 0.5 dex overestimation against {\tt StarHorse} output for giant stars, while SDSS shows the same amplitude but underestimation of logg for dwarf stars. The metallicities have excellent agreement with input differing only by 0.1 dex for most surveys except for RAVE DR6 which shows a difference up to 0.3 dex compared with {\tt StarHorse} metallicities, but they also present one of the largest uncertainties in metallicities.

\begin{figure*}
\includegraphics[width=17.5cm]{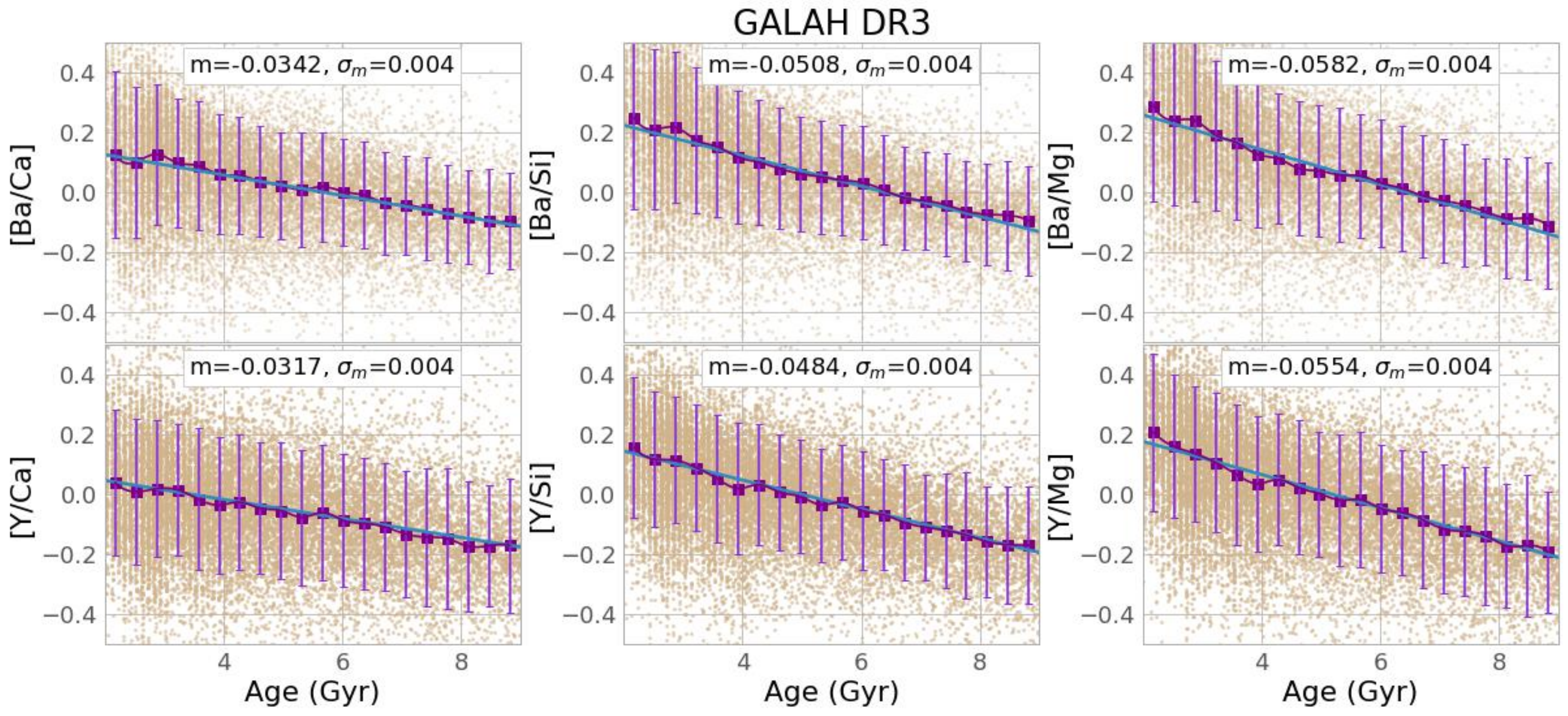}
\caption{[s/$\alpha$] abundance ratios vs. age for GALAH. The purple line shows the median abundance per age bin and the error bar represents one sigma deviation from the median.}
\label{fig:galahclock}
\end{figure*}

\begin{figure*}
\includegraphics[width=17.5cm]{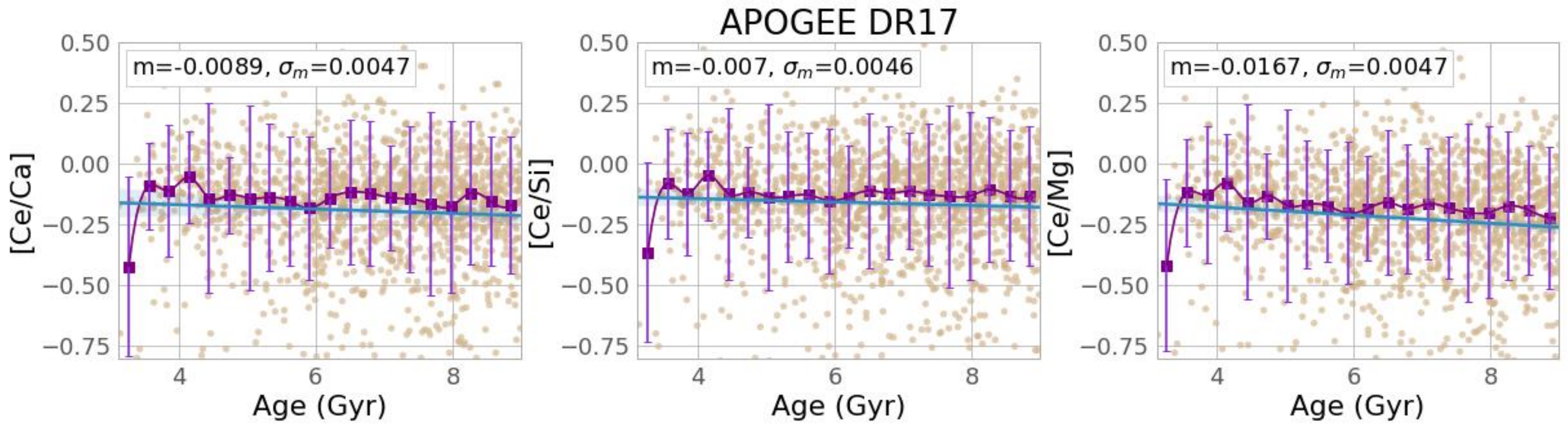}
\caption{[s/$\alpha$] abundance ratios vs. age for APOGEE. The purple line shows the median abundance per age bin and the error bar represents one sigma deviation from the median.}
\label{fig:apogeeclock}
\end{figure*}

\subsection{{\tt StarHorse} MSTO-SGB ages and masses}

For the first time, we publish ages derived with {\tt StarHorse}. Ages (and also masses) for individual stars are challenging to derive through isochrone fitting when only spectroscopic, astrometric, and photometric data are available (e.g. \citealt{Joyce2022}). In the absence of spectroscopic data, meaningful age estimates are even more complicated \citep{Howes2019}. More sophisticated methods such as asteroseismology or eclipsing binaries (where an additional constraint on the stellar mass becomes available) are much more reliable for deriving ages, and these methods can achieve uncertainties below 10\% \citep{Valle2015, Silva-Aguirre2018, Anders2017, Valentini2019, Miglio2021}. The downside is that the samples with asteroseismic and eclipsing binaries are still limited in size and pencil beams compared to spectroscopic surveys. We can achieve more statistical significance by measuring less precise ages but for larger sets. Here we do so, but we restrain ourselves to the MSTO-SGB. In these evolutionary stages, isochrone fitting methods are more reliable since the shape and duration of this stage varies strongly with the stellar mass and, therefore, the age. For SGB stars, the luminosity correlates directly with age, which makes this stage specially suitable for isochrone-based age determinations \citep[e.g.][]{Xiang2022}. 
In \citetalias[][e.g. Figure 4]{Queiroz2018} we have also shown with simulations that {\tt StarHorse} ages can achieve relative statistical uncertainties of 20\% for SGB stars.

\begin{table*}
\centering
\caption{Flags we clean for age computation in each survey}
\begin{tabular}{lrcrc}
Survey & flags  & documentation\\
\hline
APOGEE & STARFLAG \& ASPCAPFLAG $= 0$ & \citet{Jonsson2020}\footnotemark \\
GALAH & FLAG\_SP $= 0$ & \citet{Buder2021}\footnotemark \\
GES & SSP, SRP or BIN not in SFLAGS & \citet{Randich2022}\footnotemark \\
RAVE & algo\_conv\_madera $=0$ & \citet{Steinmetz2020}\\
{\it Gaia} RVS & flags\_gspspec(0:6) $= 000000$ & \citet{Recio-Blanco2022}\\
SDSS DR12 & FLAG=='nnnn' & \citet{Lee2008}\\
LAMOST LRS \& MRS DR7 & fibermak==0 & \citet{Luo2015}\footnotemark\\
\hline
\multicolumn{3}{c}{\small{$^6$\url{https://www.sdss4.org/dr14/irspec/parameters/}, $^7$\url{https://www.galah-survey.org/dr3/flags/}}}\\
\multicolumn{3}{c}{\small{$^8$\url{https://www.eso.org/rm/api/v1/public/releaseDescriptions/191}}}\\
\multicolumn{3}{c}{\small{$^9$\url{http://dr7.lamost.org/v1.3/doc/lr-data-production-description}}} \\
%\multicolumn{3}{c}{$^9$\url{https://data.sdss.org//datamodel/files/SSPP_REDUX/RERUN/PLATE4/output/param/ssppOut.html}}\\
\hline
\end{tabular}
\label{ageflagtable}
\end{table*}

Isochrone age determination can be highly uncertain and subject to biases in input temperature, surface gravity, or metallicity \cite[see][]{Queiroz2018}. To mitigate these effects, we have undertaken a cleaning procedure of the input parameters of each spectroscopic survey. Specifically, for the MSTO-SGB ages, we have applied the recommended flags from the spectroscopic pipelines and implemented a minimum signal-to-noise cut of 30 in all surveys. Table \ref{ageflagtable} provides the details of our cleaning procedure. The quality cuts are done using flags describing the quality of the spectral observations or the derivations of atmospherical parameters. These are described in more detail in the indicated papers and web pages. In our final catalogue, we have included a flag to indicate whether stars are at the MSTO or SGB stage, as well as a flag "age\_inout" to alert users of any significant differences between the input and output spectroscopic parameters (T$_{eff}$, logg, met50).% For LAMOST LRS and MRS there is no general flag we can use to their LASP results therefore we cut only in S/N.}

We display our MSTO-SGB selection in Figure \ref{fig:CMDsub}. We opt to use the output {\tt StarHorse} $T_{\rm eff}$ and $\log{g}$ for this selection, since we saw in the previous Section \ref{inout} that for some surveys there are systematic differences between input and output parameters (especially in $T_{\rm eff}$ and $\log g$). Therefore, a selection using the {\tt StarHorse} parameters is more homogeneous and, to some extent, helps eliminate systematics between the different spectroscopic surveys. {\tt StarHorse} can break degeneracies by accessing the extra information from photometry and astrometry. The selections are perfomed using the following equations adjusted by eye to comprise the MSTO and SGB regime:

\begin{equation}\label{mstosgb}
\begin{split}
 \log{g}_{SH} < -0.000005 T_{eff_{SH}}+4.6\\
\log{g}_{SH} > -0.00039 T_{eff_{SH}}+4.9\\
T_{eff_{SH}} > 500\log{g}_{SH}+3000; T_{eff_{SH}} < 8000
\end{split}
\end{equation}

And only for the SGB selection:

\begin{equation}\label{sgb}
\begin{split}
\log{g}_{SH} < -0.00013 T_{eff_{SH}}+4.7\\
\log{g}_{SH} > -0.00039 T_{eff_{SH}}+4.9\\
T_{eff_{SH}} > 500\log{g}_{SH}+3000; T_{eff_{SH}} < 8000
\end{split}
\end{equation}

In Figure \ref{fig:ageunce} we show the distributions and uncertainties in ages and masses for the selected MSTO-SGB stars. Most age distributions display two peaks: one at intermediate ages ($\approx$ 3 Gyr) and one containing an older generation ($\approx$ 9-11 Gyr). There is a noticeable depression at 10 Gyr for the higher-resolution surveys APOGEE, LAMOST MRS, and GALAH. Since the SDSS/SEGUE survey preferentially targeted the Galactic halo, the age distribution for this survey is highly skewed towards old ages, and presents a double peak at 11 and 12 Gyr. GES and LAMOST LRS only show a rise at 11 Gyr.
90\% of the MSTO stars have relative age uncertainties smaller than 50\%, and their average is below 34\% (see Table \ref{uncertaintytable}). For SGB stars, this average decreases below 20\%. From all the spectroscopic releases, APOGEE and LAMOST MRS have the smallest nominal uncertainties in age (below $10\%$), although this is strongly driven by the input parameter uncertainties of the surveys. 
We caution, therefore, about the systematic differences on the uncertainty of our derived ages from one survey to another. The spectroscopic surveys make very different choices on how to report uncertainties in their atmospheric parameters (some of them are probably underestimated in some surveys), which then will lead to underestimated age uncertainties.

Apart from ages, we also deliver mass estimates for the complete catalogues, not only the MSTO-SGB, but we remind the user to be cautious when using these values, since both statistical and systematic uncertainties can be very high (depending on the class of stars). The posterior mass distributions do not show considerable differences between surveys, besides the higher content of low mass stars in SDSS and GES. Our mass estimates have been previously validated in earlier {\tt StarHorse} versions, \citetalias{Queiroz2018}, against asteroseismic and binaries samples, which yielded relative deviations of $\approx$ 12\% and 25\%. 

In Figure \ref{fig:ageunce}, we also show the extent of heliocentric distances for the MSTO-SGB samples, which is mainly confined to an extended solar neighbourhood (0.1-3 kpc), surveys targetting the halo as GES and SDSS do reach farther distances even inside the MSTO-SGB selection. In the right panels of Figure \ref{fig:ageunce} we see the dependence of the relative age uncertainty with age, mass and distance. {\tt StarHorse} ages are more uncertain for younger, intermediate-mass stars. There is also a trend of decreasing age uncertainty with distance, which is related to older stars being found far from the disk. This effect is evident in Figure \ref{fig:ageXY}. For all surveys, we see dependence of age and Galactic height (Z$_{gal}$) more explicitly in LAMOST LRS, which has the most significant number of stars. The increasing age with Z$_{gal}$ shows the consequence of transiting between the young, thin disk (confined to the Galactic plane) to the older thick disk and halo components.

\subsection{Validation of age estimates}\label{agevalidation}

Since age estimates for field stars are highly dependent on stellar evolutionary models, it is important to identify (and, when possible, quantify) systematic biases. Although also not model-independent, asteroseismic and open cluster ages (OCs) are still our best anchor for validating field-star age estimates. Since solar-like oscillations in MSTO-SGB stars are much weaker than for the red-giant branch, large samples of MSTO-SGB benchmark ages from asteroseismology are still missing. In Figure \ref{fig:ageval1} we therefore compare our age estimates to the OC ages derived by \citet{Cantat-Gaudin2020}. Almost all considered spectroscopic surveys have observed at least some OCs with MSTO-SGB members, For SDSS and RAVE there are less than five cluster members, which therefore we chose not to show. The results of the test shown in Figure \ref{fig:ageval1} are for the most part reassuring. While the results are on average in good agreement, specially for the SGB sample only, the ages of younger OC MSTO members present some systematically overestimated results. We have checked that most of this disagreement is due to an input temperature, gravity or metallicity which is not consistent with the ages determined by \citep{Cantat-Gaudin2020}. There might be a systematic in the determination of atmospheric parameters for younger ages as well, therefore we have carefully cleaned the sample from potential problems in the spectral parameters derivation indicated by each survey in Table \ref{ageflagtable}.

Another reason for this is the dominance of the initial-mass function prior, which for massive stars will result in the preference of lower-mass (and consequently, older-age) posterior solutions. We insist, however, that this is not a genuine problem of the {\tt StarHorse} code, but a generic problem of one-fits-all isochrone-fitting codes. 

A proof of this statement (and also a secondary sanity check) is delivered in Figure \ref{fig:ageval2}, which compares our age estimates to field-star ages in the recent literature \citep{Xiang2022, Buder2021, Leung2019, Mints2020}. The figure demonstrates that our age estimates compare well with other recent attempts to derive isochrone ages, especially with the ages derived by \citet{Buder2021} for GALAH DR3 and, to a slightly lesser degree, with the results obtained by \citet{Xiang2022} for LAMOST and \citet{Mints2020} for APOGEE, RAVE, and LAMOST. The horizontal streaks in the comparison figures for the latter reference (lower panels of Figure \ref{fig:ageval2}) stem from the fact that \citet{Mints2020} used a PARSEC grid with equal spacing in log age rather than linear age (as done in this {\tt StarHorse} run). The significant scatter seen in each of the panels of Figure \ref{fig:ageval2} demonstrates that, even when similar techniques and the same input data are used, results vary systematically. To give an extreme example, some of the GALAH DR3 stars that {\tt StarHorse} indicates to be young ($<500$ Myr) are found to be old by \citet{Buder2021}, which is very likely to be a combination of a grid-edge effect and poor stochastic sampling of the posterior.

\section{Age-abundance relations}\label{agerel}

The advantages of combining {\it Gaia} with spectroscopic data are not limited to more precise distances and stellar parameters, but also opens the possibility to study detailed chemical abundances as a function of these parameters. Certain abundance ratios are strongly correlated with age in our Galaxy and can indicate the formation of different populations \citep{Chiappini1997, TucciMaia2016, Miglio2021, Morel2021}. These relations between age and chemistry are potentially of great value for understanding and constraining Galaxy evolutionary models \citep{Chiappini2014, Nissen2015, Miglio2017}. In this section, we investigate if we can recover some of the known age-chemical correlations between the {\tt StarHorse} ages, metallicity, $\alpha$-process and as {\it s}-process elements. This exercise also serves as an additional validation for the new {\tt StarHorse} MSTO-SGB ages.

\subsection{$\alpha$ abundances and metallicities}

Since $\alpha$ elements are known to be produced mainly by the massive dying stars in type-II supernovae (SNe), those elements had a larger relative contribution to the chemical evolution of the Milky Way in the past. On the other hand, the content of elements produced by type-Ia SNe increases slowly with the enrichment of the interstellar medium. Therefore ratio of $\alpha$-capture content with iron can be broadly associated with the temporal evolution of stellar populations \citep{Tinsley1980, Matteucci1989, Chiappini1997, Woosley2002}.

Diagrams of [$\alpha$/Fe] vs. [Fe/H] have also been generally used as a classification of the stellar components of our Galaxy: the chemical thick disk is mostly at high-[$\alpha$/Fe] sequence, while the thin disk can be chemically selected as the low-[$\alpha$/Fe] sequence \citep{Edvardsson1993, Fuhrmann1998, Adibekyan2012}. The morphological thin and thick disks do not coincide exactly with their chemical definitions, thick disks form from the nested flares of mono-age populations, as shown by \citet{Minchev2015}. This model explained for the first time the presence of low-$\alpha$ stars high above the disk midplane in the outer Galaxy \citep{Anders2014, Hayden2015} and the predicted strong negative age gradient the Milky Way morphological thick disk was indeed confirmed by \citet{Martig2016}.
The high-[$\alpha$/Fe] sequence or chemically defined thick disk is mostly assumed to be old, while the low-[$\alpha$/Fe] sequence is younger, but the position and shape of these sequences are known to vary across the Galaxy \citep{Bensby2011, Anders2014}. The inner disc, for example, shows a more prominent bimodality indicating different star formation paths and evolution across the Galaxy \citepalias{Queiroz2020}. The picture also gets more complex with the detection of young-$\alpha$-rich stars \citep{Chiappini2015}. Therefore we expect a clear correlation between [$\alpha$/Fe] and age but also a large spread due to the mixing of populations \citep{Anders2017, Anders2018, Miglio2021}. In Figs. \ref{fig:alphafe} and \ref{fig:agevsalpha} we show that most of the old stars populate the high-[$\alpha$/Fe] sequence, and we confirm a relation of increasing [$\alpha$/Fe] for increasing {\tt StarHorse} age for most spectroscopic surveys, but also a significant scatter (as expected).
Older ages are also visible in Figure \ref{fig:alphafe}, especially for the APOGEE and LAMOST surveys, at high metallicities and low-$\alpha$ linking the formation of the chemical thick disk and the inner thin disk in a knee where the [$\alpha$/Fe] ratio decreases at a constant rate as a function of [Fe/H] when SNIa contribution becomes important. Another set of old stars is seen in almost all surveys at low metallicity and low [$\alpha$/Fe], which is compatible with the chemical characteristics of dwarf Galaxies and the most outer parts of the Galactic thin disk.
Although the age and $\alpha$ scatter is high in Figure \ref{fig:agevsalpha} for most surveys, one can notice that spread in age is considerably smaller for high-$\alpha$ populations, suggesting that the old high $\alpha$ sequence was formed in shorter time scale and as expected from chemo-dynamical models \citep{Minchev2017}. This result is also seen by \citet{Miglio2021}, using precise asteroseismology from red giant stars with Kepler and APOGEE spectra, which showed that the old thick disk has a spread smaller than 1.2 Gyr. 

\subsection{{\it s}-process abundances}

The slow neutron-capture process ({\it s}-process) elements are produced in the asymptotic giant-branch (AGB) phase of low- and intermediate-mass stars, hence their contribution to the interstellar medium increases steadily with time \citep{Busso1999, Sneden2008, Kobayashi2020}. Studies of low-metallicity AGBs also show a strong component of {\it s}-process elements in the Galactic halo \citep{Sneden2008, Bisterzo2014}. Among the spectroscopic surveys considered in this work, APOGEE and GALAH have measured precise high-resolution abundances for a few neutron-capture elements for a significant number of stars in the SGB regime. We choose these two surveys to explore the ratio between {\it s}-process (Yttrium, Barium and Cerium) and $\alpha$-elements with age.

In Figure \ref{fig:galahclock} the ratios between [Ba/$\alpha$] and [Y/$\alpha$] show a linear dependence with age. The data points in Figure \ref{fig:galahclock} are fitted with a non-linear least mean square method, and its uncertainty is taken as a square root from the covariance matrix. It is worth mentioning that the uncertainty associated to the fits done here are probably understimated due to the large data sets and the noise it contains, which are not variables in the fitting procedure \citep{Hogg2021}, but doing a full Bayesian fit is out of the scope of the paper. 
In the GALAH data, both [Ba/$\alpha$] and [Y/$\alpha$] show strong relations with age. The [Y/Mg] chemical clock has been extensively studied in other works, from solar twins to clusters \citep{Spina2018, Tucci-Maia2019, Nissen2020, Casamiquela2021}. This relation has no apparent variation with metallicity \citep{Nissen2020}. In Table \ref{chemicalclocktable}, we compare our resulting relations for different chemical clocks with previous works. For [Y/Mg], the linear trend with age agrees very much well with \citet{Casamiquela2021}, which is a higher value compared to the other works but still close to the values found by \citet{Spina2018, Jofre2020}. For [Ba/Si] and [Ba/Mg], our results lay in between the different relations found in the literature, overall more in agreement with \citet{Jofre2020}. This shows that {\tt StarHorse} ages are, at least, meaningful in population studies and do reproduce expected chemical-clock relations. The differences between slopes found in the literature can be attributed to the different ranges in metallicity \citep{Horta2022, Viscasillas2022}, the overabundance in neutron capture elements in open clusters compared to dwarf field stars \citep{Sales-Silva2022} or still the different spectroscopic pipelines. In section \ref{chemoap} we show the same Figures \ref{fig:galahclock} and \ref{fig:apogeeclock} colour-coded by temperature and metallicity.
In Figure \ref{fig:apogeeclock} we show yet another {\it s}-process element, Cerium, derived  by the ASPCAP synspec pipeline \citep{Jonsson2020}. The precision for Cerium in APOGEE is much lower than the previously discussed {\it s}-process abundances in GALAH. It is noticeable from the figure that there is a high spread in $[Ce/\alpha]$ vs age and that most of the cerium abundances are below the solar value. In fact, a considerable shift between the Cerium derived by APOGEE and other surveys has been reported for giant stars in the Galactic bulge \citep{Razera2022} and when compared to {\it Gaia} DR3 RVS spectra \citep{Contursi2022}. In a recent work, \citet{Sales-Silva2022} shows, also using APOGEE, that [Ce/$\alpha$] has a strong dependence in metallicity and does not work as a universal chemical-clock. In light of these complexities, the relations between [Ce/$\alpha$] abundances and age derived in this work are almost flat and have lower values than other studies in the literature \citep{Jofre2020}.

\begin{table*}
\centering
\caption{Chemical clock slopes, $m_{\rm age}$, for several abundance ratios in this study (using GALAH DR3 data; see Figure \ref{fig:galahclock}) and the literature.}
\begin{tabular}{lcccc}
Publication            & [Y/Mg]           & [Ba/Si]         & [Ba/Mg]          & [Ce/Mg]\\
\hline
This work              & -0.055$\pm$0.004 & -0.050$\pm$0.004&-0.0582$\pm$0.004 & -0.017$\pm0.005$\\
\citet{Spina2018}      & -0.045$\pm$0.002 & -               &-                 & -  \\
\citet{Nissen2020}     & -0.038$\pm$0.001 & -               &-                 & - \\
\citet{Jofre2020}      & -0.042$\pm$0.002 & -0.040$\pm$0.002&-0.047$\pm$0.002  & -0.037$\pm$0.002 \\
\citet{Casamiquela2021}& -0.055$\pm$0.007 & -               &-0.098$\pm$0.003  & - \\
\citet{Viscasillas2022}& -0.036$\pm$0.011 & -0.061$\pm$0.009&-0.103$\pm$0.006  & - \\
\hline
\hline
\end{tabular}
\label{chemicalclocktable}
\end{table*}

\section{Analysing chemo-age groups of local SGB samples}\label{tsne}

As an example science case for our new catalogues, we choose three spectroscopic surveys (GALAH DR3, APOGEE DR17, and LAMOST MRS DR7) to map different populations with high-/medium-resolution spectroscopic abundances in the local sample of SGB stars. We choose only to use the SGB since this is a fast evolutionary stage where ages have an explicit dependence on its luminosity resulting in smaller {\tt StarHorse} uncertainties, see Table \ref{uncertaintytable}. We see in section \ref{agevalidation} that there is a better agreement for OCs in the case of SGB. In section \ref{agerel}, there is a clear relation between [Ba/$\alpha$], [Y/$\alpha$] and age for these stars, which all substantiate the robustness of the {\tt StarHorse} derived ages for the SGB regime.
The three surveys were chosen due to their higher-quality abundances and completeness. In this section we use the dimensionality reduction visualization technique t-distributed Stochastic Neighbor Embedding \citep[t-SNE;][]{Hinton2003, vanderMaaten2008}, in synergy with the Hierarchical Density-Based Spatial Clustering of Applications with Noise \citep[HDBSCAN;][]{Campello2013,McInnes2017}. In the following subsections we describe t-SNE and HDBSCAN as well as their application to the SGB samples.

\subsection{Methodology: t-SNE and HDBSCAN}

Finding groups of chemically similar stars aids our understanding of the formation and evolution of the Milky Way \citep{Freeman2002}. Stellar chemical abundances of most elements remain constant during most of stellar evolution, while a star's orbit can be changed radically depending on the gravitational perturbation it suffers. The composition of a star's birth cloud dictates its chemical composition, making it possible to identify stars born in similar conditions through weak chemical tagging \citep[e.g.][]{Hogg2016}. However, differences in chemical abundances can be very subtle and become masked by their observational uncertainties making strong chemical tagging or finding co-natal stars very difficult \citep{Casamiquela2021b}. It is also important to take into account the radial migration due to the dynamical effects produce by the non-axisymmetric structures (bar and spiral arms).

One way to explore this problem is by visualizing the entire complex multi-dimensional chemical abundance and age space at once to find patterns in an easier manner. t-SNE is a statistical method for visualizing high-dimensional data by giving each data point a location in a two or three-dimensional map \citep{vanderMaaten2008}. These maps are iteratively created by minimizing the Kullback–Leibler divergence between the similarity distributions of the data in the original space and the low-dimensional map, and thus preserve the proximity between similar data points. For a slightly deeper introduction focussed on a similar science case, we refer to Section 2 in \citet{Anders2018}. As in that paper, we use the python implementation of t-SNE included in {\tt scikit-learn} \citep{Pedregosa2012}.

t-SNE has demonstrated to be an effective tool to help identify peculiar groups in different parameter spaces, and has wide applications in astronomy; e.g., stellar spectral classification \citep{Matijevic2017, Traven2017, Valentini2017, Verma2021, Hughes2022}, similarities between planetary systems \citep{Alibert2019}, galaxies classification \citep{Zhang2020,Rim2022}. Finally, \citet{Anders2018,Kos2018}, and \citet{Nepal2022} show that applying t-SNE to the abundance space to perform chemical tagging confirms cluster, stream membership and different stellar populations that compose the Galactic disk. Inspired by those works, we follow a similar approach but with a few differences; we apply t-SNE to a set of chemical abundances combined with the age information of APOGEE, GALAH, and LAMOST MRS and then, instead of looking for separations in the t-SNE by eye, we apply a clustering method to identify different chemical-age groups. 

Clustering algorithms have been extensively used in astronomy to find stellar groups in the kinematical or chemodynamical space \citep{Koppelman2019, Limberg2021, Gudin2021, Hunt2021, Shank2022}. For example, HDBSCAN is an extension of the DBSCAN \citep{Ester1996} clustering method. It converts DBSCAN into a hierarchical method by extracting flat clustering based on the stability of the clusters, which leads to the detection of high-density clusters, and it is, therefore, less prone to noise clustering than DBSCAN. 

The configuration of t-SNE+ HDBSCAN for each of the samples we discuss in the following sections is displayed in Table \ref{tsneconfig}.The main hyperparameter of t-SNE is called perplexity and controls the number of nearest neighbours. We made several tests with different values for the perplexity parameter and the random state, which can influence the local minima of the cost function \citep{wattenberg2016}. These tests are summarized in the appendix \ref{extratsne}. We always choose the t-SNE configuration that visually splits the groups more clearly.

We then apply HDBSCAN to the t-SNE projections. The three relevant hyparameters described in table \ref{tsneconfig} were optimised to obtain the "best" clustering in the sense of weak chemical tagging i.e. a configuration that does not split the data into too many small groups. Since we are searching for a more global picture of the chemistry and age distribution of stellar populations, we know that a large group should be found by the method as the "thin disk" since it should dominates our samples. The HDBSCAN hyperparameter "min\_cluster\_size" controls the minimum number of stars allowed to be considered a cluster; this parameter depends on the sample size \citep[see e.g.][and Appendix \ref{extratsne}]{McInnes2017}. The hyperparameter "min\_samples" defines how conservatively the method treats noisy data. Finally, the hyperparameter "cluster\_selection\_epsilon" controls the distance between the clusters, which can change with the t-SNE projection. We also always set HDBSCAN to "eom" as cluster\_selection\_method which is optimised for larger groupings.

\subsection{SGB samples}

In this subsection we detail the exact selection of the elemental abundances and ages used for the following t-SNE and HDBSCAN analysis, separately for the APOGEE, GALAH, and LAMOST MRS samples. While this is important to understand the differences in the results for the three surveys, readers mainly interested in the overall results may consider to move on straight to Section \ref{groups} in which we discuss the chrono-chemical groups found in our analysis.

\begin{figure*}
\includegraphics[width=0.65\textwidth, height=0.5\textwidth]{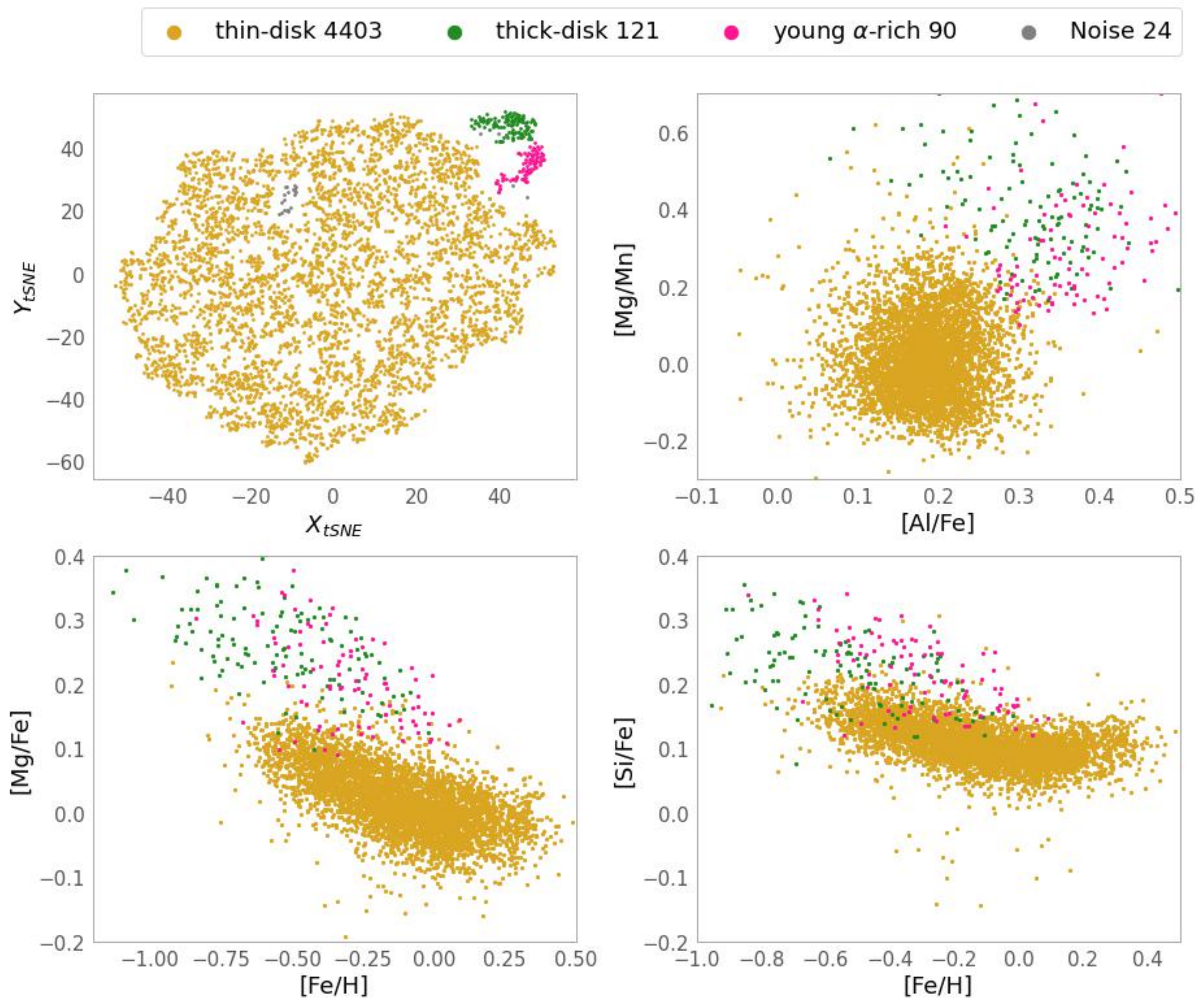}
\includegraphics[width=0.25\textwidth, height=0.5\textwidth]{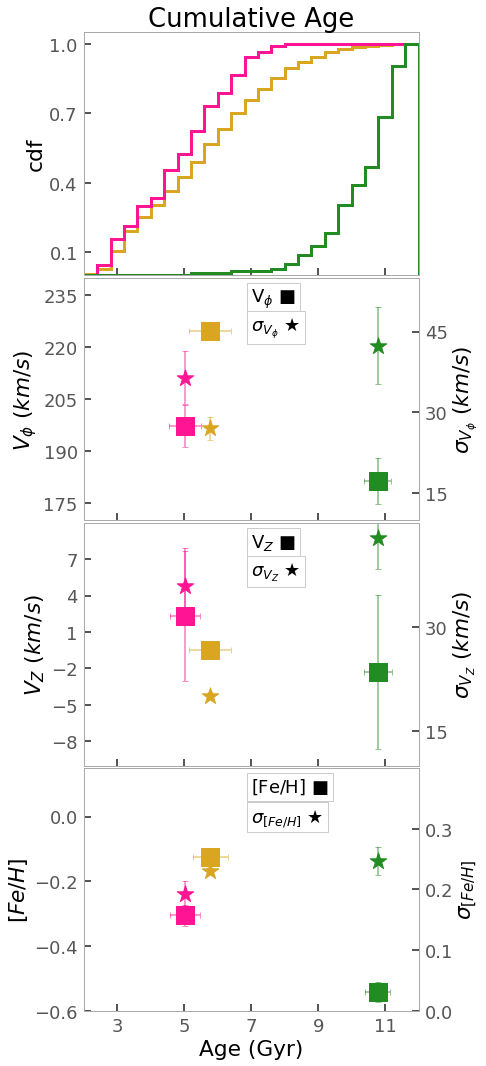}
\caption{General results of t-SNE+HDBSCAN application. Upper left panel: t-SNE projection for SGB stars in APOGEE DR17. Colours correspond to different groups found by t-SNE+HDBSCAN on the data. Lower-left and middle panels: abundance ratios of $\alpha$ elements to those of the iron group plotted against metallicity using the same colours for each identified group. Right panels from top to bottom: i) upper panel: cumulative ages distribution for each group; ii) upper middle panel: for each group, we show mean azimuthal velocity (left y-axis and square symbol) and the mean dispersion in azimuthal velocity (right y-axis and star symbol) as a function of age. iii) lower middle panel: for each group we show the mean vertical velocity (left y-axis and square symbol) and the mean dispersion in vertical velocity (right y-axis and star symbol) as a function of age. iv) bottom panel: for each group we show the mean metallicity (left y-axis and square symbol) and the mean dispersion in metallicity (right y-axis and star symbol) as a function of age. The error bars in the right panels represent the 95\% confidence interval of a bootstrap resampling.}
\label{fig:tsneapogee}
\end{figure*}

\begin{table*}
\centering
\caption{Configuration and input parameters of t-SNE+HDBSCAN}
\begin{tabular}{lrcrc}
\hline
\hline\\
\multicolumn{4}{c}{t-SNE configuration} \\\\
\hline
Survey & input &  perplexity  & random state  \\
\hline
APOGEE DR17 & [(Mg, Mn, Al, Si)/Fe] + Age$_{SH}$ & 80 & 50  \\
GALAH  DR3  & [(Mg, Al, Si, Ni, Zn, Y, Ba)/Fe] +Age$_{SH}$ & 100  & 30 \\
LAMOST DR7  & [(C, Mg, Si)/Fe]+ Age$_{SH}$ & 50 & 80  \\
\hline
\hline\\
\multicolumn{4}{c}{HDBSCAN configuration} \\\\
\hline
Survey & min\_cluster\_size &  min\_samples  & cluster\_selection\_epsilon \\
\hline
APOGEE DR17 & 38 & 1 & 0.6  \\
GALAH  DR3  & 45 & 15  & 1.7 \\
LAMOST DR7  & 207 & 8 & 1.95  \\
\hline
\hline
\end{tabular}
\label{tsneconfig}
\end{table*}

\subsubsection{APOGEE DR17}

We use APOGEE DR17 abundances from the SGB sample to find groups in the t-SNE projection with HDBSCAN. We apply the following quality cuts before executing t-SNE: SNREV $>$ 70, ASPCAP\_CHI2 $<$ 25, VSCATTER$<$1, ASPCAPFLAG$=$0, STARFLAG$=$0, "NEGATIVE" not in StarHorse\_OUTPUTFLAGS, and "CLUSTER", "SERENDIPITOUS", "TELLURIC" not in TARGFLAGS. And finally, we also make a strict cut in temperature 5500 K $<T_{\rm eff}<$ 6000 K. 

In Table \ref{tsneconfig}, we list the abundance ratios chosen as input for the t-SNE method, we only select elements with a relatively small uncertainty, as seen in Figure \ref{fig:snrsyns}. In the abundance group, we have iron-peak, odd-Z, and $\alpha$ elements. The {\it s}-process element cerium is not included because of the large uncertainties and poor statistics for the SGB sample.
We need to be cautious when using APOGEE abundances since its pipeline is optimised for giants \citep{Jonsson2020}. In the case of subgiants, there might still be many artefacts \citep[e.g.][]{Souto2021,Souto2022,Sales-Silva2022} that could possibly lead t-SNE and HDBSCAN to find an unphysical clustering in the chemo-age space. In Figure \ref{fig:snrsyns}, we see some drastic differences when one uses different spectral analysis codes in the APOGEE pipeline. Even for abundances with minimal errors like [Al/Fe] and [Si/Fe], there is a significant difference at temperatures below $T_{\rm eff}<$5500 K, thus our strict temperature cut. After a further cleaning per each abundance flag, elem\_flag$=$0, we are left with 4\,638 stars to which we apply t-SNE and HDBSCAN.

In the case of APOGEE the method finds at least three different groups (see Fig. \ref{fig:tsneapogee} and tests with t-SNE parameters in Fig. \ref{fig:perpapogee}). To check if there is any dependence of the t-SNE clustering with the abundance pipeline, we also show in the appendix Figure \ref{fig:dependencyapogee}, which displays the final t-SNE projections colour-coded by $T_{\rm eff}$, $\log{g}$, [Fe/H] and signal-to-noise ratio. Some areas of the resulting projections are predominantly found at a metallicity and $T_{\rm eff}$ range. This indicates that the clustered groups have a certain dependence on those parameters.
The groups differ in [$\alpha$/Fe] content, metallicity, and also in their age distribution. While most of the stars belong to a (chemically defined) "thin-disk" component $\approx$ with high dispersion in metallicity and age, two other groups are found with chemical characteristics of "thick disk" and "transition"/"high-$\alpha$ metal-rich" stars \citep{Fuhrmann2008, Adibekyan2011, Anders2018, Ciuca2021}. We discuss these groups in more detail together with the ones found in GALAH and LAMOST in Sect. \ref{groups} and in Appendix \ref{extratsne}.

\begin{figure*}
\includegraphics[width=0.68\textwidth, height=0.7\textwidth]{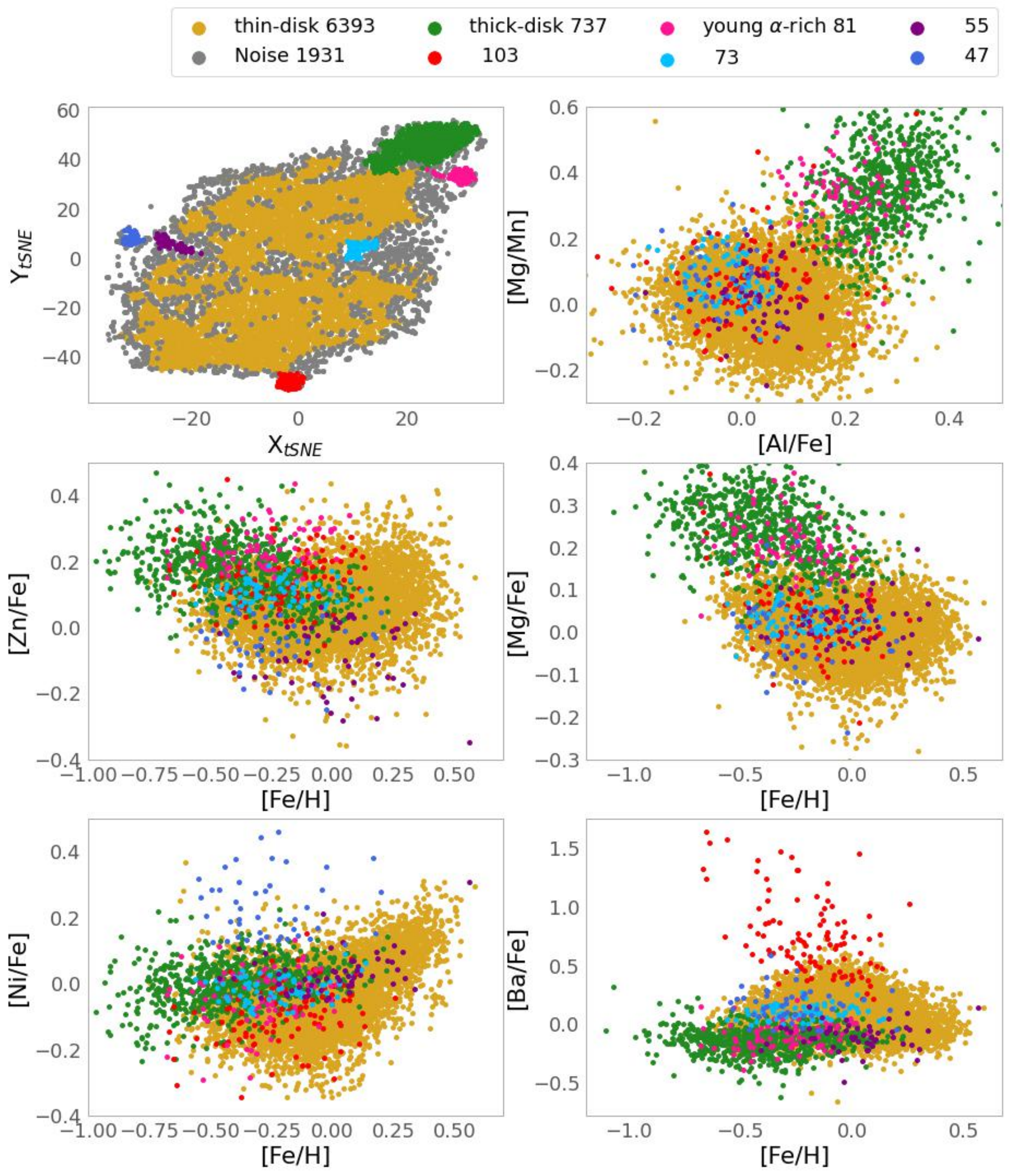}
\includegraphics[width=0.32\textwidth, height=0.65\textwidth]{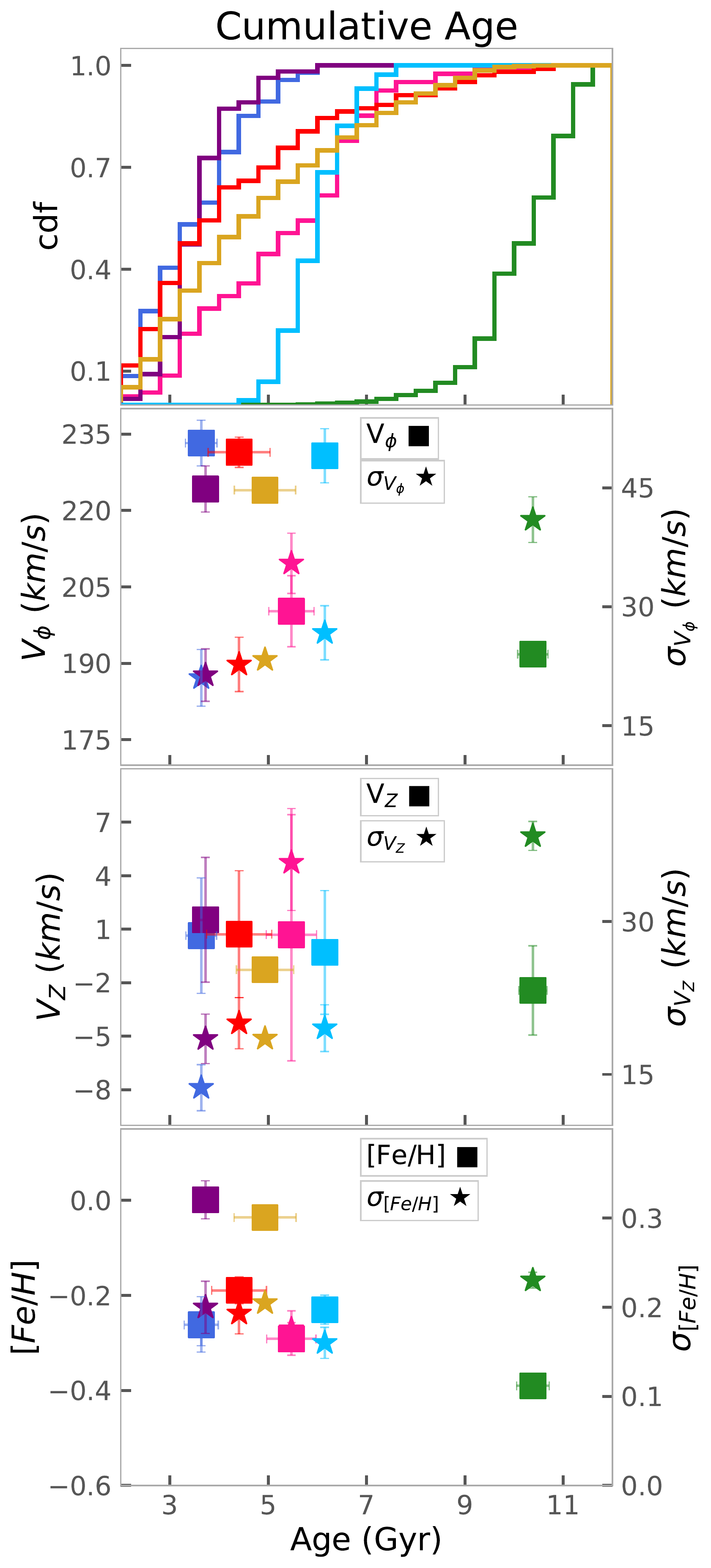}
\caption{t-SNE projection for SGB stars in GALAH DR3, colours show the groups that the HDBSCAN method identified. Abundance ratios are also show coloured by the same groups. Colours correspond to each group found by t-SNE+HDBSCAN on GALAH data. Upper pannel: Cummulative ages distibuion for each group; middle panel left axis squares: mean azimuthal velocity for each group; middle panel right axis stars: mean dispersion in azimuthal velocity for each group; lower panel left axis squares: mean vertical velocity for each group; lower panel right axis stars: mean dispersion in vertical velocity for each group. The error bars in the right panels represent the 95\% confidence interval of a bootstrap resampling.}
\label{fig:tsnegalah}
\end{figure*}

\subsubsection{GALAH DR3}

Similarly to APOGEE, we use GALAH DR3 abundances for the SGB sample to find groups in the t-SNE projection with HDBSCAN. Before performing the analysis, we made the quality cuts suggested by \citet{Buder2021}: snr\_c3\_iraf$<$30, flag\_sp$=$0, flag\_fe\_h$=$0 and 'other' not in survey\_name, as well as anything with negative extinctions in {\tt StarHorse}.
The chemical abundances that we choose for the analysis are described in Table \ref{tsneconfig}. The set contains iron peak elements as well as $\alpha$ and neutron capture elements, covering different nucleosynthetic paths.  From this group of abundances, we select only stars for which the flag\_elem $=$ 0. The GALAH DR3 SGB sample that satisfy all the mentioned flag conditions is reduced from 47\,524 to 9\,420  stars. We did not choose all abundances available in GALAH since this reduces the sample size even more drastically.
We then combine the chosen abundance ratios from Table \ref{tsneconfig} together with the ages from {\tt StarHorse} as t-SNE input. Here we also experiment with the different test parameters on t-SNE seen on Figure \ref{fig:perpgalah}. For the GALAH sample we select the case for perplexity = 100 and random  state=30. To check if there is any dependence of the t-SNE clustering with the abundance pipeline, we also show in the appendix Figure \ref{fig:dependencygalah} the final projections colour-coded by $T_{\rm eff}$, $\log{g}$, $[Fe/H]$ and Signal to noise, again here as expected there are some dependencies in temperature and therefore metallicity.

We then apply HDBSCAN to the t-SNE projection, with the parameters described in table \ref{tsneconfig}.
The results from the t-SNE projection and HBDSCAN clustering groups are shown in Figure \ref{fig:tsnegalah} along with various abundance relations for the different coloured groups. We discuss a possible interpretation for each of the groups in \ref{groups}, with a particular focus on the chemical thick disc.

\begin{figure*}
\includegraphics[width=0.7\textwidth, height=0.5\textwidth]{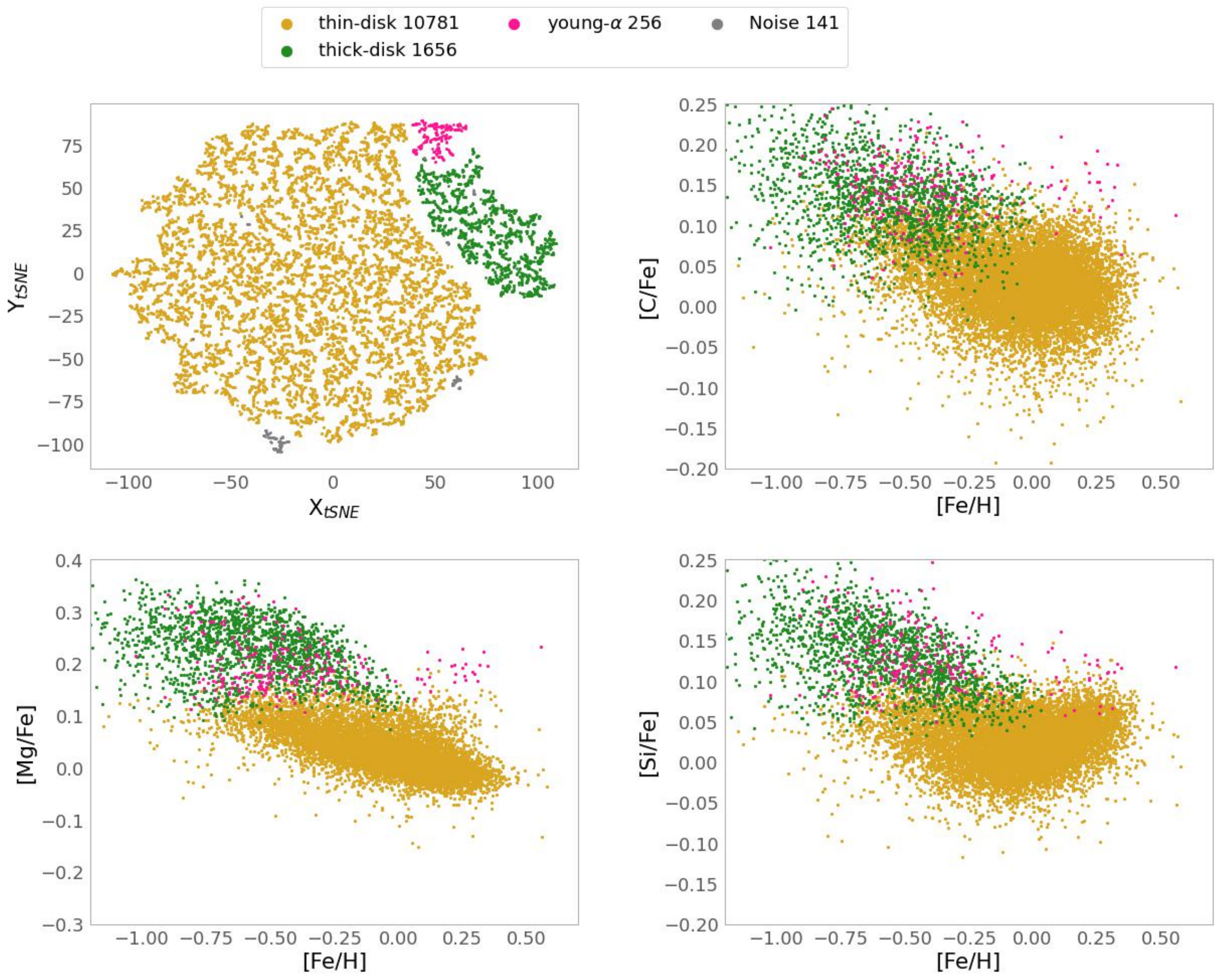}
\includegraphics[width=0.25\textwidth, height=0.5\textwidth]{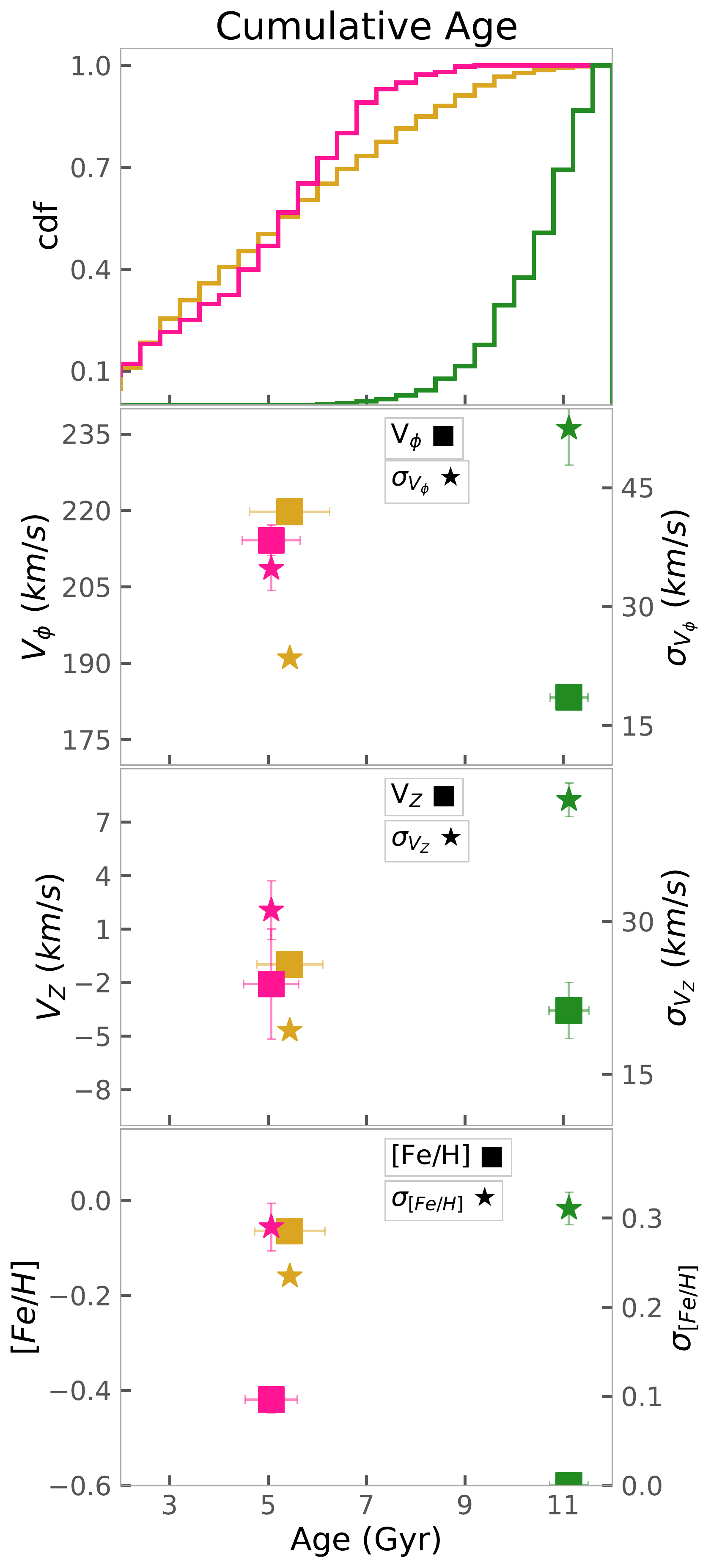}
\caption{t-SNE projection for SGB stars in LAMOST MRS. Same as Figure \ref{fig:tsneapogee} for LAMOST MRS data.}
\label{fig:tsnelamost}
\end{figure*}

\subsubsection{LAMOST MRS}
As a final exploration of the t-SNE+HDBSCAN method, we choose the medium-resolution survey from LAMOST. It is essential to keep in mind that this has a lower resolution than the previously discussed surveys APOGEE and GALAH. LAMOST MRS has 12 individual element abundances, derived through a label-transfer method based on convolutional neural network (CNN) using as training set APOGEE spectra \citep{Xiang2019}. There are many caveats in such methodologies, e.g. incompleteness and noise in the training data and unavailability of uncertainties. Therefore, we must be aware of these problems when analysing the results.
To proceed with the method, we made the following quality cuts to LAMOST MRS SGB sample: S/N $>$ 30, fibermask$=$0 and 3500$<T_{\rm eff}<$6500. We decide on the chemical abundances from LAMOST shown in table \ref{tsneconfig}. The choice is mostly based on the mutual availability of the abundances since we do not have uncertainties or flags to control in this sample. With these choices we are left with 12\,834 stars in LAMOST DR7 SGB sample. Combining the set of abundances with ages into the multidimensional space t-SNE+HDBSCAN can find three different groups as seen in Figure \ref{fig:tsnelamost} together with some abundance ratios, age and kinematical properties. We discuss a possible interpretation in the next subsections.

\subsection{Chrono-chemical groups}\label{groups}
In this subsection, we will describe the physical properties of the groups discovered in APOGEE, GALAH, and LAMOST surveys with the t-SNE+HDBSCAN method. Given the difficulty of accurately determining the uncertainties associated with each group association, we place greater emphasis on those groups that were identified in all three surveys, namely the genuine thick disk, thin disk, and young $\alpha$-rich groups. In Figure \ref{fig:mdf}, we present the average distribution of their properties (metallicity, age, and $\alpha$-enhancement). 
%In the following subsections, we will examine this distribution in greater detail to compare and contrast the findings from each survey.
It is worth noting that due to the significant differences in the quality of the stellar parameters and sample selection of each survey, the distributions of the same populations can present some differences in their properties, such as extended tails and number of peaks.

\subsubsection*{Thin disk stars (dark yellow)}
Traditionally the disk in the Milky Way and in external galaxies \citep{Dalcanton2002} can be divided into geometric thin and thick disks. The thin disk dominates in density in the solar neighbourhood since its geometrical component is denser and mostly confined to the Galactic plane. In contrast, the thick disk has a more extensive scale height \citep{Juric2008}. Since our selection of SGB stars are limited to the solar neighbourhood, we expect that our samples have a strong dominance of thin-disk-like stellar populations. However, the disks defined chemically and geometrically are not identical (\citealt[See][]{Kawata2016, Minchev2015, Anders2018} for a discussion).
Our method for recovering chrono-chemical groups finds mostly stars similar to a chemical thin-disk highlighted with the dark yellow colour on Figures \ref{fig:tsneapogee}, \ref{fig:tsnegalah} and \ref{fig:tsnelamost}. Chemically, the thin disk is much more complex and less well-mixed than the chemically defined thick disk. A metallicity gradient with radius has been long reported as a characteristic of the thin-disk \citep{Anders2014,Hayden2015} and radial migration is efficient in circular orbits, which can bring stars born in a certain inner radius to the solar neighbourhood \citep{Minchev2011,Minchev2013}. With all these complexities, we expect that a technique to search chrono-chemical groups could find multiple systems in the thin disk.
Our results show that the thin-disk population has a broad age and metallicity distribution, and multiple systems are found in the GALAH DR3 sample. The chemical composition of the thin disk does not extend much beyond 0.1 dex in $\alpha$-abundances, and their enrichment in {\it s}-process elements is higher than the detected thick disk (green dots), but there is also considerable overlap. The thin disk has orderly rotation, with smaller velocity dispersion and high rotational velocities, as seen in the right panels of Figures  \ref{fig:tsneapogee}, \ref{fig:tsnegalah} and \ref{fig:tsnelamost}. All those characteristics are consistent with the chemical and kinematical "thin disk" populations defined in multiple works in the literature \citep[e.g.][]{Adibekyan2013, Anders2018}.
The mean age of the thin disk detected in this work lies between 5 to 6 Gyr, depending on the survey. We notice from Figure \ref{fig:mdf} that the thin disk age distributions have slight differences from survey to survey; all surveys show a prominent peak at about 3 Gyr, with GALAH having a higher proportion of those young stars. While in GALAH, the thin disk stars steadily decrease in proportion with age, APOGEE shows a secondary peak at 6 Gyr, and LAMOST mainly presents a flat distribution from four to eight Gyr. Curiously on APOGEE and LAMOST, the thin-disk extends to ages larger than ten Gyr; these stars are older than one would expect for standard thin-disk formation scenarios, even though recently \citet{Prudil2020} also found evidence for a population of RR Lyrae stars older than 10 Gyr with chemo-kinematical thin disk characteristics. We can attribute the cause of the differences in the thin disk's age distribution between surveys due to their different selection functions or the breakage into subpopulations, another issue is the case for the different solar scales utilised through the suveys which can led to differences in the chemical distributions. Furthermore, the consistent result of a broad age distribution throughout the surveys is in line with a slow and inside out formation of the chemical thin-disk component \citep[e.g.][]{Chiappini1997, Chiappini2001, Minchev2013, Minchev2014}.
In Figure \ref{fig:amrsthindisk}, we show the thin disk's age vs. metallicity in the three surveys. We see that for the thin disk populations, there is a clear relation of increasing ages to decreasing metallicities until about 3 Gyr, which corresponds to the prominent young peak seen in the age distributions of Figure \ref{fig:mdf}. After 3 Gyr, the relation between age and metallicity becomes more complex. Still, there is an apparent change in the relation, suggesting an overall flat relation in age with metallicity but with high dispersion. As other works have shown it is complicated to reach strong conclusions from currently available age-metallicity relations, still affected by substantial age errors and important and difficult to correct selection effects. \citep{Feltzing2001, Casagrande2011, Bergemann2014}. Even though we can separate the thin disk via the chrono-age groups, we need to correct for selection effects, which is out of the scope of this paper. We refer to future works for a proper analysis of the age metallicity relation of these samples.

\subsubsection*{Genuine thick disk stars (green)}
We find stars compatible with the abundance pattern of chemically defined thick disk stars \citep{Reddy2006,Adibekyan2011,Bensby2014, Anders2014, Nidever2014, Mikolaitis2014, Hayden2015}, which present high alpha abundances in the three different SGB samples for which we run t-SNE+HDBSCAN. Here we will refer to this population as the "genuine thick disc". These stars clearly occupy the high-[Mg/Fe] sequence in the classical Tinsley-Wallerstein diagram \citep{Wallerstain1962} (lower left panel of  Figures \ref{fig:tsneapogee} ; \ref{fig:tsnelamost} and central panel of Figure \ref{fig:tsnegalah}) and show elevated [Mg/Mn] and [Al/Fe] abundances \citep{Das2020}. In the GALAH sample,  where we have the {\it s}-process abundance ratios like [Ba/Fe], it shows a slightly lower location than the bulk of the populations. At the same time, [Zn/Fe] is mildly enhanced compared to the thin disk at the same metallicity in agreement with previous measurements \citep{Delgado-Mena2017, Friaca2017}. 
In the right panels of Figs. \ref{fig:tsnegalah},\ref{fig:tsneapogee} and \ref{fig:tsnelamost} we show the cumulative age distribution as well as the age-$V_{\Phi}$, age-$\sigma_{V_{\Phi}}$, age-$V_Z$ age-$\sigma_{V_{Z}}$ relations, binned by HDBSCAN population. These plots show that the genuine thick disk is relatively old ($\gtrsim 10.9$ Gyr), has lower rotation and is kinematically hotter than the thin disk populations, in line with the recent analysis of \citet{Rendle2019, Miglio2021}. This result also agrees with observations of disk galaxies at redshifts $\approx$ 2, \citet{Ubler2019} measured velocity dispersions of about 45 km/s for thick disks observed at that look-back time.
Is also very clear that the genuine thick disk found by t-SNE+HDBSCAN has a contrasting mean age difference and a noticeable jump in velocity dispersion compared with all the other populations in the SGB samples, suggesting that it has indeed a very different formation path in agreement with \citealt{Chiappini1997, Chiappini2001, Reddy2006, Miglio2021}. Recent self-consistent dynamical models of the Milky Way, also show distinctive characteristics in the kinematics and composition of the thin and thick disks \citep{Robin2022}. The age distribution of genuine thick disk stars in Figure \ref{fig:mdf} shows a double peak. The second prominent peak at ages between 9-10 Gyr (very clear in the GALAH sample) is possibly the contribution of transition or bridge stars, previously detected by other works \citep{Anders2018, Ciuca2021}. The transition stars were probably formed in the inner Galaxy and extend from the high $\alpha$ abundances to low $\alpha$ and high metallicities, filling the gap between the thin and thick disks [$\alpha$/Fe] diagram. Further analysis of bridge stars with t-SNE+HDBSCAN is a matter for a forthcoming paper  \citep{Nepal2022}.
The metallicity distribution shown in the left panels of Figure \ref{fig:mdf} is reasonably similar for the three samples of SGB stars, ranging from -1.5 to 0.0 and with a clear peak at around -0.5 dex. For APOGEE and LAMOST the metallicity distribution shows a second prominent peak at -0.8 dex. The [Mg/Fe] distribution is also analogous throughout the different surveys showing a smooth distribution from 0.1 to 0.4 dex. 
In Table \ref{gthicktable} we compare the mean age, azimuthal velocity and their dispersion values. We see that the age values agree very well between the surveys, varying from 10.4-10.9 Gyr, and the highest age dispersion being that of LAMOST DR7, of 1.35 Gyr. The age and age dispersion might be higher for LAMOST and APOGEE due to minor debris contamination as we see some stars in the green group to extend to the very metal-poor side.

\subsubsection*{Young $\alpha$-rich (magenta)}

\begin{figure*}
\centering
\includegraphics[width=0.85\linewidth]{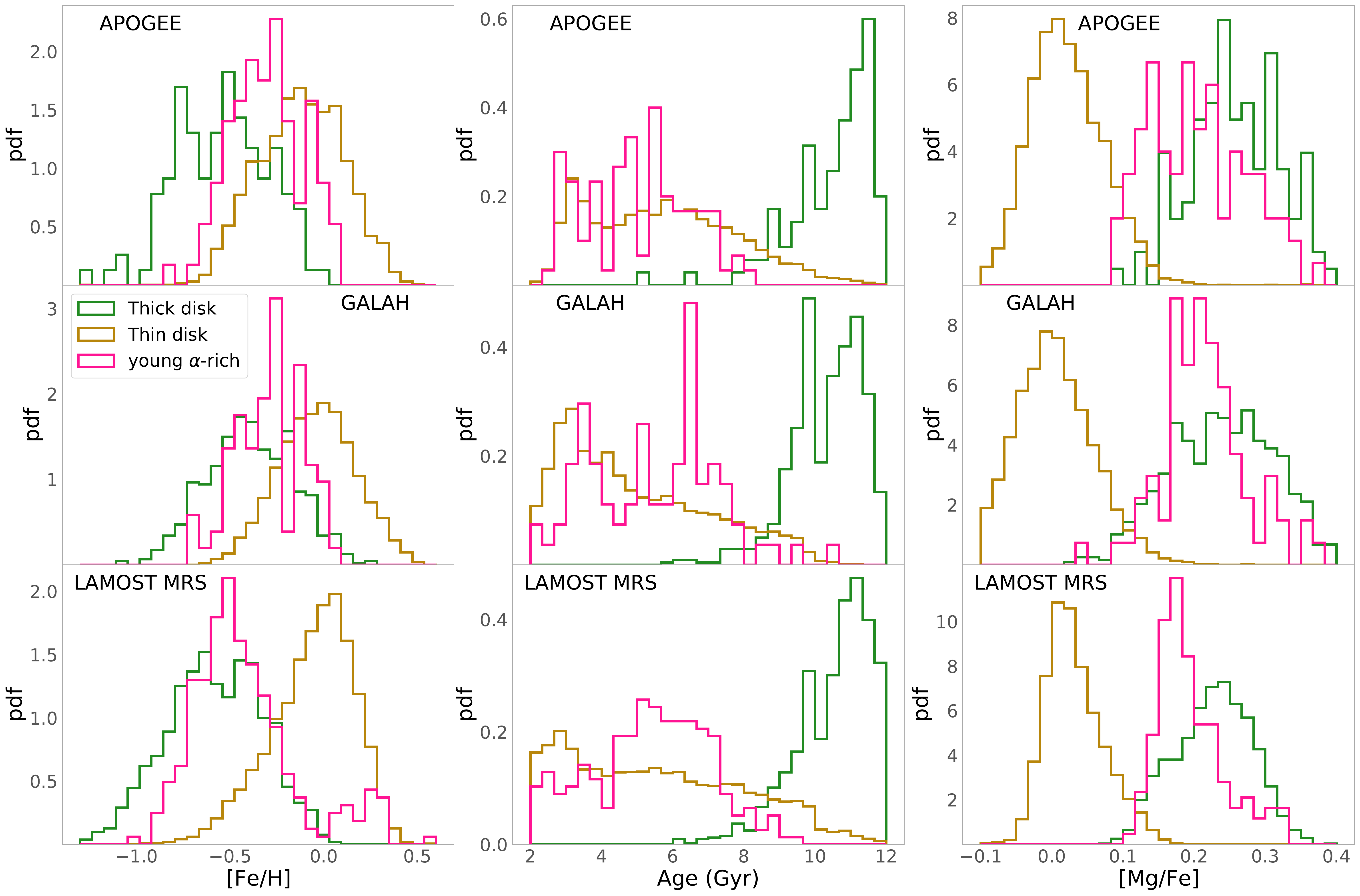}
\caption{Probability density function of metallicity, age and [Mg/Fe] for the main populations founded in APOGEE, GALAH and LAMOST with t-SNE HDBSCAN.}
\label{fig:mdf}
\end{figure*}

\begin{figure*}
\includegraphics[width=0.31\textwidth, height=0.35\textwidth]{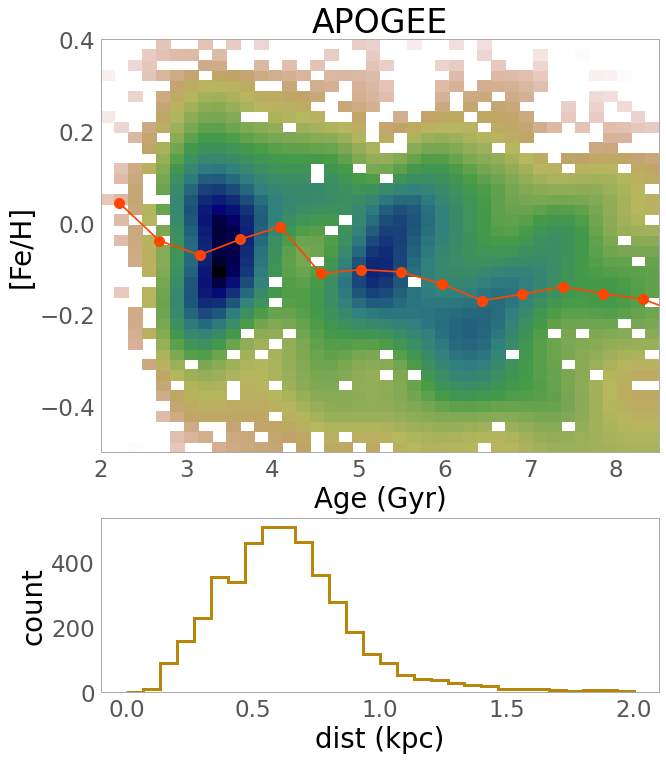}
\includegraphics[width=0.31\textwidth, height=0.35\textwidth]{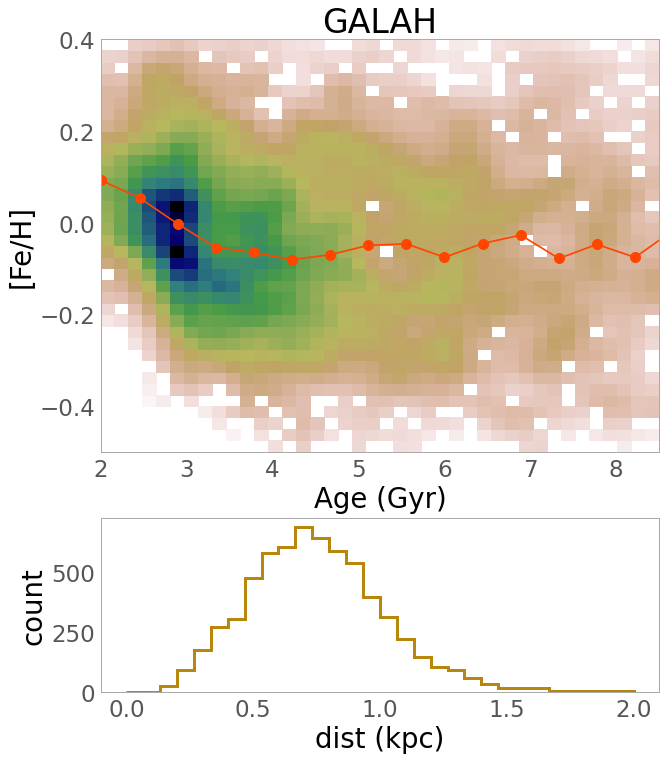}
\includegraphics[width=0.31\textwidth, height=0.35\textwidth]{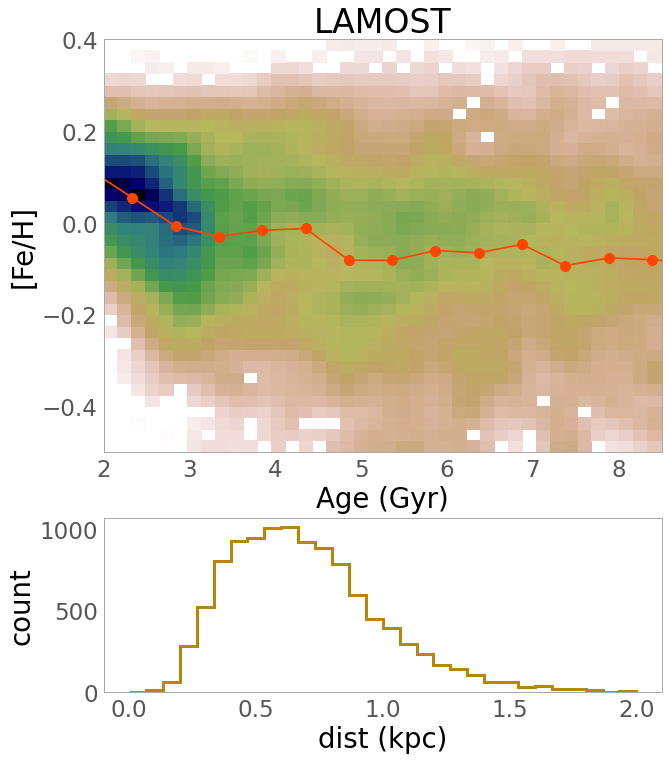}
\caption{Age metallicity relations for the thin disk. Upper panels: metallicity vs age for the detected thin-disk component in three surveys. The map shows the mean density per pixel where we have also applied gaussian smoothing. The red dotted line represents the median metallicity for the given age. The lower panels show the heliocentric distance coverage for each sample.}
\label{fig:amrsthindisk}
\end{figure*}

\begin{figure}
\centering
\includegraphics[width=0.85\linewidth]{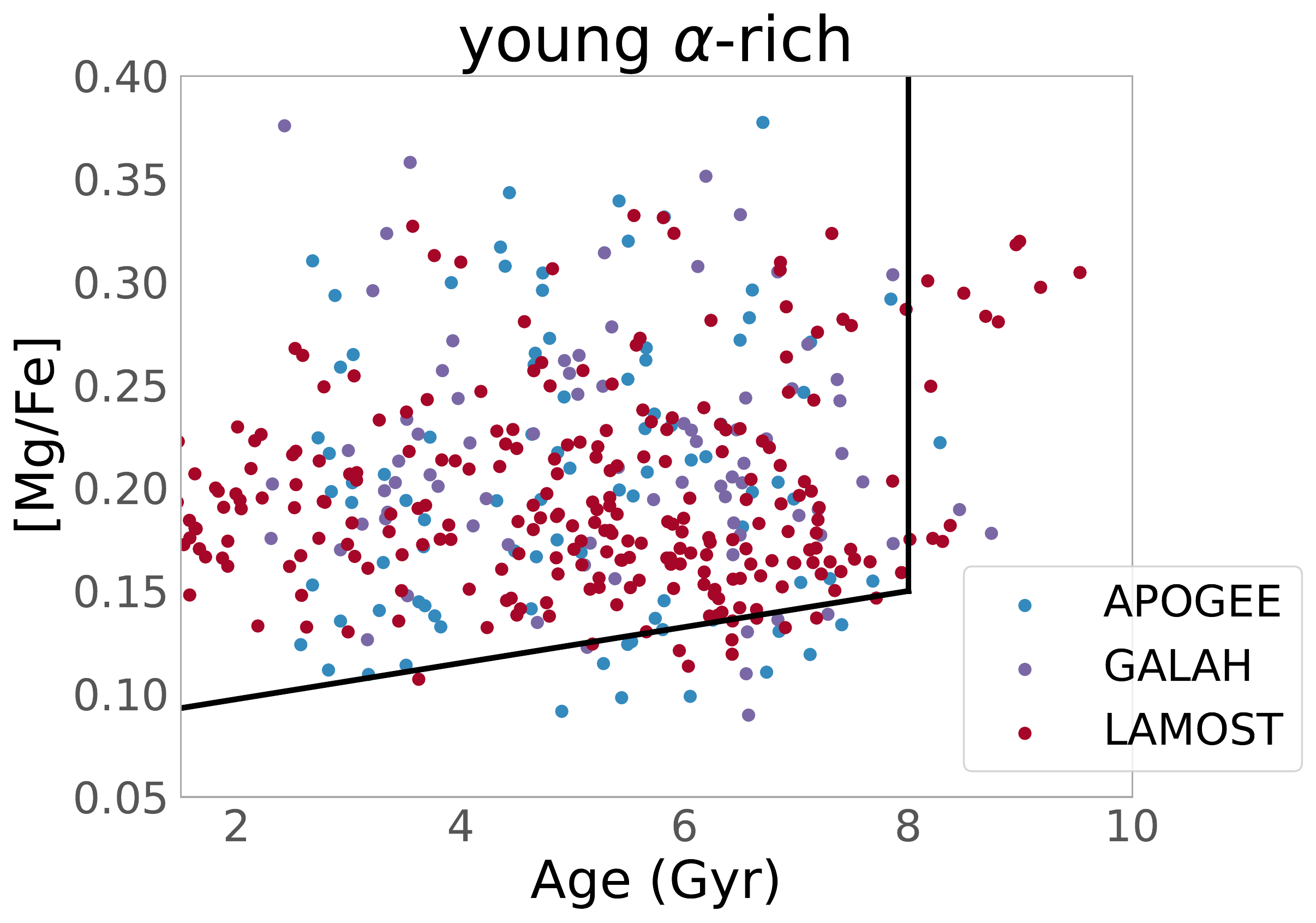}
\caption{Age vs. [Mg/Fe] for the young $\alpha$-rich groups found in APOGEE, GALAH and LAMOST. The black delineation shows the region where chemical evolutionary models cannot explain, as in \citet{Chiappini2014}.}
\label{fig:yalpha}
\end{figure}

Although young $\alpha$-enhanced stars cannot be explained by standard chemical evolution models, a significant number of them have been previously detected by diverse works in the literature \citep{Chiappini2015,Martig2015,Jofre2016, Silva-Aguirre2018,Ciuca2021, Miglio2021}. Our method also recovers stars with such characteristics in the SGB samples of GALAH, APOGEE and LAMOST, which we indicate by the magenta colour in Figures \ref{fig:tsneapogee}, \ref{fig:tsnegalah}, \ref{fig:tsnelamost}, and \ref{fig:mdf}. In Figure \ref{fig:yalpha}, we show that most of the stars detected as the magenta group fall in the area delimited by the black curves. It is hard to explain with chemo-evolutionary models of the Milky Way stars that fall in this area \citep{Chiappini2015}. 
The young $\alpha$-rich populations detected here show a mean age of about 5 Gyr for the three different surveys. The cumulative age distributions have truncation at about 7 Gyr, but we see an extension to older ages in GALAH and LAMOST. Those older stars could perhaps again be part of the transition or bridge stars which have intermediate chemical characteristics between thin and thick disks \citep{Anders2018}.
The range of abundances of the young-$\alpha$-rich stars is similar to the genuine-thick-disk (green-dots), except that their metallicity and [$\alpha$/Fe] distribution is more concentrated at intermediate values between thin-disk (dark yellow) and genuine-thick disk stars, see Figure \ref{fig:mdf}. For LAMOST the metallicity distribution of the magenta stars is overall poorer than in APOGEE and GALAH, although their [Mg/Fe] is lower which could classify some of those stars as debris \citep{Hasselquist2021, Limberg2022a}, outer disk, or a pipeline problem. Is worth mentioning that the [Mg/Fe] measure through CNN algorithm in LAMOST DR7 MRS is not always consistent with the [$\alpha$/Fe] measured from LASP pipeline \citep{Wu2014}.
We also notice differences between the green and magenta stars in the [Mg/Mn] vs [Al/Fe] diagram: contradictorily in APOGEE these stars are richer in [Al/Fe], ranging from 0.2 to 0.5 dex, while in GALAH they range in [Al/Fe] from 0 to 0.2 dex. The kinematics of the young-$\alpha$-rich stars is mostly hot  from the right panels of Figures \ref{fig:tsneapogee}; \ref{fig:tsnegalah}; \ref{fig:tsnelamost} the magenta groups show velocities dispersion both vertically and azimuthally of about 35 km/s. Although their mean V$_{\phi}$ is similar to the one of the thin disk,  the hot kinematics agrees with previous works \citep{Silva-Aguirre2018,Miglio2021,Ciuca2021} suggesting that these stars formed from the same gas as the genuine-thick disk but they appear young because they are probably mergers from binary stars \citep{Jofre2016}. %Another possible explanation for 

\subsubsection*{Other populations found in Galah DR3}

{\it Enriched {\it s}-process stars (red)}:

For the GALAH SGB sample, we can identify enhanced {\it s}-process stars since [Ba/Fe] is available and input to t-SNE. This chrono-chemical group shares very similar properties to the thin disk, showing low-$\alpha$ enhancement and a large dispersion in age and iron-peak elements,  but a significant difference in barium enhancement , [Ba/Fe]$>$ 0.5. The right panels of Figure \ref{fig:tsnegalah} show that these stars are slightly younger than the thin disk group, with a mean age of about 4.5 Gyr. Their rotation is similar to the thin disk, showing high azimuthal velocities and low-velocity dispersion. Despite broad distribution in [Fe/H], these stars almost do not present any higher metallicities than Solar, and the most barium-enhanced stars are on the metal-poor side. One can also notice a considerable fall of [Ba/Fe] for metal-rich in the thin-disk \citep{Israelian2014, Bensby2014}. However, it could be challenging to measure barium at higher metallicities as mentioned before in \citet{Delgado-Mena2017, Buder2019}. We also checked that the high [Ba/Fe] stars are also enriched in [Y/Fe] and [Zr/Fe].
The stars detected here do not belong to the enhanced barium stars seen as Am/Fm stars \citep{Fossati2007, Xiang2020, Buder2021}, since those are much younger, with high temperatures, $T_{\rm eff}$ > 6500 K and extremely low $\alpha$ abundances.
These enriched {\it s}-process stars could be the outcome of binary stars systems that have accreted mass from a dim white dwarf companion enriching them with heavy elements \citep{McClure1983}. Two of the 103 stars detected here are also in the binary catalogue from \citet{Traven2020}. Another possibility is that these stars come from an accreted dwarf galaxy, since in those systems stars can present a different chemical evolution than in the Milky Way. \\

{\it Outer thin disk (Cyan)}:

The stars marked as cyan are mainly concentrated at the metal-poor end of the chemical thin disk, occupying the locus of outer disk stars in an [alpha/Fe] and [Fe/H] plot \citep{Hayden2015, Queiroz2020}. The different populations overlap in [Zn/Fe] and [Ni/Fe]. The cyan points show systematically larger [Ba/Fe] ratios than the thick disk stars (green), but also lower than the rest of thin disk stars. This is into slight contradiction with the inside-out picture formation \citep{Chiappini2001} in which the outer disk star formation history proceeds on longer timescales than the in the inner parts of the Galaxy leading to a larger Ba enrichment by low and intermediate mass stars. 
Perhaps this population is related to recent works finding metal-poor stars with thin disk rotation \citet{Fernandez-Alvar2021}. 
We see from the right panels of Figure \ref{fig:tsnegalah},  that the cyan population presents a steep age distribution at about 6.0 Gyr and has an older mean age than the thin disk. The [Mg/Mn] vs. [Al/Fe] diagram, upper panel second column of Figure \ref{fig:tsnegalah}, shows that the cyan population is at low [Al/Fe] and intermediate [Mg/Mn] borderline to the region occupied by dwarf galaxies \citep[e.g.][]{Limberg2022a, Hawkins2015, Das2020}.
The chemical characteristics and the older age attributed to this group could also indicate that these stars have been formed by gas polluted by the accretion of a dwarf Galaxy, e.g. {\it Gaia}-Enceladus dwarf, similar to what \citet{Myeong2022} recently suggests as {\it EoS} system which would chemically evolve to resemble the outer thin disk. Another possible interpretation for the characteristics of the cyan group is that they are the outcome of the perturbation caused by one of the passages of the Sagittarius Dwarf. In the star formation history reconstructed by \citet{Ruiz-Lara2020}  there is a clear peak at about 5.7 Gyr which coincides with the mean age of the cyan group found by the t-SNE+HDBSCAN method. 
The robustness of this group is not very strong when we introduce noise to the t-SNE method as seen in appendix \ref{extratsne}, which means this populations has a weak signal in the data and needs further investigation. The tests with the random\_state parameter in appendix C also show that this group is not robust. Although it presents an interesting peaked age distribution, this grouping might be an artificiality introduced by the t-SNE+HDBSCAN method uncertainties.

{\it Young chemically peculiar stars (Navy blue, purple)}:

These two groups of stars are amongst the youngest stars detected by our method. They present high rotational velocities and very low dispersion indicating these stars were probably formed within the thin disk. The navy-coloured stars show high [Ni/Fe] and low [Zn/Fe] content and are also at hotter temperatures, T$_{\rm eff}$>6100K see Figure \ref{fig:dependencygalah}. The metallicity of the navy stars is concentrated in the metal-poor end of the thin disk, not extending further than Solar metallicities, similar to the previously discussed outer disk but significantly younger. The contribution of nickel is higher than iron in supernova (SNe) type Ia \citep{Tsujimoto1995, Sneden1991}. As we see in the left lower panel of Figure \ref{fig:tsnegalah} thin disk stars gradually become more [Ni/Fe] enriched for higher metallicities where  SNe type Ia contribution dominates the interstellar medium. Therefore is puzzling that the stars rich in [Ni/Fe] have metallicities lower than solar. Since these stars are at a similar temperature range, it could also indicate a problem in the pipeline. 
In contrast, the purple stars cover almost the whole range of metallicities as the thin disk, but they present low [Ba/Fe], similar to thick disk stars and very low [Zn/Fe]. Trends of low [Zn/Fe] for higher metallicities are seen in the direction of the Galactic bulge \citep{Barbuy2015, Duffau2017}, although the stars that we discuss here appear too young to have migrated from the Galactic centre. 
It is also true that for these populations the robustness of the groups is heavily perturbed when we introduce noise to the t-SNE as seen in appendix \ref{extratsne}.

\begin{table}
\caption{Mean parameters of the genuine thick disk found in the different surveys}
\begin{tabular}{lrcrc}
Survey & $age$ &  $\sigma_{age}$  & $V_\phi$  & $\sigma_{V_\phi}$ \\
       & (Gyr) & (Gyr) & (km/s) & (km/s)  \\
\hline
LAMOST DR7 MRS & 11.12 & 1.35 & 192.66 & 52.50  \\
GALAH  DR3     & 10.38 & 1.03 & 191.78 & 40.98 \\
APOGEE DR17    & 10.77 & 1.33 & 181.31  & 42.37 \\
\hline
\hline
\end{tabular}
\label{gthicktable}
\end{table}

%amr_thindisk_lamost.png

\section{Conclusions}\label{discuss}

We present new {\tt StarHorse} catalogues for eight (past and ongoing) spectroscopic surveys, coupled with {\it Gaia} DR3 and multi-wavelength photometry. We deliver a total of 10\,998\,676 distances, extinctions, masses, temperatures, surface gravity, metallicities, as well as  $\approx$ 2.5 million age estimates for MSTO+SGB stars. For APOGEE DR17 results are also available, except for ages, as a value-added catalogue of the survey \citep{sdss2021}. 

Compared to the \citetalias{Queiroz2020}-{\tt StarHorse} release, we have included new results for more than 4 million stars from {\it Gaia} RVS spectra and additional data from LAMOST, GALAH, APOGEE and GES. For RAVE DR6 and SDSS/SEGUE we have updated our results to include {\it Gaia} DR3 parallaxes. We also make available {\tt StarHorse} ages for the first time by selecting stars in the MSTO+SGB evolutionary stages, since the age determination by isochrone fitting methods is not reliable outside of this regime. 

Validation against other methods \citep{Xiang2022, Buder2022, Mints2020} and OCs \citep{Cantat-Gaudin2020} have shown that our ages are trustworthy for stars older than 2 Gyr. {\tt StarHorse} parameters have also been extensively validated in \citetalias{Santiago2016, Queiroz2018, Queiroz2020}.

We deliver typical distance uncertainties of about 4-10\% and SGB age uncertainties of about 8-20\%, depending on the survey's spectroscopic resolution. For distances, our results are about 5 to 10\% better than when spectroscopic information is not used \citep{Anders2022, Bailer-Jones2021}. The inclusion of {\it Gaia} DR3 astrometry, along with improvements in the spectroscopic pipelines of GALAH, APOGEE, LAMOST and GES, allowed us to determine more precise parameters than in our earlier papers (\citetalias{Queiroz2018, Queiroz2020}).

By combining the chemical abundances and radial velocities from the spectroscopic releases with the final {\tt StarHorse} data products, we were able to make the following findings:

\begin{itemize}

\item We have shown classical diagrams of [$\alpha$/Fe] vs [Fe/H] colour-coded by age for each of the eight spectroscopic surveys. The results manifest the old thick disk population at high [$\alpha$/Fe], the old accreted metal-poor stars with low $\alpha$ abundances, and a transition population extending from the thick disk to the high metallicity inner thin disk stars with intermediate ages \citep{Anders2018, Ciuca2021, Nepal2022}. We see a non-linear relation between $\alpha$ abundances and age for surveys with typical uncertainties below 30\%. We also notice in Figure \ref{fig:agevsalpha} that the age dispersion decreases with increasing [$\alpha$/Fe], in LAMOST LRS, the age dispersion is of about 3 Gyr for an [$\alpha/Fe]=$ -0.1 and of only 1.4 Gyr for an [$\alpha$/Fe]$\sim 0.3$. The statistics are similar for the other surveys, indicating that old stars (mostly thick disk high-$\alpha$) had a fast formation history  \citep{Miglio2021}.

\item The dependence of {\it s}-process/$\alpha$-process abundances ratios against age (chemical-clocks) for the local sample of SGB stars reveals a linear correlation in most cases. The correspondence is strong for several abundance ratios, especially for  [Ba/$\alpha$]. A comparison with literature results of \citet{Spina2018, Jofre2020, Casamiquela2021} for the same chemical clocks shows a similar effect, demonstrating that the {\tt StarHorse} ages are sensitive enough to the abundance variations. The chemical clock's determination also covers a large number of stars in the local volume of GALAH DR3 ($\approx$ 18\,000 stars).

\item Using an unsupervised machine learning approach coupled with a clustering algorithm, we can map different populations into their unique chemical age properties. For this exercise, we have collected a set of abundances spanning distinct nucleosynthetic paths and the SGB ages for three different surveys APOGEE DR17, GALAH DR3 and LAMOST MRS. In all samples, we recovered the same three populations: chemical thin disk, genuine thick disk and young-$\alpha$-rich stars corroborating the method's robustness. We stress that our method avoids pre-assumptions on the chemistry or kinematics of the thick and thin disk components of the Milky Way.

\item Our results show that the stars we obtained from t-SNE+HDBSCAN cluster method and that follow the chemical pattern of the thin disk have low $\alpha$ abundances, span a broad distribution in metallicity, have a mean age of about 5.0 Gyr, prominent peaks at 3 Gyr and a flattened distribution from 4 to 9 Gyr. These chemical characteristics and the flattened distribution of ages are in line with the slow and inside-out formation of the thin disk  \citep{Chiappini1997, Minchev2013}. At the same time, the younger counterpart shows the influence of mergers in the star formation history of the thin disk  \citep{Ruiz-Lara2020}. In APOGEE and LAMOST, a small portion of stars also extends to ages larger than 10 Gyr, indicating clumpy star formation scenarios in the early disk \citep{Beraldo-e-Silva2021}.

\item Stars marked as green in Figures  \ref{fig:tsneapogee}, \ref{fig:tsnegalah}, and \ref{fig:tsnelamost} represent genuine thick disk stars. They have high [$\alpha$/Fe] and lower metallicity, as seen by many works in the literature \citep{Adibekyan2011, Bensby2014, Anders2014}. We find mean age values in this group ranging from 10.38-11.77 Gyr depending on the survey, although all age distributions exhibit a double peak at  $\approx$ 11.5 Gyr and $\approx$ 9.5 Gyr. The younger counterpart of the genuine thick disk is probably a contribution of another population described in the literature as transition or bridge stars \citep{Anders2014, Ciuca2021}, further analysis of transition stars with high-resolution samples is part of a forthcoming paper \citep{Nepal2022}. These results corroborate a formation scenario for the thick disk that happened at lookback times of z $\approx$ 2 (lookback time of 10-12 Gyr), and according to it, the small age dispersion of 1.05-1.35 Gyr indicates that the thick disk was fully formed before the interaction with {\it Gaia} Enceladus  \citep{Miglio2021, Montalban2021}.

\item The genuine thick disk dispersion in velocity is strikingly different from the thin disk, with values of standard deviation in vertical and azimuthal velocity of about 50 km/s which is in agreement with recent self-consistent dynamical models of the Milky Way \citep{Robin2022}. This result also agrees with the kinematics of extragalactic thick disks at redshift $\approx$ 2. Based on KMOST integral field spectroscopy \citet{Ubler2019, Schreiber2020} suggest that gravitational instabilities power the large velocity dispersions observed in thick disks. This suggests the chemical bimodality \citep{Queiroz2020} to be linked to a kinematical bimodality \citep{Miglio2021}, a clear signature of stellar populations formed during different star formation regimes.

\item We find a significant number of young $\alpha$-rich stars in all surveys studied with t-SNE and HDBSCAN (427 stars). These stars have chemical enrichment and kinematics very similar to the genuine thick disk but a contrasting younger age that cannot be explained by any Milky way evolutionary models \citep{Chiappini2015, Martig2015}. The fact that these stars present large velocity dispersions suggests that they were formed in the same gas as the genuine-thick disk \citep{Silva-Aguirre2018, Miglio2021, Lagarde2021}. They appear to be younger because they potentially are the outcome of binary stars mergers \citep{Jofre2016}.

\item Besides the chemical thin disk, thick disk and young $\alpha$-rich stars, we find in the GALAH DR3 SGB sample another four groups within the low-$\alpha$ regime. Some of these stars show high {\it s}-process enrichment (red), some show characteristics similar to outer disk stars (cyan), and some are young and show peculiar enrichments in iron-peak elements (purple and navy blue). These population singularities can be caused by mass accretion in binary interactions and consequent passage and perturbation from dwarf galaxies. The stars marked as the outer disk (cyan) and peculiar (purple) have a low significance and can be easily perturbed by noise, according to our tests in appendix C. Therefore, complementary information is needed to confirm their reality. 
\end{itemize}

In summary, we deliver catalogues with precise astrophysical parameters for public spectroscopic surveys and for the first time, we provide {\texttt{StarHorse}} age estimates on a large scale. These catalogues are fundamental for Galactic archaeology and work as optimal training sets for machine-learning algorithms that extend these results to larger samples. The new approach we presented here by joining t-SNE+HDBSCAN to detect different chrono-chemical populations in the solar neighbourhood has shown to be robust across surveys of various pipelines and resolution quality, sampling a variety of chemical elements. The method is ideal for disentangling the overlapping properties of stellar populations in our Galaxy. We also make available a catalogue with the IDs of all the groups we found. In two accompanying papers, we use this technique, applied to high-resolution samples, to study the age and chemical structure of the local disk (revealing clearly distinct thin disc, thick disc, and high-alpha metal-rich components; \citealt{Nepal2022}, and the Galactic bulge population also in comparison to local samples, but without age information; \citealt{Queiroz2022}). Two recent publications make use of our datasets to successfully investigate and characterize halo substructures \citep{Perottoni2022,Limberg2022}.
All the samples published here in conjunction with the first release of ages will play a vital role in the future. With 4MOST \citep{DeJong2019}, we can extend the volume for which this will be possible.

\begin{acknowledgements}

We thank the referee for the suggestions and contructive report.
The {\tt StarHorse} code is written in python 3.6 and makes use of several community-developed python packages, among them {\tt astropy} \citep{AstropyCollaboration2013}, {\tt ezpadova}\footnote{\url{https://github.com/mfouesneau/ezpadova}}, {\tt numpy} and {\tt scipy} \citep{Virtanen2019}, and {\tt matplotlib} \citep{Hunter2007}. The code also makes use of the photometric filter database of VOSA \citep{Bayo2008}, developed under the Spanish Virtual Observatory project supported from the Spanish MICINN through grant AyA2011-24052.

Funding for the SDSS Brazilian Participation Group has been provided by the Minist\'erio de Ci\^encia e Tecnologia (MCT), Funda\c{c}\~ao Carlos Chagas Filho de Amparo \`a Pesquisa do Estado do Rio de Janeiro (FAPERJ), Conselho Nacional de Desenvolvimento Cient\'{\i}fico e Tecnol\'ogico (CNPq), and Financiadora de Estudos e Projetos (FINEP).

Funding for the Sloan Digital Sky Survey IV has been provided by the Alfred P. Sloan Foundation, the U.S. Department of Energy Office of Science, and the Participating Institutions. SDSS-IV acknowledges support and resources from the Center for High-Performance Computing at the University of Utah. The SDSS web site is \url{www.sdss.org}.

SDSS-IV is managed by the Astrophysical Research Consortium for the Participating Institutions of the SDSS Collaboration including the Brazilian Participation Group, the Carnegie Institution for Science, Carnegie Mellon University, the Chilean Participation Group, the French Participation Group, Harvard-Smithsonian Center for Astrophysics, Instituto de Astrof\'isica de Canarias, The Johns Hopkins University, 
Kavli Institute for the Physics and Mathematics of the Universe (IPMU) / University of Tokyo, Lawrence Berkeley National Laboratory, 
Leibniz-Institut f\"ur Astrophysik Potsdam (AIP),  
Max-Planck-Institut f\"ur Astronomie (MPIA Heidelberg), 
Max-Planck-Institut f\"ur Astrophysik (MPA Garching), 
Max-Planck-Institut f\"ur Extraterrestrische Physik (MPE), 
National Astronomical Observatory of China, New Mexico State University, 
New York University, University of Notre Dame, 
Observat\'ario Nacional / MCTI, The Ohio State University, 
Pennsylvania State University, Shanghai Astronomical Observatory, 
United Kingdom Participation Group,
Universidad Nacional Aut\'onoma de M\'exico, University of Arizona, 
University of Colorado Boulder, University of Oxford, University of Portsmouth, 
University of Utah, University of Virginia, University of Washington, University of Wisconsin, 
Vanderbilt University, and Yale University.

Guoshoujing Telescope (the Large Sky Area Multi-Object Fiber Spectroscopic Telescope LAMOST) is a National Major Scientific Project built by the Chinese Academy of Sciences. Funding for the project has been provided by the National Development and Reform Commission. LAMOST is operated and managed by the National Astronomical Observatories, Chinese Academy of Sciences.

Funding for RAVE has been provided by: the Australian Astronomical Observatory; the Leibniz-Institut f\"ur Astrophysik Potsdam (AIP); the Australian National University; the Australian Research Council; the French National Research Agency; the German Research Foundation (SPP 1177 and SFB 881); the European Research Council (ERC-StG 240271 Galactica); the Istituto Nazionale di Astrofisica at Padova; The Johns Hopkins University; the National Science Foundation of the USA (AST-0908326); the W. M. Keck foundation; the Macquarie University; the Netherlands Research School for Astronomy; the Natural Sciences and Engineering Research Council of Canada; the Slovenian Research Agency; the Swiss National Science Foundation; the Science \& Technology Facilities Council of the UK; Opticon; Strasbourg Observatory; and the Universities of Groningen, Heidelberg and Sydney. The RAVE web site is at \url{https://www.rave-survey.org}.

This work has also made use of data from {\it Gaia}-ESO based on data products from observations made with ESO Telescopes at the La Silla Paranal Observatory under programme ID 188.B-3002.\\

This work has made use of data from the European Space Agency (ESA) mission \textit{Gaia} (www.cosmos.esa.int/gaia), processed by the \textit{Gaia} Data Processing and Analysis Consortium (DPAC, www.cosmos.esa.int/web/gaia/dpac/consortium). Funding for the DPAC has been provided by national institutions, in particular the institutions participating in the \textit{Gaia} Multilateral Agreement. 

This work was partially supported by the Spanish Ministry of Science, Innovation and University (MICIU/FEDER, UE) through grant RTI2018-095076-B-C21, and the Institute of Cosmos Sciences University of Barcelona (ICCUB, Unidad de Excelencia ’Mar\'{\i}a de Maeztu’) through grant CEX2019-000918-M. FA acknowledges financial support from MICINN (Spain) through the Juan de la Cierva-Incorporcion programme under contract IJC2019-04862-I.

A.P.-V. acknowledges the DGAPA-PAPIIT grant IA103122.

TM acknowledges financial support from the Spanish Ministry of Science and Innovation (MICINN) through the Spanish State Research Agency, under the Severo Ochoa Program 2020-2023 (CEX2019-000920-S) as well as support from the ACIISI, Consejer\'{i}a de Econom\'{i}a, Conocimiento y Empleo del Gobierno de Canarias and the European Regional Development Fund (ERDF) under grant with reference  PROID2021010128."

CG and EFA acknowledge support from the Agencia Estatal de Investigación del Ministerio de Ciencia e Innovación (AEI-MCINN) under grant "At the forefront of Galactic Archaeology: evolution of the luminous and dark matter components of the Milky Way and Local Group dwarf galaxies in the Gaia era"  with reference PID2020-118778GB-I00/10.13039/501100011033 

CG also acknowledge support from the Severo Ochoa program through  CEX2019-000920-S

EFA also acknowleges support from the 'María Zambrano' fellowship from the Universidad de La Laguna.

\end{acknowledgements}


\bibliographystyle{aa}
\bibliography{releases_DR3_v2}
%%%%%%%%%%%%%%%%% APPENDICES %%%%%%%%%%%%%%%%%%%%%
\begin{appendix}

\section{Chemical clocks dependency}\label{chemoap}

\begin{figure*}
\includegraphics[width=17.5cm]{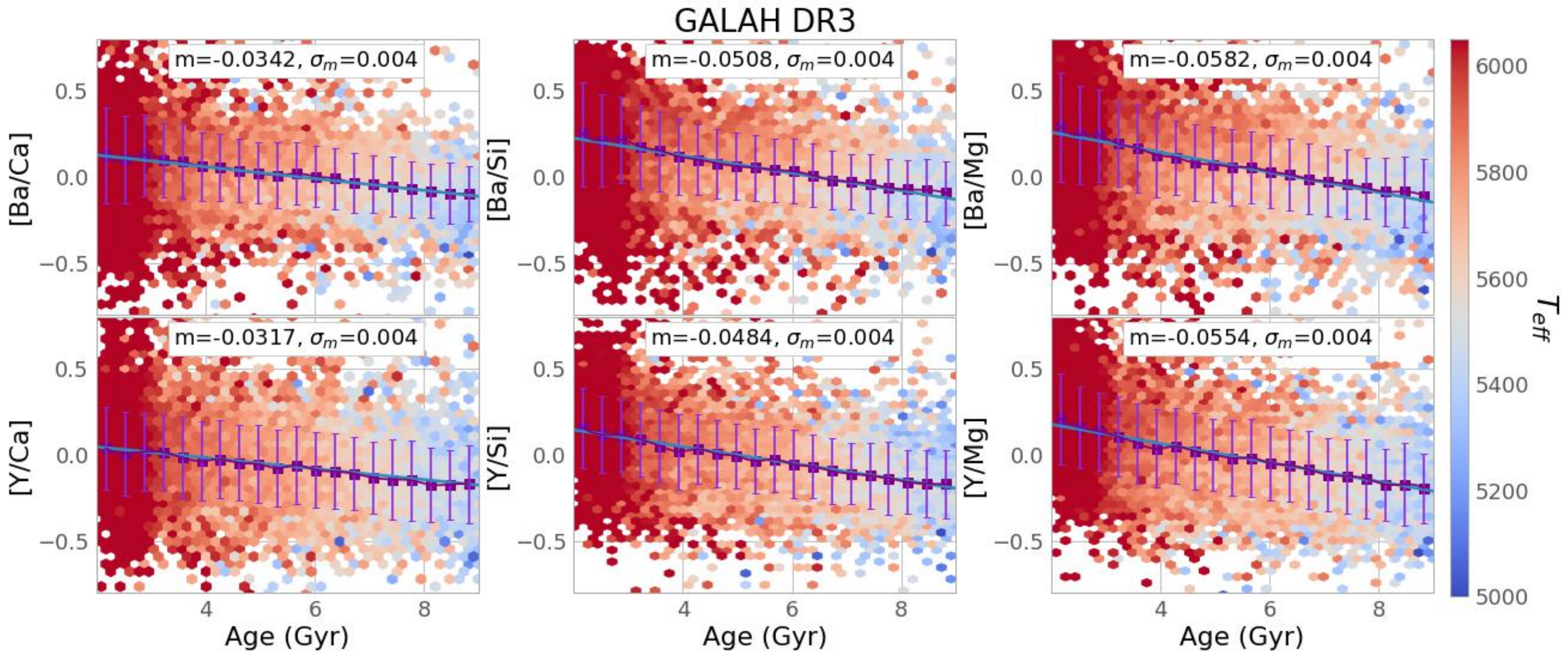}
\includegraphics[width=17.5cm]{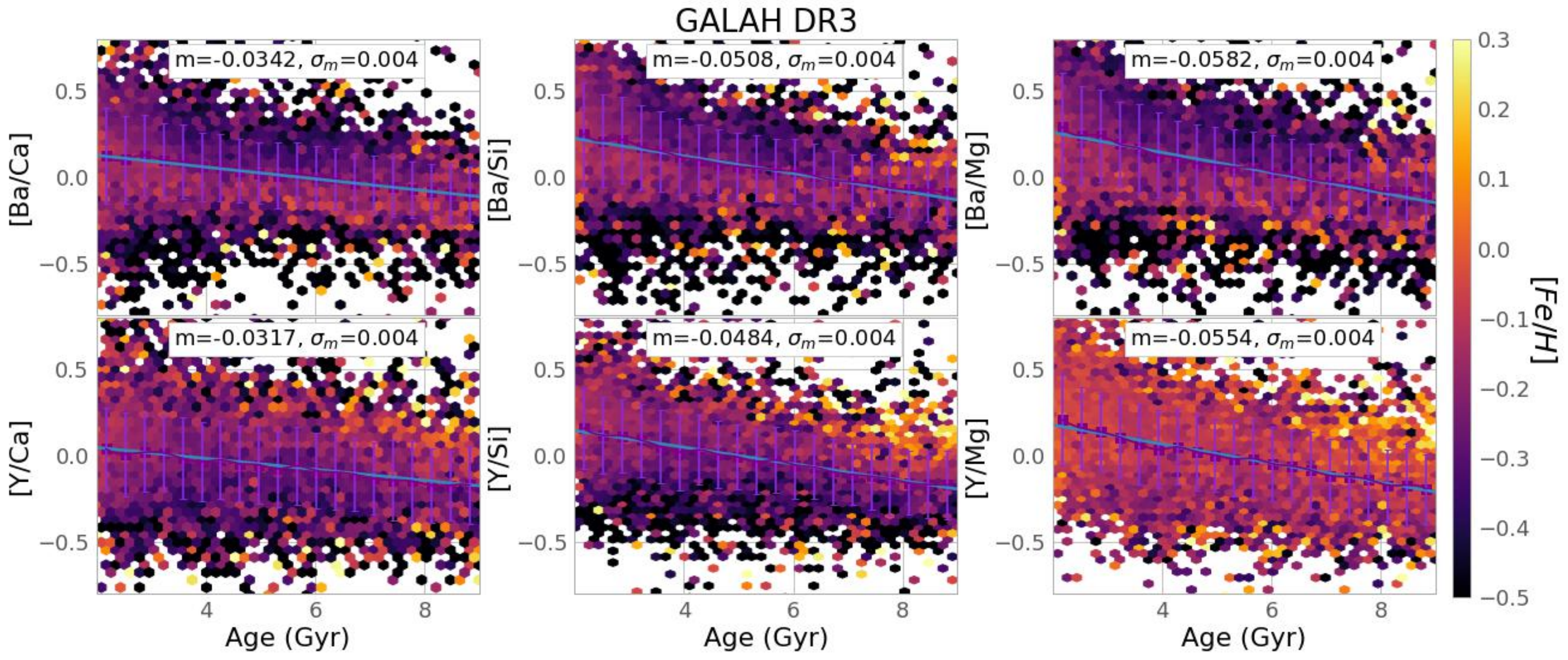}
\caption{[s/$\alpha$] abundance ratios vs. age for GALAH. The purple line shows the median abundance per age bin and the error bar represents one sigma deviation from the median as in Figure \ref{fig:galahclock} but now colour-coded by temperature upper panels and metallicity lower panels.}
\label{fig:galahclockdep}
\end{figure*}

\begin{figure*}
\includegraphics[width=17.5cm]{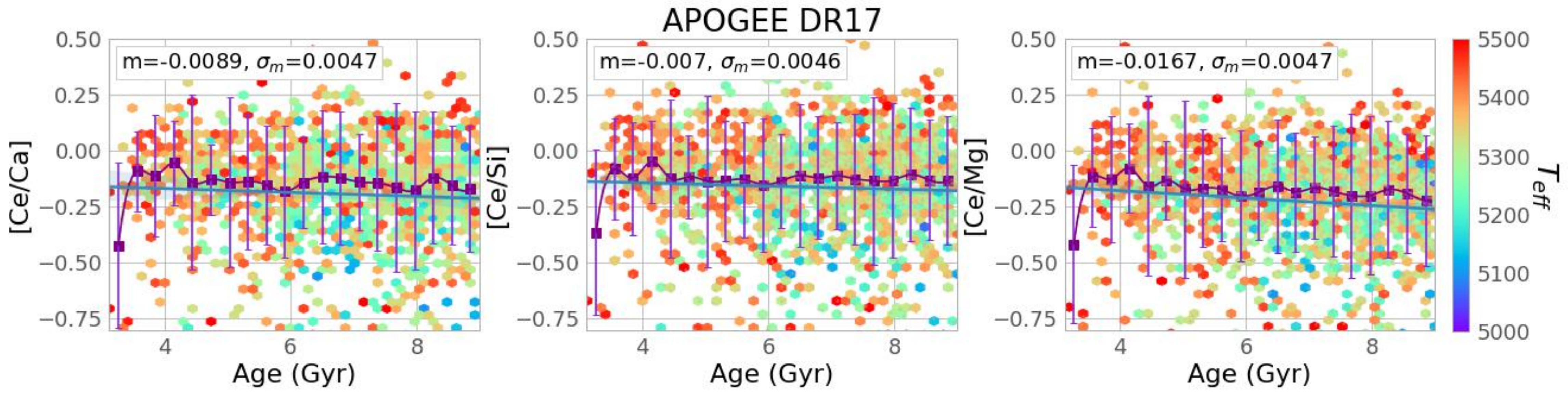}
\includegraphics[width=17.5cm]{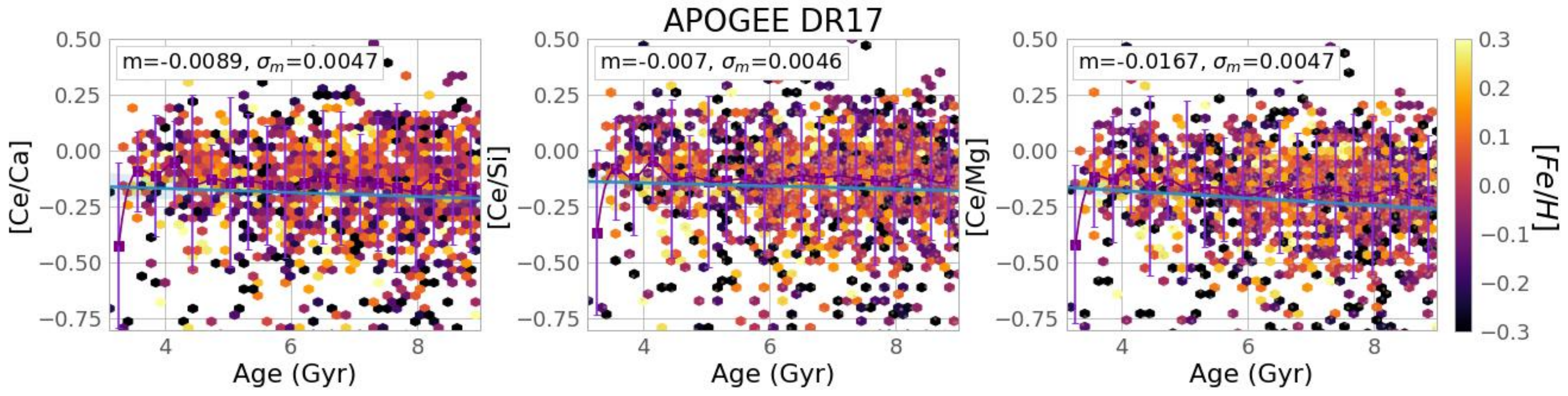}
\caption{[s/$\alpha$] abundance ratios vs. age for APOGEE. The purple line shows the median abundance per age bin and the error bar represents one sigma deviation from the median as in Figure \ref{fig:apogeeclock} but now colour-coded by temperature upper panels and metallicity lower panels.}
\label{fig:apogeeclockdep}
\end{figure*}

In Section \ref{agerel} we discussed the trends between ages and different abundances, and we derived chemical-clock relations based on a linear fit to GALAH and APOGEE [{\it s}-process/{$\alpha$}-process] abundance ratios. Temperature and metallicity can also influence the spectroscopic pipeline and therefore will have a impact in the uncertainty and precision of the derived age. We can expect that the dispersion around the trend we detected in Figures \ref{fig:galahclock} and \ref{fig:apogeeclock} would be an increasing function as temperature increases and metallicity decreases, since those stars have lines harder to detect. We show in Figures \ref{fig:galahclockdep} and \ref{fig:apogeeclockdep} that indeed the more metal-poor stars have a larger spread around the mean trend while this effect is not as clear in Temperature since this parameter has a stronger dependence with age. We refer the reader to other chemical-clock analysis where metallicity is also considered in the fitting procedure or only a certain range of metallicity is taking into account \citep[e.g.][]{Casamiquela2021, Viscasillas2022}. One can also notice that the metallicity dependence is way less strong for the [Y/Mg], which was already noticed by \citet{Nissen2020}.

\section{APOGEE DR17 abundances}\label{apoabu}

\begin{figure*}
\includegraphics[width=17.5cm]{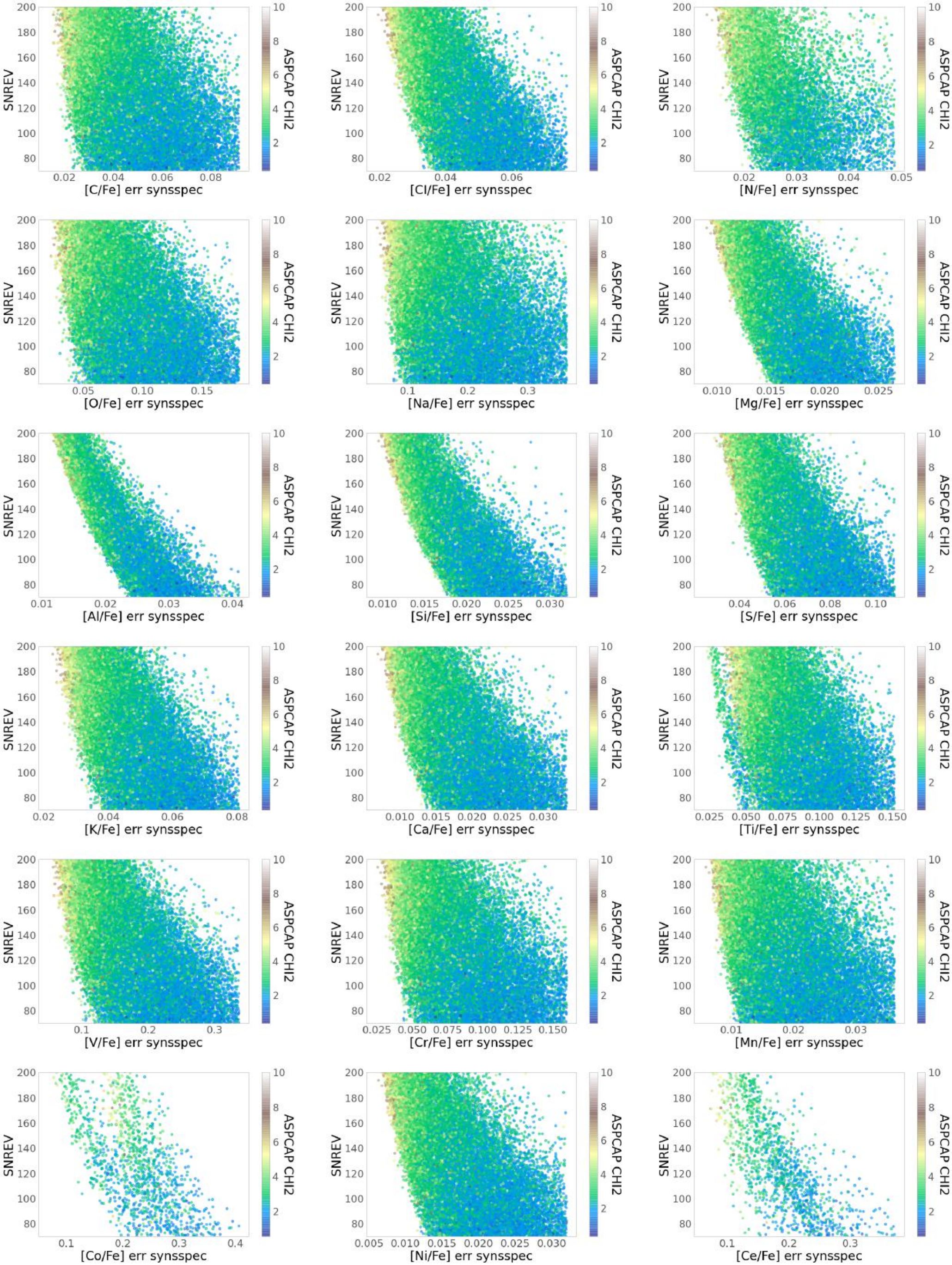}
\caption{Abundance uncertainty vs. signal-to-noise ratio ({\tt SNREV}) for each chemical species published in APOGEE DR17.}
\label{fig:snrsyns}
\end{figure*}

\begin{figure*}
\includegraphics[width=17.5cm]{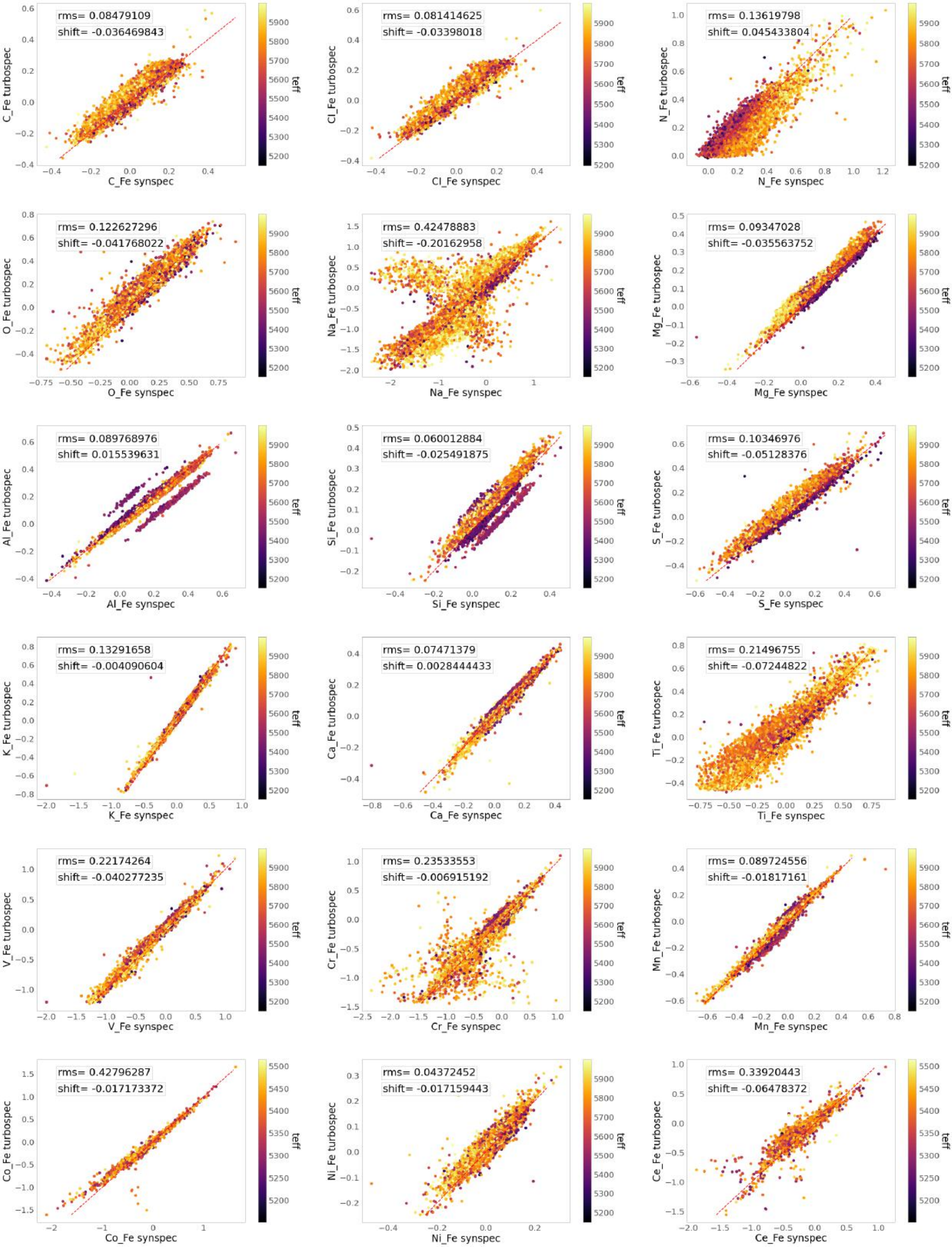}
\caption{Comparison between turbospec and synspec APOGEE DR17}
\label{fig:turbosyns}
\end{figure*}

In this section, we investigate the abundance uncertainties and the different stellar synthesis approaches used by the ASPCAP APOGEE DR17 pipeline \citep{Garcia-Perez2016, Jonsson2020}. Since the ASPCAP pipeline is primarily focused on and optimised for giant stars, we want to investigate how reliable the abundances used in this work are for the MSTO-SGB stars. We only show figures for the MSTO-SBG regime between temperatures of 5000K-6000K and cleaned by: SNREV $>$ 70, ASPCAP\_CHI2 $<$ 25, ASPCAPFLAG$=$0, STARFLAG$=$0, ELEM\_FE\_FLAG=0. We see that the uncertainties show in Figure \ref{fig:snrsyns} are mostly bellow 0.3 dex for [Mg/Fe],[Si/Fe],[Al/Fe], [Ca/Fe], [Mn/Fe] and [Ni/Fe], statistics is very low for [Co/Fe] and [Ce/Fe],  $< 3\,000$. The signal-to-noise, SNREV, is higher for smaller uncertainties, as expected. Still, the quality of the match with synthetic spectral models, ASPCAP\_CHI2, is worse for stars with low uncertainty, which might be an effect caused by the temperature range of these stars. 
Figure \ref{fig:snrsyns} compare the results from two different spectral synthesis codes available on the APOGEE DR17 release. The official release from APOGEE uses a new spectral synthesis code, Sysnpec \citep{Hubeny2017}, that can accommodate the effects of non-local thermodynamical equilibrium (non-LTE) for Na, Mg, K, and Ca \citep{Osorio2020}. Although Synspec allows for non-LTE calculations,  it uses the assumption of plane parallel geometry which is not entirely valid for large giant stars. While the previous synthesis code used in the APOGEE pipelines, TurboSpectrum \citep{Alvarez1998}, can use spherical geometry but cannot consider non-LTE effects. The figure \ref{fig:turbosyns} shows non-negligible differences for several elements. We see high spreads for [Na/Fe], [Ti/Fe] and [Cr/Fe], but since those have significant uncertainties, we did not include them in the scientific analysis of this manuscript. Non-LTE effects might be able to explain the differences between the codes, especially for [Na/Fe]. Still, a clear shift is seen for abundances such as [Si/Fe] and [Al/Fe] for temperatures colder than 5500 K, which could be an artefact in the derivation of the abundances. We, therefore, abstain from using [Si/Fe] and [Al/Fe] for temperatures cooler than 5500. [Ce/Fe] shows no concerning differences between the two spectral synthesis codes.

\section{Additional t-SNE analysis}\label{extratsne}

In this appendix, we add some additional illustrative plots to the combined t-SNE + HDBSCAN analysis performed in Sect. \ref{tsne} for each of the three SGB samples (GALAH, APOGEE, LAMOST).

In principle, there are a number of hyperparameters both in t-SNE and in the HDBSCAN methods that need to be chosen wisely. Apart from those, the main important choice for our work is the set of input parameters and the selection cuts. We optimised the number of input chemical abundances such that as many chemical elements as possible are used without significantly diminishing the total number of stars with useful abundances. Since t-SNE cannot treat missing data, all chosen chemical abundances have to be mutually available for each star in the final dataset.

\begin{figure*}
\includegraphics[width=14.5cm]{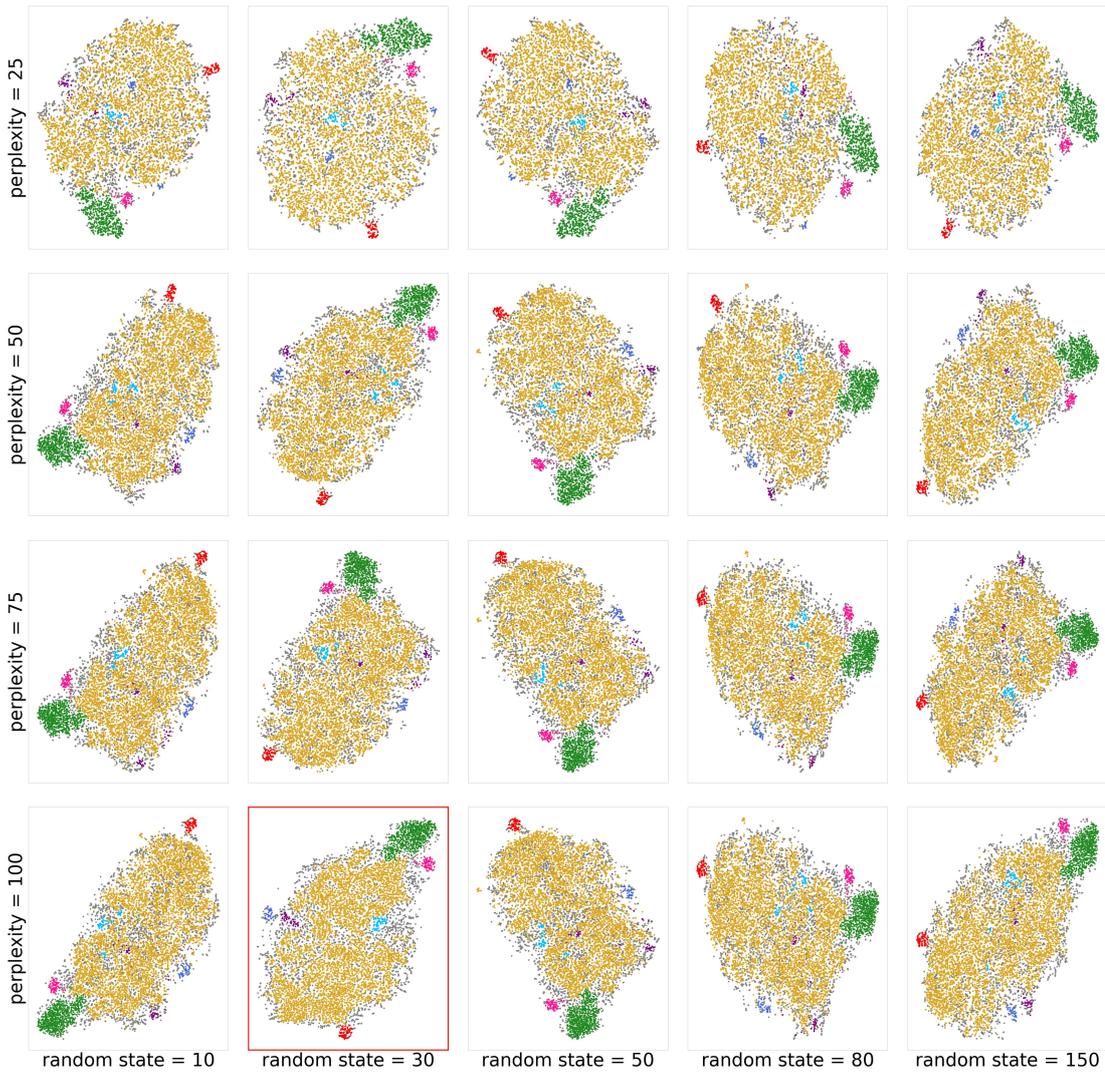}
\caption{The t-SNE projection for the GALAH DR3 SGB sample for different perplexity and random state values. The colours indicate the original groups we found previously and which are represented in Figure \ref{fig:tsnegalah}.}
\label{fig:perpgalah}
\end{figure*}

\begin{figure*}
\includegraphics[width=14.5cm]{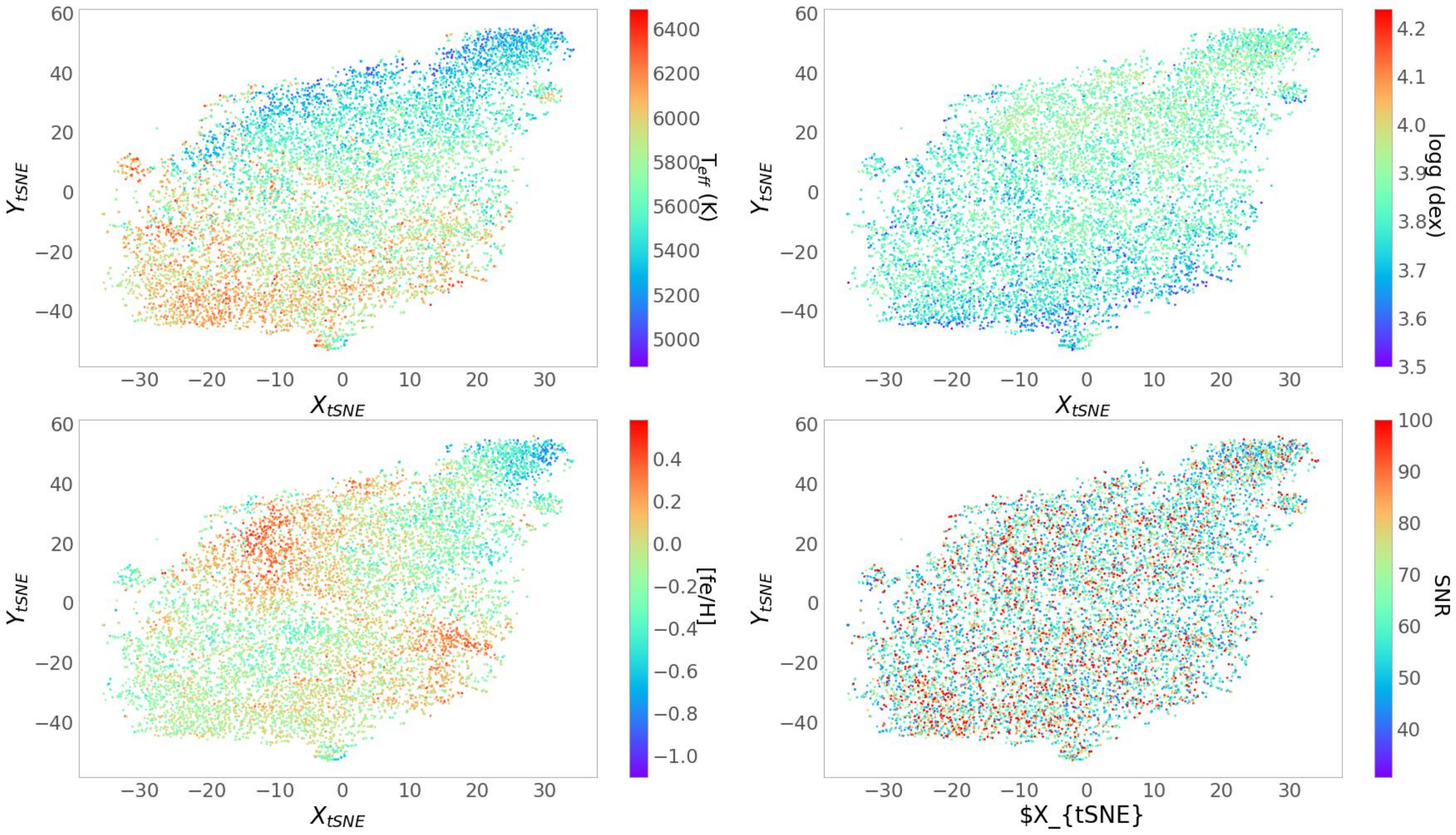}
\caption{t-SNE projection for GALAH DR3 SGB data colour-coded by $T_{\rm eff}$,$\log{g}$, metallicity and signal to noise.}
\label{fig:dependencygalah}
\end{figure*}

Secondly, we needed to choose a sensible configuration of hyperparameters both for t-SNE and HDBSCAN for each survey. 
In Figs. \ref{fig:perpgalah} through \ref{fig:dependencylamost} we show a few plots to explain the robustness of the groupings found when using the unsupervised machine learning technique, t-SNE \citep{Hinton2003, vanderMaaten2008}, in section \ref{tsne}. We have experience with the "perplexity" and the "random\_state" parameter to find the optimal t-SNE projection. The perplexity controls the number of nearest neighbours while the random state only influences the local minima of the cost functions, therefore, having a minor impact on the final projection. The random initialisation factor ("random\_state") plays a crucial role in validating the robustness of groupings in t-SNE space. We want to ensure that the groups we identify are stable against random variations. The results from Figures \ref{fig:perpapogee}, \ref{fig:perpgalah}, and \ref{fig:perplamost} demonstrate that the young $\alpha$-rich, thick disk, and thin disk stars remain tightly grouped together in the t-SNE space for different random\_state choices, even when the perplexities vary widely. This suggests that the groupings are robust and not affected by random variations.

However, the GALAH dataset exhibits less robustness as the cyan and purple marked groups of stars do not remain together in the t-SNE space across different random\_states and perplexities. On the other hand, the high-Ba stars marked in red and high-Ni peculiar stars marked in royal blue remain consistently grouped together in the t-SNE space when different t-SNE parameters are used. These findings suggest that care must be taken when interpreting the results we found as "cyan" and "purple" groupings since their validity is less robust across changes in t-SNE parameters.

After testing several perplexity values, we selected the one that best accentuates the separation between the two largest populations in the solar neighbourhood, namely the thick and thin disks. 

By choosing the optimal perplexity value, we were able to obtain a clear separation in a multi-dimentional chemical analysis. Our results show that mostly high values of perplexity are better suited for our datasets, especially for APOGEE and GALAH. For LAMOST high perplexity values do not improve the visualisation of the different populations.

We similarly chose the final HDBSCAN hyperparameters, first fixing a t-SNE configuration that visually shows two or more overdensities and then we experienced with different values for min\_cluster\_size, which controls the minimum size of the groupings, the min\_samples, which controls how conservative the clustering is and the cluster\_selection\_epsilon, which controls the separation distance between the groups, for more information on the HDBSCAN parameters we recommend the reader to the following page\footnote{\url{https://hdbscan.readthedocs.io/en/latest/parameter_selection.html}}. By testing HDBSCAN hyperparameters with a fixed t-SNE configuration and vice-versa, we have settled for the values described in Table \ref{tsneconfig}, which optimally separates the chemical thin and thick disks and also find other structures. The chrono-chemical groups found are also reproducible for many other hyperparameter combinations, which supports the groupings' robustness.

In Figures \ref{fig:dependencygalah}, \ref{fig:dependencyapogee}, \ref{fig:dependencylamost} we show the t-SNE projections colour-coded by different parameters. Since this method could project false groups due to artefacts in the chemical abundances derivation. In all cases, the projected t-SNE has no clear dependency on signal-to-noise, which controls the quality of the spectra. For LAMOST and GALAH there is  also no clear connection between the projected density and the surface gravity, while for APOGEE Figure \ref{fig:dependencyapogee} shows that low $\log{g}$ is preferentially at the bottom left of the map. The temperature and metallicity seem to influence the projections for all surveys. The metallicity is information given to the t-SNE method in the form of abundance ratios; therefore, we expect to find different clumps of metallicity across the new t-SNE dimension. This might also influence the temperature distribution since stars with a certain metallicity are easier to detect at certain temperatures.

\begin{figure*}
\includegraphics[width=14.5cm]{images/tsne_APOGEE_hdbscan.pdf}
\caption{t-SNE projection for the APOGEE DR17 SGB sample using different values of perplexity and random state. We use the HDBSCAN method for each panel on top of the t-SNE projection. The colours indicate the original groups we found previously and which are represented in Figure \ref{fig:tsneapogee}.}
\label{fig:perpapogee}
\end{figure*}

\begin{figure*}
\includegraphics[width=14.5cm]{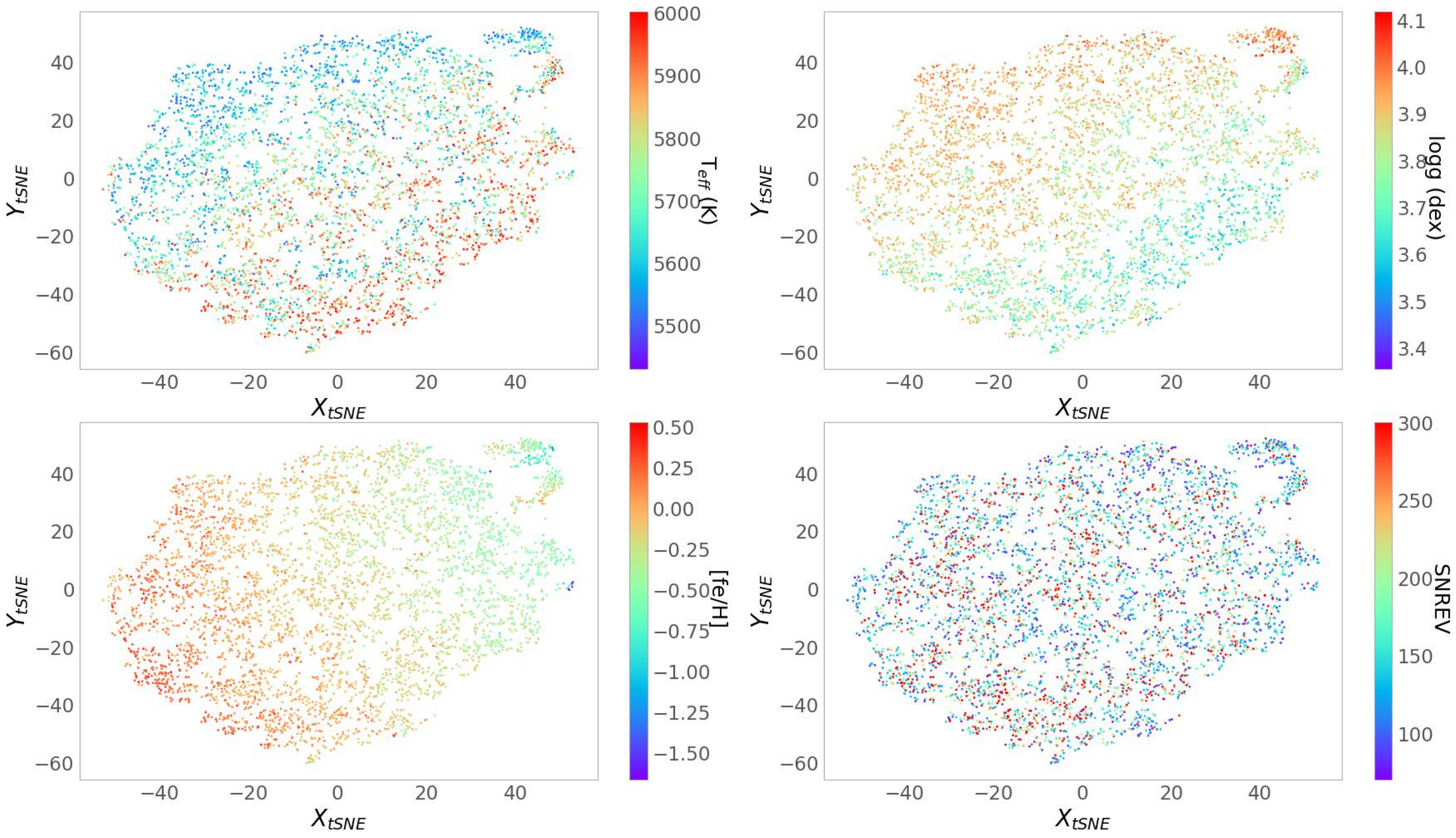}
\caption{Same as Figure \ref{fig:dependencygalah} but for APOGEE DR17 SGB sample.}
\label{fig:dependencyapogee}
\end{figure*}

\begin{figure*}
\includegraphics[width=14.5cm]{images/TSNE_LAMOST_DR7.pdf}
\caption{t-SNE projection for the LAMOST DR7 SGB sample using different values of perplexity and random state. We use the HDBSCAN method for each panel on top of the t-SNE projection. The colours indicate the original groups we found previously and which represented in Figure \ref{fig:tsnelamost}.} 
\label{fig:perplamost}
\end{figure*}

\begin{figure*}
\includegraphics[width=14.5cm]{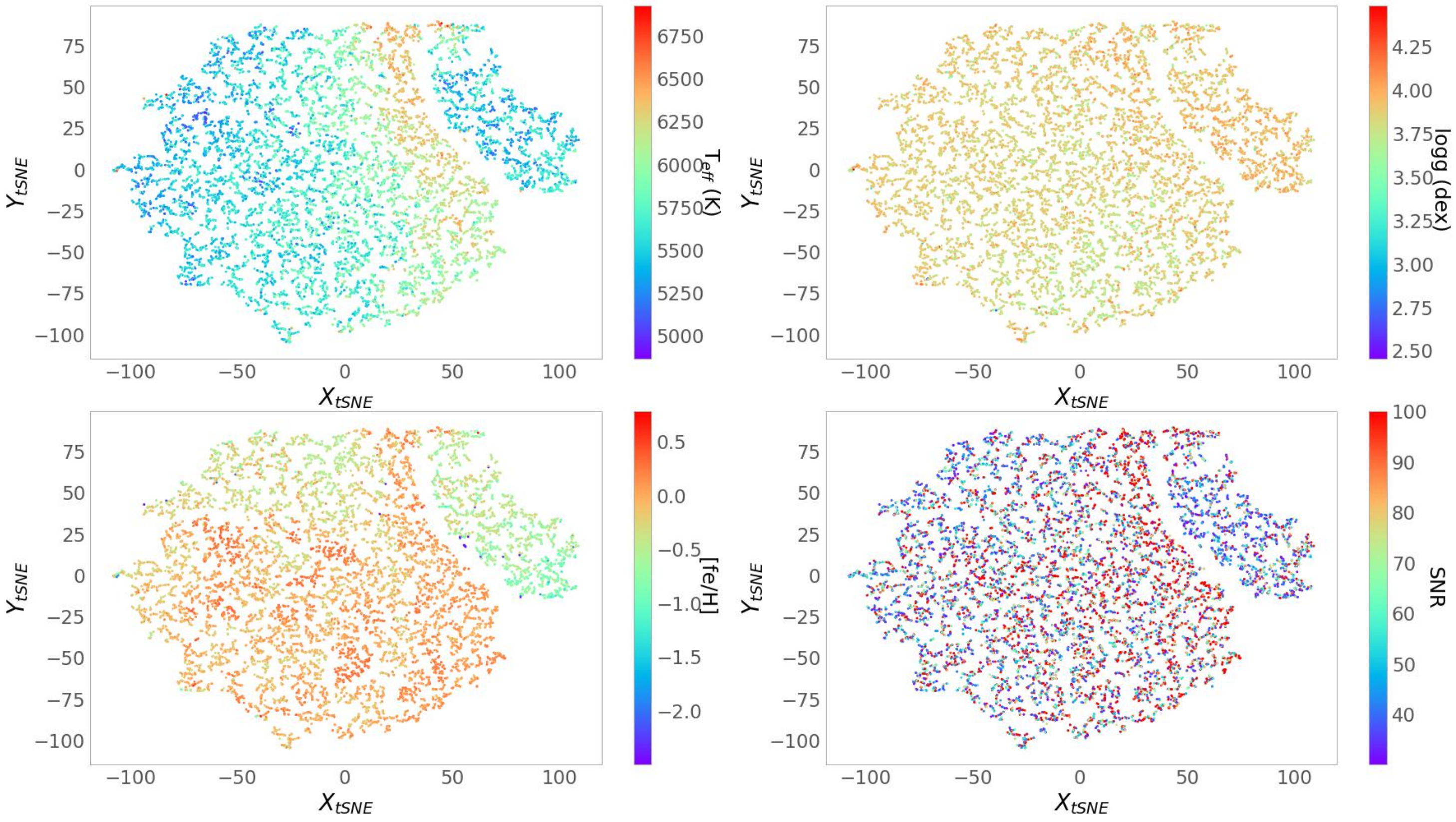}
\caption{Same as Figure \ref{fig:dependencygalah} but for LAMOST DR7 SGB sample.}
\label{fig:dependencylamost}
\end{figure*}

\subsection{t-SNE uncertainties}

\begin{figure}
\includegraphics[width=10cm]{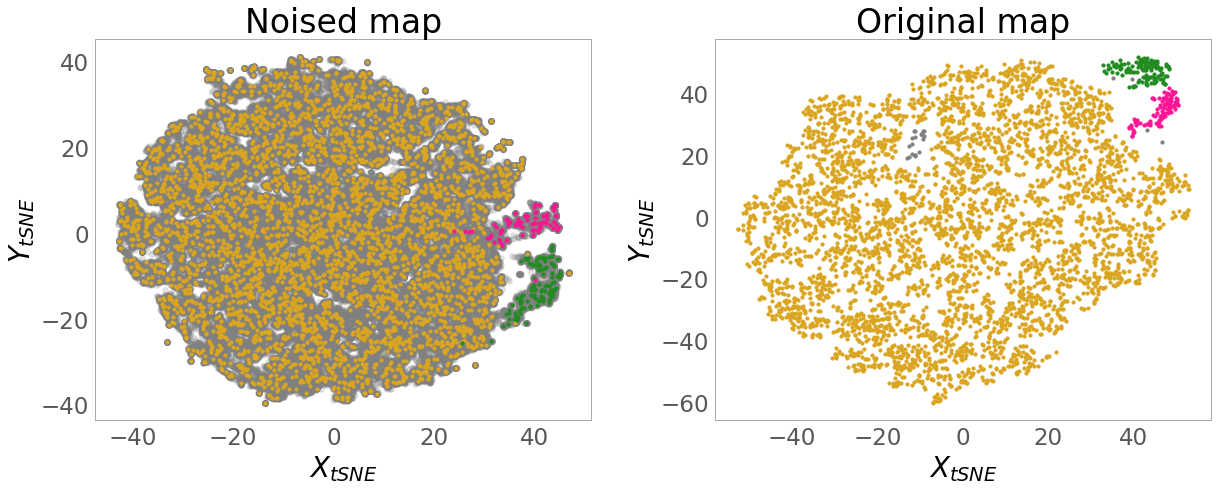}
\includegraphics[width=10cm]{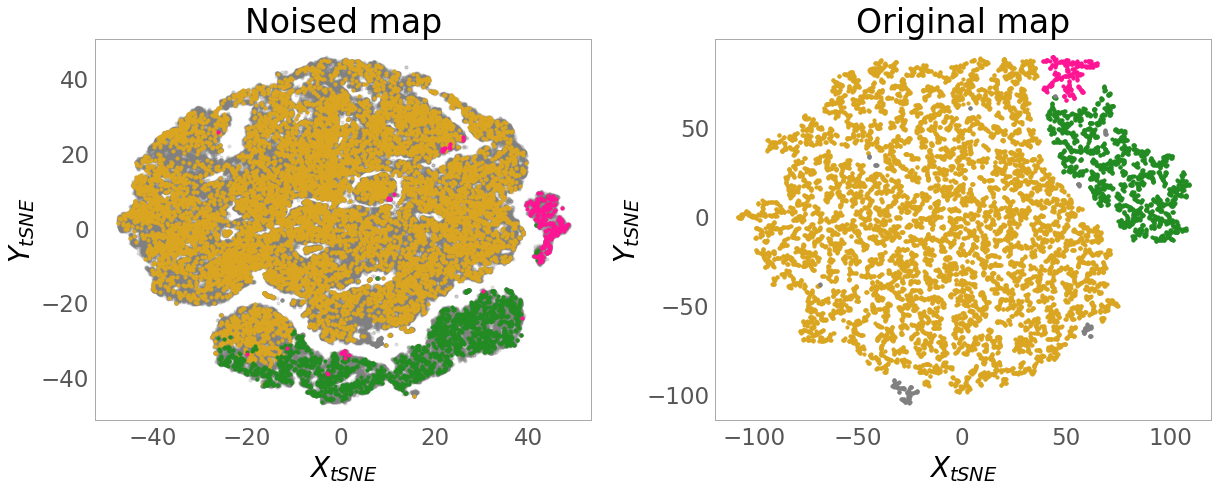}
\includegraphics[width=10cm]{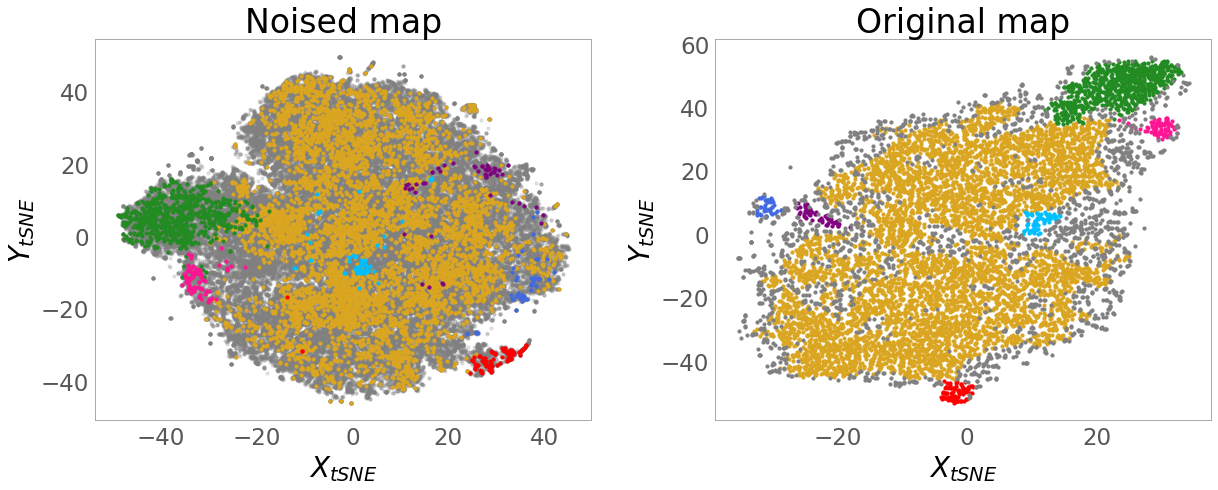}
\caption{t-SNE projections of noised data (produced by randomly adding Gaussian uncertainties to the abundances and ages of each star) vs. original data. Top row: APOGEE. middle row: LAMOST. botton row: GALAH.}
\label{fig:noiseapo}
\end{figure}

Manifold learning algorithms cannot treat observational uncertainties and also do not provide uncertainties associated to the produced mapping. They merely provide a projection of a high-dimensional dataset into a lower-dimensional space. We reinforce that the primary method behind our analysis is the dimensionality reduction technique, collapsing the multiple chemical-age spaces into a 2D visualization. We then use HDBSCAN to avoid delineating the overdensities by eye. 
Throughout the analysis, we have already had a critical validation of the method: We recover the thin disk, thick disk and young-$\alpha$-rich in three completely different surveys. \citet{Limberg2021,2022Ou} have used HDBSCAN alone to find groupings in kinematics space - and Monte-Carlo experiments were able to assign a certain probability of pertinence to the clusters. In our case, this is much more difficult since each random re-sampling of our data will result in a new projection space for t-SNE. Here we do a small exercise to test the robustness of the groups found by t-SNE+HDBSCAN. As in \citet{Anders2018}, see their Figure 5, we introduce noise to the data in a Monte-Carlo experiment test. We have sampled 20 random abundances and ages for each star using a Gaussian distribution centred in the abundance and age using its uncertainties as standard deviation. We have then run t-SNE on this increased random sample. In Figure \ref{fig:noiseapo}, we show the result of the new t-SNE projection in this "noisy" data as grey points and with the original tagged groups in their respective colours (See Figures \ref{fig:tsnelamost}, \ref{fig:tsneapogee} and \ref{fig:tsnegalah}) versus the original t-SNE map. For the thick, thin disk and young $\alpha$-rich groups, the overdensities are preserved even with the introduced noise in all three surveys. For GALAH, some populations get dispersed by the experiment, especially the young peculiar "navy-blue" group, while the outer disk "cyan" is in the middle of the thin disk group. The high barium stars are an overdensity that remains visibly separable from the thin disk main cloud. We want to stress also that adding noise to the data may artificially blur real signals. Despite the significance of the "Cyan", "Purple" and "Navy blue" groups being less robust, these findings are still an essential first step to the investigation of these populations in the Milky Way since they present some interesting features as the peaked age of the "Cyan" group.

\end{appendix}
\end{document}